\documentclass[a4paper,prd,twocolumn,showpacs,superscriptaddress,floatfix,nofootinbib]{revtex4-1}
\usepackage{graphicx}
\usepackage[normalem]{ulem}
\usepackage{float}
\usepackage{amsmath}
\usepackage{amssymb}
\usepackage{booktabs}
\usepackage{hhline}
\usepackage{multirow}
\usepackage[rightcaption]{sidecap}
\usepackage{hyperref}
\usepackage{xcolor}
\usepackage{wrapfig}
\usepackage{braket}
\usepackage{lineno}
\usepackage{orcidlink}
\usepackage{comment}
\usepackage{tabularx}
\usepackage{array}
\usepackage{times}

\def\rhs{\rho_s}
\def\rs{r_s}
\def\a{\alpha}
\def\b{\beta}
\def\g{\gamma}
\def\d{\delta}
\def\s{\sigma} 
\def\o{\omega}
\def\O{\Omega}
\def\m{\mu}
\def\n{\nu}
\def\ph{\phi} 
\def\non{\nonumber}
\def\n{\noindent}

\begin{document}
\title{Extreme-Mass-Ratio Inspirals Embedded in Dark Matter Halo: Existence of Homoclinic Orbit and Horizon-Induced Chaos}
\author{Surajit Das\,\orcidlink{0000-0003-2994-6951}}
\email{surajitdas@mail.ustc.edu.cn, surajit.cbpbu20@gmail.com (corresponding author)}
\affiliation{%
Department of Astronomy, School of Physical Sciences, \textcolor{blue}{University of Science and Technology of China}, Hefei, Anhui 230026, China}
\affiliation{%
CAS Key Laboratory for Researches in Galaxies and Cosmology, School of Astronomy and Space Science, \textcolor{blue}{University of Science and Technology of China}, Hefei, Anhui 230026, China}
\author{Surojit Dalui\,\orcidlink{0000-0003-1003-8451}}
\affiliation{%
Department of Physics, \textcolor{blue}{Shanghai University}, Baoshan District, Shanghai 200444, China}
\author{Bum-Hoon Lee\,\orcidlink{0009-0008-3322-2087}}
\affiliation{%
Department of Physics, \textcolor{blue}{Sogang University}, Seoul 04107, Korea}
\affiliation{%
Center for Quantum Spacetime, \textcolor{blue}{Sogang University}, Seoul 04107, Korea}
\affiliation{%
Department of Physics, \textcolor{blue}{Shanghai University}, Baoshan District, Shanghai 200444, China}
\author{Yi-Fu Cai\,\orcidlink{0000-0003-0706-8465}}
\email{yifucai@ustc.edu.cn (corresponding author)}
\affiliation{%
Department of Astronomy, School of Physical Sciences, \textcolor{blue}{University of Science and Technology of China}, Hefei, Anhui 230026, China}
\affiliation{%
CAS Key Laboratory for Researches in Galaxies and Cosmology, School of Astronomy and Space Science, \textcolor{blue}{University of Science and Technology of China}, Hefei, Anhui 230026, China}
\begin{abstract}
We study the existence of homoclinic orbit and the onset of chaotic motion for a massive particle moving around a Schwarzschild-like black hole embedded in a Dehnen-$(1,4,5/2)$ type dark matter halo, within the extreme-mass-ratio limit $q=m/M<<1$, where $m$ and $M$ are the masses of the particle and the central black hole, respectively. The presence of the halo modifies the spacetime curvature and consequently deforms the effective potential governing the particle’s motion. Using the Hamiltonian formulation, we derive the conditions under which unstable circular orbit and the associated homoclinic trajectory arise, marking the separatrix between bound and plunging motion. By analyzing the effective potential and the corresponding phase-space structure, we identify the transition from regular to chaotic dynamics in the near-horizon region. Numerical analyses through Poincaré sections and Lyapunov exponents calculations demonstrate that increasing the halo density, scale radius along with energy amplifies nonlinear effects which leads to chaos eventually. We demonstrate that within a dark matter halo environment, the dynamical stability of particle motion can be significantly altered without violating the universal surface gravity bound on chaos. This work provides a deeper understanding of horizon-induced chaos in astrophysically realistic environments and serves as a theoretical basis for exploring its possible imprints on gravitational wave signals in extreme-mass-ratio inspirals system.
\end{abstract}
\pacs{04.20.-q, 04.25.-g, 04.25.Nx, 04.30.Db, 04.70.-s}
\keywords{Black hole, Dark matter halo, Chaos}
\maketitle
\section{Introduction}\label{s1}
Black holes (BHs), once purely theoretical predictions of general relativity, have become firmly established astrophysical realities through a series of spectacular observations. The direct detection of gravitational waves (GWs) from binary black hole mergers by the LIGO and Virgo collaborations  \cite{LIGOScientific:2016emj,TheLIGOScientific:2017qsa,LIGOScientific:2018mvr,LIGOScientific:2020ibl,LIGOScientific:2021usb,LIGOScientific:2025rid,KAGRA:2021vkt} and the imaging of the black hole shadow at the centers of M87* and SgrA* by the Event Horizon Telescope \cite{EventHorizonTelescope:2019dse,EventHorizonTelescope:2019ggy,EventHorizonTelescope:2019ths,EventHorizonTelescope:2022wkp} have transformed our understanding of gravity in the strong-field regime. These discoveries resulted in a period of remarkable precision in the study of the near-horizon structure of spacetime and the dynamics of matter and radiation in its vicinity. Such observations now motivate theoretical efforts to understand the complex motion of particles and fields near BHs, where the inherent nonlinearity of general relativity gives rise to rich and often chaotic phenomena.\\
One of the direct ways for investigating the geometry of spacetime is the study of geodesic motion around BHs. Test particle trajectories not only encode the gravitational field structure but also help to reveal subtle dynamical transitions such as stable and unstable circular orbits, separatrices, and plunge behavior. In the case of Schwarzschild or Kerr geometries, the equations of motion are completely integrable due to the existence of conserved quantities associated with time-translation, axial, and hidden symmetries \cite{Chandrasekhar:1985kt,Carter,Frolov:2017kze}. However, in the realistic astrophysical scenarios such as in the presence of external fields, dark matter distributions or deviations from the exact symmetry can break the integrability. Once the symmetry structure is perturbed, the dynamics of the particle motions become nonlinear and chaos can naturally emerge. Within the classical framework, a black hole (BH) horizon's influence on particle trajectories and inducing chaos has a rich history in these preceding works \cite{Suzuki:1999si,Kiuchi:2004bv,Bombelli:1991eg,Sota:1995ms,Vieira:1996zf,Suzuki:1996gm,Cornish:1996ri,deMoura:1999wf,Hartl:2002ig,Han:2008zzf,Takahashi:2008zh,Hashimoto:2016dfz,Li:2018wtz,Lei:2020clg,DeFalco:2020yys,DeFalco:2021uak,Dalui:2018qqv,Dalui:2019umw,Bera:2021lgw,Das:2024iuf,Lu:2026kcm}, which shows that this topic has promising ground for huge exploration.\\
Before entering into the stage of chaos, one particular important feature in this context is the investigation of the homoclinic orbit, which serves as a critical boundary between bound and plunging trajectories \cite{Bombelli:1991eg,Levin:2008yp,Perez-Giz:2008ajn,Jeong:2023hom,Li:2023bgn,Ciou:2025ygb}. Such an orbit asymptotically approaches the same unstable circular orbit (hyperbolic fixed point) both in the infinite past and future and therefore corresponds to the separatrix in the phase space. The existence of a local maximum in the effective potential is a necessary condition for such a homoclinic closed trajectory. Therefore without any external disturbance, this orbit forms a single, well-defined loop. However, one of the consequences for the introduction of a small perturbation near the unstable equilibrium causes a deformation this structure, indicates the loss of integrability and marks the onset of chaotic dynamics \cite{Bombelli:1991eg,Jeong:2023hom,Polcar:2019kwu} (For a more comprehensive details of this topic, the reader is directed to \cite{KAM,Wigg}). Therefore, the study of the homoclinic orbit provides a natural framework for identifying the transition from regular to chaotic motion in curved spacetime.\\
In relativistic regime, this chaotic dynamics is not merely of mathematical interest, it carries direct physical implications. For instance, in extreme-mass-ratio inspiral (EMRI) systems, where a stellar mass compact object orbiting a supermassive BH, even slight variations in orbital motion can cause detectable modulations in the emitted GWs \cite{Hughes:2000ssa,Babak:2017tow,Amaro-Seoane:2012lgq}. Since EMRIs are among the prime targets for future space-based GW detectors such as LISA \cite{LISA:2017pwj}, TianQin \cite{TianQin:2020hid}, Taiji \cite{Hu:2017mde} and others \cite{Wang:2024tnk,CSST:2025ssq}, understanding the phase-space structure and the possible chaotic transitions near the horizon is essential for accurate waveform modeling. The presence of environmental structures like accretion disks, magnetic fields, or dark matter (DM) halos further enriches this dynamical landscape. In particular, the inclusion of a DM halo surrounding the central BH, provides a more realistic galactic setting and offers a natural perturbation that can alter the stability of orbits, the morphology of homoclinic separatrices.\\
Astrophysical and cosmological observations strongly suggest that most galaxies, including our Milky Way, are embedded in massive DM halos \cite{Bertone:2004pz,Bovy:2013raa,deSwart:2017heh,Wechsler:2018pic,Bertone:2018krk}. Such halos can be modeled by various density profiles \cite{Dubinski:1991bm,Hernquist:1990be,Navarro:1995iw,Navarro:1996gj,Burkert:1995yz,Salucci:2000ps,Jaffe:1983iv,Tremaine:1993qb,Dehnen,Zhao:1995cp,Dutton:2014xda,Graham:2005xx,Urena-Lopez:2002ptf,Harko:2011xw,Begeman:1991iy,Moore:1999gc,Retana-Montenegro:2012dbd,Navarro:1994hi,Schive:2014dra,Shen:2009my}. Among these models, the Dehnen profile \cite{Dehnen} represents a family of spherical mass distributions that are characterized by the $(1,4,\gamma)$ model. This profile offers a highly adaptable framework for modeling both the inner and outer structural properties of DM halos in spherical galaxies and bulges. Its generalized formulation is capable of reproducing a variety of density distributions, including both cuspy and cored profiles. Consequently, it is well-suited for fitting a broad spectrum of observational data, ranging from dwarf galaxies to massive galaxy clusters \cite{Graham:2005xx,Dehnen,shukirgaliyev2021bound}. Several composite black hole-dark matter (BH-DM) halo solutions exist under the consideration of Dehnen-$(1,4,\g)$ model. For example, in Ref.~\cite{Pantig:2022whj} a BH solution with the Dehnen-$(1,4,\g)$ profile is constructed for the study of an ultrafaint dwarf galaxy. Furthermore, in Refs.~\cite{Gohain:2024eer,Al-Badawi:2024asn,UktamjonUktamov:2025emm} a combined BH-DM halo spacetimes are constructed, surrounded by $(1,4,0),~(1,4,5/2)$ and $(1,4,2)$ type DM halo profiles, respectively. On the other hand, recently Konoplya et al. has been proposed a regular BH geometries corresponding to Einasto and Dehnen-type DM profiles with the phenomenological freedom in halo modeling \cite{Konoplya:2025ect}. In this work, we consider a particular DM halo model for which $\g=5/2$ \cite{Al-Badawi:2024asn}. In particular, the Dehnen-$(1,4,5/2)$ profile, exhibits a steep central cusp and provides an excellent fit to luminous elliptical galaxies \cite{Dehnen,shukirgaliyev2021bound}. In recent research, the influence of Dehnen-$(1,4,5/2)$ type DM halos on BHs has been extensively investigated from a variety of distinct perspectives \cite{Alloqulov:2025ucf,Liang:2025vux,Luo:2025xjb,Rani:2025esb,Xamidov:2025prl,Ashoorioon:2025ezk,Li:2025ver,Alloqulov:2025edn}. Therefore the presence of a Schwarzschild-like BH surrounded by DM as a halo of a BH-DM combined system modifies the surrounding spacetime through additional gravitational potential terms, thereby changing the locations of horizons, photon spheres, and stable circular orbits. Consequently, the effective potential experienced by test particles develop richer structures, often containing extrema conditions favorable for the appearance of unstable orbits and chaotic motion.\\
In this work, we investigate homoclinic orbit and the onset of chaos for a massive probe particle moving near a Schwarzschild-like BH immersed in a $(1,4,5/2)$ type of Dehnen DM halo within the extreme mass ratio limit. Our goal is to understand how the DM halo parameters (core density $\rho_s$ and scale radius $r_s$) influence the structure of the effective potential, hence the existence of homoclinic orbit, as well as to investigate the DM halo distribution around supermassive black holes (SMBHs) through a detailed examination of the transition from regular to chaotic orbital dynamics. However, the principal objective of this work is not to explore the emergence of chaos in the vicinity of a homoclinic orbit resulting from frequency perturbations of the background spacetime. Instead, our focus is to examine how a probe particle, orbiting within this BH and DM halo system, reacts to the presence of the event horizon. To investigate the horizon's influence, we analyze the dynamics of a relativistic particle that is subject to an external potentials \cite{Hashimoto:2016dfz,Lei:2020clg,Das:2024iuf,Dalui:2019umw,Bera:2021lgw} and is moving around a BH embedded in a DM halo environment. As a consequence, this allows us to explore bounded motion near the BH horizon and identify how small perturbations evolve under the combined influence of BH and the surrounding DM field. We analyze the phase-space dynamics through Poincaré sections, quantify sensitivity to initial conditions using Lyapunov exponents, and discuss how chaos emerges as a function of the halo parameters.\\
The significance of this study lies in several aspects. First, it provides a dynamical characterization of the existence of homoclinic orbit and chaos in a realistic galactic BH environment, bridging the gap between idealized vacuum solutions and astrophysical systems. Second, it establishes the connection between geodesic stability and chaotic evolution, thereby identifying the parameter thresholds where the system transitions from integrable to chaotic regimes. Finally, since the same near-horizon region governs the emission of gravitational radiation in EMRI systems, the results presented here serve as a theoretical foundation for our companion work, where we explore the signature of chaos in gravitational wave in such BH-DM halo configuration \cite{partII}.\\
This paper is organized as follows. In Sec.~\ref{sec:background}, we review the construction of the Schwarzs\-child-like BH solution surrounded by a $(1, 4, 5/2)$ type DM halo. In Sec.~\ref{sec:homoclinic}, we analyze the geodesic motion and derive the conditions for the existence of homoclinic orbit. We have derived the dynamical equations of motion of a test particle analytically in Sec.~\ref{sec:eqn}, followed by a numerical investigation on chaotic dynamics through Poincaré maps and Lyapunov exponents, highlighting how halo parameters affect the onset of chaos in Subsec.~\ref{sec:numerical}. Finally, Sec.~\ref{sec:conclusion} summarizes our findings and discusses their astrophysical implications, especially in connection with the dynamics of EMRIs and potential observational consequences in future space-based GW missions.

\section{Schwarzschild black hole immersed in a dark matter halo}\label{sec:background}
Considering a static spherically symmetric (SSS) metric immersed in a Dehnen-$(1,4,5/2)$ type DM halo in 4D spacetime having the form,
    \begin{equation}
		ds^2=-f(r)dt^2+\frac{dr^2}{f(r)}+r^2d\Omega^2~,\label{3}
	\end{equation}
with $d\Omega^2=d\theta^2+\sin^2{\theta}d{\phi}^2$ is the line element of a unit $2D$ sphere. For the above equation, the metric solution of $f(r)$ up to a leading order is given by \cite{Al-Badawi:2024asn}
    \begin{equation}
        f(r)=1-\frac{2M}{r}-32\pi\rhs\rs^2\sqrt{1+\frac{\rs}{r}}~,\label{4}
    \end{equation}
where $M$ is the typical mass of the BH. Recently, the author of Ref.~\cite{Bolo} argues that the aforementioned solution does not represent a consistent solution to the Einstein field equations, and introduces a new BH solution embedded within a Dehnen-$(1,4,5/2)$-type DM halo. In the present work, we confine our analysis to the solution given in Eq.~\eqref{4}, primarily due to the fact that this solution (Eq.~\eqref{4}) is indeed an exact solution of Einstein field equations, as already established by Al-Badawi et al. (see \cite{Al-Badawi:2024asn} and the references therein) and also due to the existence of a number of well-established literature based on this particular solution (Eq.~\eqref{4}) at the leading order approximation \cite{Xie:2025udx,Alloqulov:2025ucf,Liang:2025vux,Luo:2025xjb,Rani:2025esb,Xamidov:2025prl,Ashoorioon:2025ezk,Li:2025ver,Alloqulov:2025edn}. For a brief technical details required to derive the BH solution, given in Eq.~\eqref{4}, along with a brief discussion on the DM halo density profile, readers refer to see the Supplemental Material.\\
On the other hand, several important and relevant points must be highlighted when comparing the  Schwarzschild-like BH solution, embedded within a Dehnen-$(1,4,5/2)$-type DM halo in this paper, with other distinct classes of solutions available in the literature that are based on different methodologies. In this regard, readers are directed to the section titled ``A brief discussion of our considered model with others existing in the literature'' in our Supplemental Material, where the key differences between the model examined in this paper and other existing solutions in the literature are presented.\\
As an additional comment, let us mention that the Dehnen-type DM halo profile in our context, which is especially characterized by an inner slope for the parameter $\g=5/2$ as a steep cusp, modifies the spacetime geometry around the BH, leading to observable deviations from the vacuum Einsteins' general relativity predictions. Therefore it will be interesting to explore how DM halo profile parameters (central density and scale radius) affect the dynamics of a probe particle from an EMRI system, which we will forward to discuss in the later sections.

\section{Homoclinic orbits around a Schwarzschild black hole-dark matter halo combined spacetime}\label{sec:homoclinic}
\begin{figure*}[htbp]
    \centering
    \includegraphics[width=0.9\linewidth]{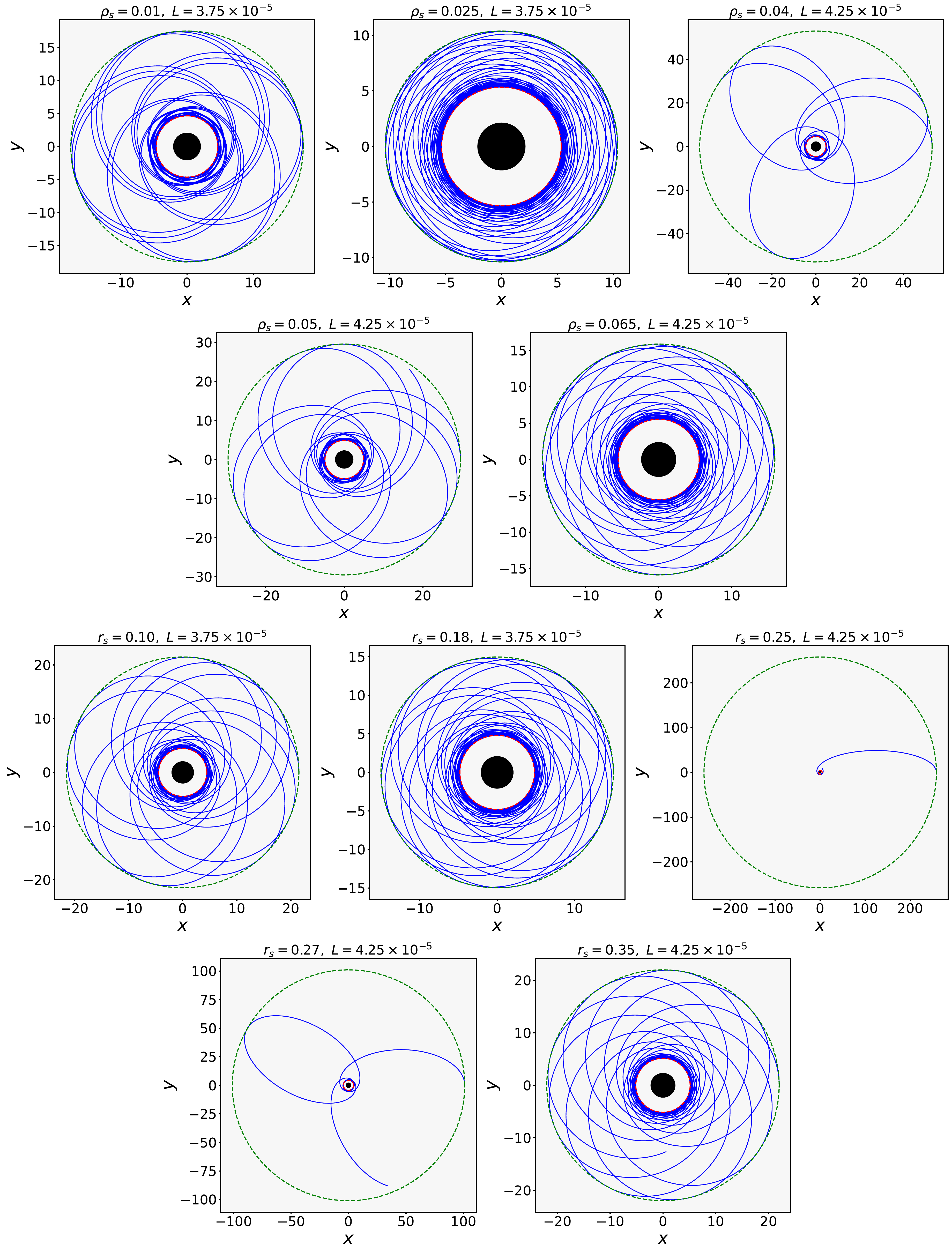}
    \caption{A segment of a homoclinic trajectories is illustrated for various DM halo parameters ($\rhs$, $\rs$) in the spacetime of a Schwarzschild BH surrounded by a Dehnen-type DM halo, within the EMRI limit. The analysis is conducted for two distinct values of angular momentum: $L = 3.75\times10^{-5}$ and $L = 4.25\times10^{-5}$. The boundaries of the unstable circular orbits, $r_{\rm un}$, and the maximum attainable radii, $r_{\rm max}$, are indicated by the red and green dotted circles, respectively. The central solid black circle denotes the location of the Schwarzschild BH surrounded by the DM halo. Note that the radius of the BH is depicted individually for each specific case in the parameter set.}
    \label{f:homo_orbits}
\end{figure*}

In this section, we analyze particle motions as the homoclinic orbit around a Schwarzschild BH embedded within a Dehnen-type DM halo, considering an EMRI system. Within the reduced two-dimensional phase space that describes the radial motion of a relativistic particle with a conserved energy and angular momentum, we explore that a homoclinic orbit is present. We explicitly derive the form of this orbit for various central density and scale radius of the DM halo. Starting from the covariant dispersion relation, we introduce $x:\equiv 2M/r$ to rewrite the effective potential and energy. We then locate the extremal points of this potential in a Schwarzschild-like BH–DM halo spacetime. Because angular momentum $L$ must lie between the values defining the innermost stable circular orbit (ISCO) and the marginally bound orbit (MBO) for a homoclinic orbit to exist, we derive the conditions for the ISCO and MBO. This yields the allowed ranges of $L$ and $E$ as the DM halo parameters vary. Detailed derivations, analysis, and figures are given in the Supplemental Material.\\
In Particular, we display a complete set of numerical results detailing the unstable points, maximum points, stable points, and the associated energies corresponding to homoclinic orbits of a Schwarzschild-like BH with a Dehnen-type DM halo in Table~S1 and Table~S2 (see the Supplemental Material). These values are computed for various DM halo parameters, $\rhs$ and $\rs$, for a given angular momentum $L$. The data in Table~S1 is organized for parameter pairs of $(\rhs, L)$, while Table~S2 presents results for pairs of $(\rs,L)$.\\
Now using the data from Table~S1 and Table~S2, we numerically integrate Eq.~(s20) to generate homoclinic orbits in the $(x=r\cos\phi,~y=r\sin\phi)$ plane. These orbits are plotted for various values of the DM halo's central density $\rhs$ and scale radius $\rs$, while the angular momentum is held fixed at $L=3.75$ and $L=4.25$ (in units of $10^{-5}$). The resulting plot is displayed in Fig.~\ref{f:homo_orbits}. On the other hand, the segment of a homoclinic orbit in the absence of DM halo parameters (i.e., $\rhs=\rs=0$) is plotted in Fig.~S4 (see the Supplemental Material).\\
For a comprehensive analysis of homoclinic orbits in our context, see the ``Detailed analysis on homoclinic orbits" in the Supplemental Material.

\vspace{-20pt}
\section{Horizon-induced chaotic dynamics of a massive object around a Schwarzschild black hole-dark matter halo combined spacetime}\label{sec:dynamics}
To focus on the dynamics of a massive probe particle near the event horizon of a Schwarzschild BH that is embedded in a Dehnen-type DM halo, we begin by defining its Lagrangian, subject to an external potential $V(x^\alpha)$ of a massive test particle of mass $m$ (see ``Lagrangian of a massive probe in curved background" of our Supplemental Material for details). 

\noindent
We are particularly interested to observe the effects of the DM halo parameters on a massive probe particle, which is located very close to the event horizon of a Schwarzschild BH immersed in a DM halo in galaxies. Therefore we employ a better choice of coordinates, known as Painlev\'e-Gullstrand coordinates \cite{PG}, in the SSS BH-DM spacetime (Eq.~\eqref{3}) (see the section ``A better choice of coordinates in our analysis" of the Supplemental Material for more details).

\subsection{Equations of motion of a massive probe particle}\label{sec:eqn}
We derive the equations of motion governing a massive test particle moving in the background of a combined BH-DM halo spactime (Eq.~\eqref{3}), but in a choice of Painlev\'e-Gullstrand coordinates, which is discussed in the section ``Analytical equations of motion" of the Supplemental Material.\\
Correspondingly following Eq.~(s35) from the Supplemental Material, we obtain the following dynamical equations of motion:
    \begin{eqnarray}
        \frac{dr}{dt} &=& \frac{\partial E}{\partial p_r} = -\sqrt{1-f(r)} + \frac{p_r}{\sqrt{p^2_r + \frac{p^2_\phi}{r^2}+m^2}}~,\label{3.8} \\
        \frac{dp_r}{dt} &=& -\frac{\partial E}{\partial r} = -\frac{\partial_rf(r)}{2\sqrt{1-f(r)}}p_r + \frac{p_\phi^2/r^3}{\sqrt{p_r^2 + \frac{p_\phi^2}{r^2}+m^2}} \nonumber \\
        && - \partial_ra(r)~,\label{3.9} \\
        \frac{d\phi}{dt} &=& \frac{\partial E}{\partial p_\phi} = \frac{p_\phi/r^2}{\sqrt{p_r^2 + \frac{p_\phi^2}{r^2}+m^2}}~,\label{3.10} \\
        \frac{dp_\phi}{dt} &=& -\frac{\partial E}{\partial \phi} = -\partial_\phi b(\phi)~.\label{3.11}
    \end{eqnarray}    
The equations presented above will serve as the foundation for our subsequent numerical analysis in the later sections.  For simplicity in our calculations, we have made the reasonable approximation that coupling effects between the external potentials $a(r),b(\phi)$ (see the Supplemental Material for the choices of $a(r),b(\phi)$) and the BH-DM halo spacetime geometry are sufficiently minor to be neglected.

\subsection{Numerical analysis on chaotic dynamics}\label{sec:numerical}
Now we explore how the event horizon of a Schwarzschild BH embedded in a Dehnen-type DM halo along with the halo parameters influence the emergence of chaotic dynamics for a massive probe particle. To achieve this, we study the phase space dynamics along with its orbital trajectories of a probe particle within the background framework, mentioned in Sec.~\ref{sec:background}. This investigation primarily involves analyzing the nonlinear dynamics governed by Eqs.~\eqref{3.8}, \eqref{3.9}, \eqref{3.10}, and \eqref{3.11}, along with the associated Poincar\'{e} sections, orbital evolutions and Lyapunov exponents of the system. For the convenience of the readers, we present here only the results of our analyzed Poincar\'{e} sections and Lyapunov exponents. The detailed theoretical background on Poincar\'{e} sections, orbital evolutions, and Lyapunov exponents can be found in our Supplemental Material.

\subsubsection{Analysis of Poincaré sections}\label{sec:Poincare}
\begin{figure*}[htbp]
    \centering
    \scalebox{1.0}{\includegraphics[width=0.9\textwidth, height=0.9\textheight]{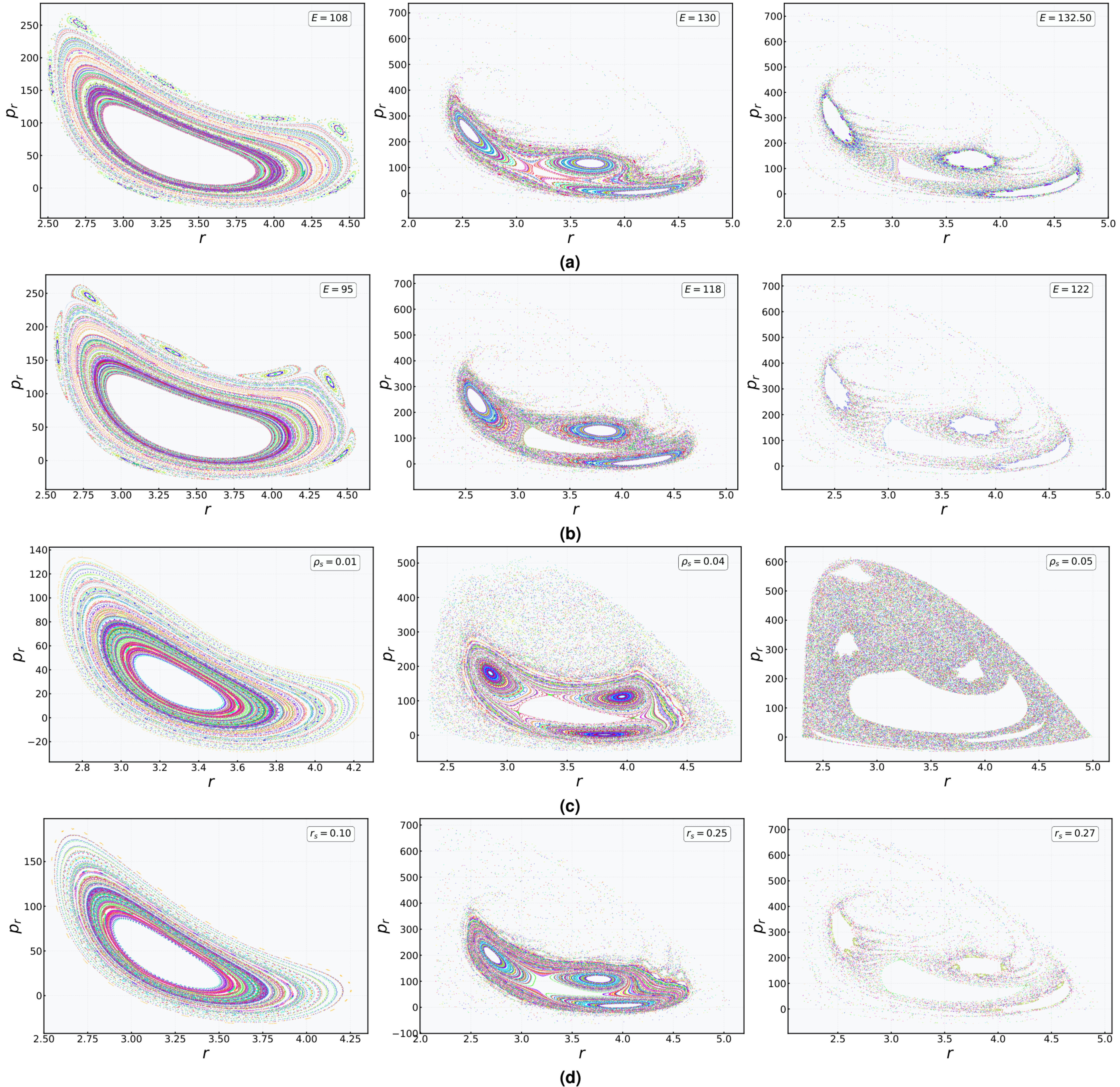}}
    \caption{The Poincar$\Acute{e}$ sections in the $(r-p_r)$ phase plane with $\phi=0$ and $p_{\phi}>0$ for different values of the energy and DM halo parameters.\\
    (a): For different values of energy with fixed DM halo parameters $\rs=0.15,~\rhs=0.02$.\\
    (b): With fixed DM halo parameters $\rhs=0.01,~\rs=0.25$ for another different values of energy.\\
    (c): For different values of the DM halo central density with fixed DM halo scale radius $\rs=0.15$ and energy $E=90$.\\
    (d): For different values of scale radius with fixed DM halo central density $\rhs=0.01$ and energy $E=115$.}
    \label{f2}
\end{figure*}

In Fig.~\ref{f2} ((a), (b), (c) and (d)), we display the Poincar\'{e} sections of a massive particle orbiting a Schwarzschild BH surrounded by a Dehnen-$(1,4,5/2)$ type DM halo. These sections are projected onto the $(r-p_r)$ phase plane, with $\phi = 0$ and $p_{\phi} > 0$, for various values of the energy $E$ and DM halo parameters $\rhs$ and $\rs$, as specified in the corresponding figure captions. For a thorough understanding and anlysis of the Poincar$\Acute{e}$ maps, see the Supplemental Material.

\subsubsection{Chaotic orbital evolution from EMRI}\label{sec:orbits}
See the section ``Trajectory analysis" of the Supplemental Material for a thorough understanding and visualization of representative particle trajectories for two distinct, but very close initial conditions in the case of an EMRI system.

\subsubsection{Analysis of Lyapunov exponents}\label{sec:Lyapunov}
\begin{figure*}[htbp]
    \centering
    \scalebox{0.95}{\includegraphics[width=0.85\textwidth, height=0.85\textheight]{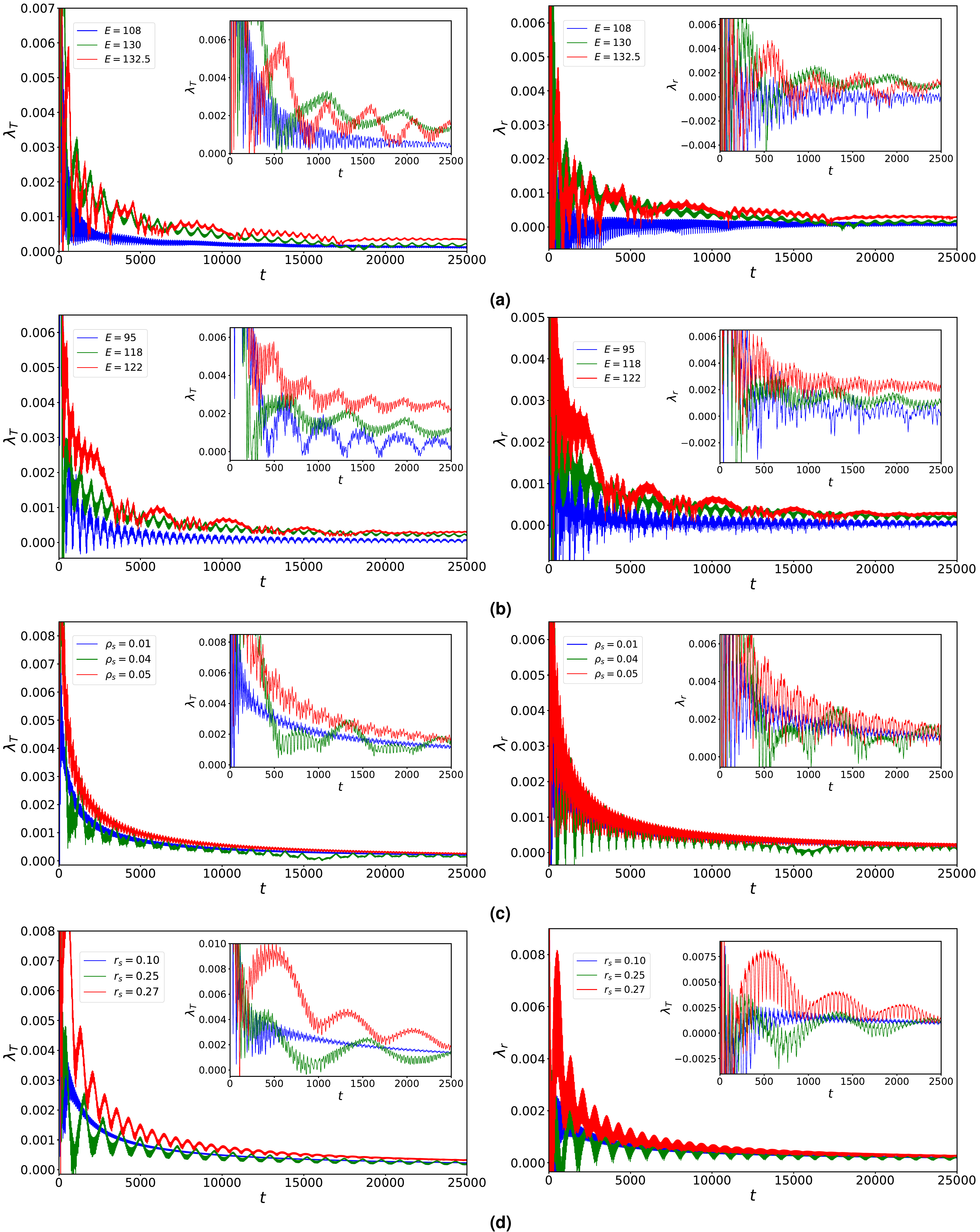}}
    \caption{{\it Left column:} These represent the total Lyapunov exponent $(\lambda_T)$ with $t$. The inset panel of each figures show the maximum positive exponent value. {\it Right column:} These represent the radial Lyapunov exponent $(\lambda_r)$ with time $t$. The inset panel of each figures show the maximum positive radial exponent value.\\
    (a): Case-I: Fixed DM halo parameters $\rs=0.15,~\rhs=0.02$ at different energy values, in which both, the total maximum positive exponent and radial maximum positive exponent values are for $E=132.5$.\\
    (b):Case-II: Fixed DM halo parameters $\rs=0.25,~\rhs=0.01$ at different energy values, in which both, the total maximum positive exponent and radial maximum positive exponent values are for $E=122$.\\
    (c):Case-III: Fixed DM halo parameters $\rs=0.15,~E=90$ at different $\rho_s$ values, in which both, the total maximum positive exponent and radial maximum positive exponent values are for $\rhs=0.05$.\\
    (d):Case-IV: Fixed DM halo parameters $E=115,~\rhs=0.01$ at different $\rs$ values, in which both, the total maximum positive exponent and radial maximum positive exponent values are for $\rs=0.27$.}
    \label{f3}
\end{figure*}

Figures~\ref{f3}(a) and \ref{f3}(b) show the time evolution of the total ($\lambda_T$) (left) and radial ($\lambda_r$) (right) Lyapunov exponents for different energies, with fixed DM halo parameters: Case-I $(\rs=0.15, \rhs=0.02)$ and Case-II $(\rhs=0.01, \rs=0.25)$. Figure~\ref{f3}(c) presents $\lambda_T$ and $\lambda_r$ over time for varying central densities $(\rhs)$ at fixed $\rs=0.15$ and $E=90$ (Case-III). Figure~\ref{f3}(d) does the same for varying scale radii $(\rs)$ at fixed $\rhs=0.01$ and $E=115$ (Case-IV). See the Supplemental Material for details and physical interpretation of all cases for both Lyapunov exponents.

\subsubsection{Analysis of the MSS chaos bound}\label{sec:bound}
See the Supplemental Material for a detailed discussion on the MSS chaos bound violation, in which our conclusion is that while the DM halo parameters $\rhs,\rs$ significantly influence the chaotic dynamics of a massive test particle, they do not violate the surface gravity bounds—whether in the framework of general relativity or in the environment of a galactic BH, surrounded by a Dehnen-$(1,4,5/2)$ type DM halo.

\section{Discussions and conclusion}\label{sec:conclusion}
In this work, we have investigated the existence of homoclinic orbit and the emergence of chaotic motion for a massive probe particle moving around a Schwarzschild-like BH embedded in a Dehnen-$(1, 4, 5/2)$ type DM halo by considering an EMRI system. Starting from the Einstein's field equation with the particular Dehnen density profile, we reviewed the modified metric function and examined how the halo parameters, i.e. the central density ($\rhs$) and the scale radius ($\rs$), alter the horizon structure and the effective potential experienced by a massive probe particle. In the effective potential structure, we found out extrema points whose unstable branches naturally give rise to a homoclinic trajectory separating the bound and the plunging motion.\\
We have established that the existence of a homoclinic orbit provides a geometric indicator of the transition from regular to chaotic behavior in phase space. Through a combination of analytical treatment and extensive numerical integration of the full dynamical equations in Painlevé–Gullstrand coordinates, we analyzed the phase-space structure using Poincaré sections, orbital evolution, and Lyapunov exponent's behavior under the consideration of an EMRI configuration. Our computations show that with the increment of energy along with either the halo's density $(\rho_s)$ or its scale radius $(r_{s})$, the nonlinearity in the system increases. As a result at some parameter range it leads to the disruption of KAM tori and the eventful onset of chaos. In the chaotic regime, however, both the maximal and separate (radial) Lyapunov exponents remain bounded by the surface gravity of such BH-DM halo combined spacetime, confirming that the MSS bound is respected even in the presence of an extended DM halo environment.\\
From an astrophysical perspective, these findings have direct relevance to EMRIs in galactic nuclei, where stellar-mass compact objects spiral into supermassive BHs possibly surrounded by dense halos or accretion structures.  Small perturbations in such systems can imprint subtle phase modulations in the emitted GWs signals detectable by space-based interferometers such as LISA, TianQin, and Taiji. Hence, our analysis provides a theoretical foundation for linking near-horizon chaotic dynamics with potential gravitational-wave observables.\\
Additionally, let us mention that Cardoso et al., in their Ref.~\cite{Cardoso:2021wlq}, proposed a novel Schwarzschild BH solution surrounded by a Hernquist-type DM distribution within the Einstein cluster framework. This solution has recently been systematically generalized by Shen et al. for any given DM halo density profile, incorporating a model-independent inner boundary radius (where the density vanishes), as detailed in Refs.~\cite{Shen:2024qbb,Shen:2023erj}. In all such cases the density profile of the DM halo is based on the following Refs.~\cite{Gondolo:1999ef,Sadeghian:2013laa,Speeney:2022ryg}, in which the existence of SMBHs at the center of the DM halo cloud can create over-dense regions as DM spikes structures. It would be interesting to explore such a Schwarzschild BH solution with a Dehnen-type density distribution; however, it is not guaranteed whether the resulting solution will take a closed-form analytic expression or will be entirely numerical in nature. This lies beyond the scope of our current work, and we plan to investigate it in future studies.\\
In summary, the present study explores the interplay between DM halo distribution, homoclinic instability, and horizon-induced chaos in a realistic galactic BH-DM halo background. It demonstrates that while the DM halo significantly reshapes the effective potential and strengthens the chaotic response of particle motion, it does not lead to any violation of the universal chaos bound. The results presented here form the first part of our investigation. In the companion paper \cite{partII}, we extend this framework to explore the GWs signatures associated with these chaotic trajectories and assess their observational imprints on future space-based detectors.

\section*{Supplementary Information} 
Readers are encouraged to consult the Supplemental Material, where the complete derivations, thorough analysis, and all detailed interpretations are presented with some additional References.
\\
\\
\textbf{Data Availability:} 
The analytical computations supporting the findings of this study are presented within the article and the Supplemental Material. The numerical code will be made available upon reasonable request from the corresponding authors.
\\
\\
\textbf{Conflict of interest:}
The authors declare that they have no conflict of interest.

\section*{Acknowledgements}{We are grateful to Xian-Hui Ge, Rickmoy Samanta, Yu-Qi Lei, Jia-Geng Jiao and Yu-Min Hu for various discussions. The authors would like to acknowledge the anonymous Reviewers for a constructive and insightful reports that has significantly improved this paper. S. Das thanks to Surhud More for a valuable discussion on dark matter halos during his visit to USTC and also acknowledge the support of USTC Fellowship Level A--CAS-ANSO Scholarship 2024 for PhD candidates (formerly the ANSO Scholarship for Young Talents). S. Dalui thanks the Department of Physics, Shanghai University, for providing postdoctoral funds and support from the Super Postdoctoral Fund (Grant No. 2023307) during the period of this work. BHL thanks APCTP, KIAS and USTC for the hospitality during his visit, where a part of this project was done. This work was supported in part by the National Key R\&D Program of China (2021YFC2203100), by the National Natural Science Foundation of China (12433002, 12261131497, 125B1023), by CAS young interdisciplinary innovation team (JCTD-2022-20), by 111 Project (B23042), by CSC Innovation Talent Funds, by USTC Fellowship for International Cooperation, and by USTC Research Funds of the Double First-Class Initiative. BHL is supported by the National Research Foundation of Korea (NRF) grant RS-2020-NR049598, RS-2026-25473640 and Overseas Visiting Fellow Program of Shanghai University. Numerical computations were performed on the computer facilities ``LINDA $\&$ JUDY" in the Particle Cosmology group (COSPA) at USTC.}

\newpage
\onecolumngrid
\renewcommand{\theequation}{s\arabic{equation}}

\renewcommand{\thefigure}{S\arabic{figure}}

\renewcommand{\thetable}{S\arabic{table}}

\hypersetup{
	colorlinks   = true, 
	urlcolor     = blue, 
	linkcolor    = blue, 
	citecolor    = blue    
}


\hypersetup{
	colorlinks   = true, 
	urlcolor     = blue, 
	linkcolor    = blue, 
	citecolor    = blue 
}
\maxdeadcycles=1000 

\def\rhs{\rho_s}
\def\rs{r_s}
\def\a{\alpha}
\def\b{\beta}
\def\g{\gamma}
\def\d{\delta}
\def\s{\sigma} 
\def\o{\omega}
\def\O{\Omega}
\def\m{\mu}
\def\n{\nu}
\def\ph{\phi} 
\def\non{\nonumber}
\def\n{\noindent}

\definecolor{lightblue}{RGB}{173,216,230}
\definecolor{lightgreen}{RGB}{144,238,144}
\definecolor{magenta}{RGB}{255,0,255}
\definecolor{myolive}{RGB}{128,128,0}
\definecolor{masteredyellow}{RGB}{255,255,102}
\definecolor{mymaroon}{RGB}{128,0,0}
\newcommand{\hb}[1]{\textcolor{orange}{\textit{\textbf{HB:} #1}}}

\section*{Supplemental Material for ``\textit{Extreme-Mass-Ratio Inspirals Embedded in Dark Matter Halo: Existence of Homoclinic Orbit and Horizon-Induced Chaos}"}

\section*{Double power-law density distribution}
\noindent
Let us now consider a static spherically symmetric (SSS) BH solution in galaxies surrounded by a DM halo. The double power-law density distributions for DM halos and elliptical galaxies was originally introduced by Hernquist \cite{Hernquist:1990be}. This distribution is expressed as \cite{Hernquist:1990be,Van}
    \begin{equation}  
        \rho = \rho_s \left( \frac{r}{r_s} \right)^{-\gamma} \left[ \left( \frac{r}{r_s} \right)^{\alpha} + 1 \right]^{\frac{\gamma - \beta}{\alpha}}~,\label{2c}  
    \end{equation} 
where $\a,\b,\g$ are free parameters and $\rhs$, $\rs$ are known as the typical central density and scale radius of the DM halo, respectively. In the context of the double power-law density profiles for DM halos, as shown in Eq.~\eqref{2c}, the parameter $\alpha$ controls the smoothness of the transition between the inner and outer slopes, $\beta$ denotes the outer power-law slope, and $\gamma$ determines the specific form of the density profile \cite{Zhao:1995cp}. Subsequently, Dehnen \cite{Dehnen} demonstrated that the spherical subset with $(\alpha,\beta,\gamma)\equiv(1,4,\gamma)$ possesses several advantageous analytical properties, including a closed-form gravitational potential and intrinsic velocity dispersion for all real values of $0\leq\gamma<3$. This $(1,4,\gamma)$ model is now widely recognized as the Dehnen model, which constitutes a subset of the $\gamma/\eta$-models and relates to the $\eta$-models of Tremaine et al. \cite{Tremaine:1993qb} through the relation $\gamma\equiv 3-\eta$ \cite{Zhao:1995cp, Van}. Moreover, for the specific cases $\gamma = 0, 1, 2$, both the projected mass density and velocity dispersion can be derived analytically \cite{Dehnen,Tremaine:1993qb}. These models are frequently used to fit the surface brightness distributions of elliptical galaxies, as they closely approximate the de Vaucouleurs $r^{1/4}$ profile \cite{Shakeshaft} when $\gamma = 3/2$. It is also worth noting that the parameter sets $(\alpha,\beta,\gamma) \equiv (1,4,1)$ and $(1,4,2)$ correspond to the well-known Hernquist \cite{Hernquist:1990be} and Jaffe \cite{Jaffe:1983iv} density profiles, respectively.\\
In this work, we consider a SSS BH solution embedded in a galaxy with $(1,4,\gamma)\equiv(1,4,5/2)$ type DM halo, which we refer as the Dehnen-$(1,4,5/2)$ type DM halo. In the later sections, we will first study on the existence of a homoclinic orbit in the background geometry discussed in this section and then proceed further explorations on chaotic dynamics.\\
The generic action in Einstein gravity is given by
	\begin{equation}
		\mathcal{A}=\frac{1}{2\kappa}\int d^4x\sqrt{-g}~R+\mathcal{A}^{tot}_{matter}~,\label{1}
	\end{equation}
where $R$ is the Ricci scalar and $g$ is the determinant of  metric $g_{\mu\nu}$ in a 4D spacetime. $\mathcal{A}^{tot}_{matter}$ refers to the total effective matter action, which includes the contributions from DM halo as well as BH. Taking the variation of the generic action Eq.~\eqref{1} with respect to metric, we get the field equations (setting $\kappa=8\pi G=1$) as
    \begin{equation}
        R_{\mu\nu}-\frac{1}{2}Rg_{\mu\nu}=T^{tot}_{\mu\nu}~,\label{2}
    \end{equation}
where the total energy-momentum tensor is defined as $T^{tot}_{\mu\nu}\equiv T^{(BH)}_{\mu\nu}+T^{(DM)}_{\mu\nu}$ for a BH-DM halo combined system.

\section*{Technical details on the derivation of a Schwarzschild-like black hole embedded in a Dehnen-$(1,4,5/2)$ type dark matter halo}
\noindent
For the completeness, we will now take a moment to review a brief technical details required to derive the BH solution, given in Eq.~(2), following the work in Refs.~\cite{Al-Badawi:2024asn,Xu:2018wow,Matos:2003nb,Azreg-Ainou:2014pra}. To derive the Schwarzschild-like BH solution embedded in DM halo characterized by a particular density profile, one can use the procedures introduced in Refs.~\cite{Xu:2018wow,Matos:2003nb}. In this approach, one can first construct the DM spacetime metric using the particular DM density profile in general relativity and then by solving Einstein's field equations in the presence of DM, one can derive an approximate BH solution. A purely relativistic DM-dominated halo can be described by the SSS line element (which describes a global matter distribution and particles follow constant rotational trajectories within it) as \cite{Al-Badawi:2024asn,Xu:2018wow}  
    \begin{equation}  
        ds^2 = -A(r) dt^2 + \frac{dr^2}{B(r)} + r^2 d\O^2~,\label{2a}  
    \end{equation}  
where $A(r)$ represents the redshift function depends on the DM halo profile and $B(r)$ is known as the shape function.\\
On the other hand, for a BH-DM halo combined spacetime, one can define the following SSS spacetime metric as \cite{Al-Badawi:2024asn,Xu:2018wow}  
    \begin{equation}  
        ds^2 = -\Big( A(r) + f_1(r) \Big) dt^2 + \frac{dr^2}{\Big(B(r) + f_2(r)\Big)} + r^2 d\O^2~,\label{2b}  
    \end{equation}  
where $f_1(r)$ and $f_2(r)$ are correction terms determined by the BH and DM halo parameters.\\
In addition one can now define the mass distribution of the DM density profile under SSS consideration as the following \cite{Van}:
    \begin{equation}  
        M_{DM} = 4\pi \int_{0}^{r} \rho(x) x^2 dx~.\label{2d}
    \end{equation} 
Using Eqs.~\eqref{2c} and \eqref{2d}, the constant tangential velocity of a stellar object within the particular Dehnen-$(1, 4, 5/2)$ type DM halo is given by
    \begin{equation}  
        v_D^2 = \frac{M_{DM}}{r} = \frac{8\pi \rho_s r_s^3}{r \sqrt{1 + \frac{r_s}{r}}}\label{2e}~.  
    \end{equation} 
Now employing the relation between tangential velocity and redshift function in the background of DM halo spacetime \cite{Matos:2000ki} as $v_D^2 = r\frac{d}{dr}(\ln\sqrt{A(r)})$, we can find the expression of $A(r)$ \cite{Al-Badawi:2024asn}:
    \begin{equation}  
        A(r) = \exp\left[ -32\pi \rho_s r_s^2 \sqrt{1+\frac{r_s}{r}} \right]\label{2f}~.
    \end{equation}  
Now one can obtain the BH solution for the leading order term only in the exponential series and therefore from Eq.~\eqref{2f}, one can find the leading order redshift function of a Dehnen-$(1,4,5/2)$ type DM halo as follows.
    \begin{equation}
        A(r)\approx 1 - 32\pi \rho_s r_s^2 \sqrt{1+\frac{r_s}{r}}~.\label{2g}
    \end{equation}
For the spacetime metric of a Schwarzschild BH immersed in DM halo, as mentioned in Eq.~\eqref{2b}, the Einstein's field equations yield the equation mentioned in Eq.~\eqref{2}. Since Schwarzschild BH solution is a SSS vacuum solution of Einstein's field equations which has vanishing energy-momentum tensor \Big(i.e., $T^{(BH)}_{\mu\nu}=0$\Big), one can consider the DM energy-momentum tensor as the only contribution for the BH-DM halo combined spacetime. Therefore using the combined metric ansatz Eq.~\eqref{2b}, the Einstein's field equations (Eq.~\eqref{2}) can be simplified as
    \begin{eqnarray}  
        &\left( B(r) + f_2(r) \right) \left(\frac{B'(r) + f_2'(r)}{r\left(B(r) + f_2(r)\right)} + \frac{1}{r^2}\right)-\frac{1}{r^2}&\nonumber\\
        &= T^{0~(DM)}_0~,\label{2h0}\\
        &\left( B(r) + f_2(r) \right) \left(\frac{A'(r) + f_1'(r)}{r\left(A(r) + f_1(r)\right)} + \frac{1}{r^2} \right) -\frac{1}{r^2}&\nonumber\\
        &= T^{1~(DM)}_1~,\label{2h}
    \end{eqnarray}          
where the energy-momentum tensor of the Dehnen-type DM halo spacetime has the form: $T^{\mu~(DM)}_{\nu} = \text{diag}[-\rho, p_r, p, p]$. Here $\rho,~p_r$ and $p$ represent the energy density, radial pressure and cross-radial pressure of the DM halo energy-momentum tensor, respectively.\\
Now for a pure Dehnen-type DM halo profile with the corresponding energy-momentum tensor $T^{\mu~(DM)}_{\nu}$, the Einstein’s field equations read,
    \begin{equation}  
        R_{\mu \nu} - \frac{1}{2} R g_{\mu \nu} = T^{(DM)}_{\mu \nu}~.\label{2i}
    \end{equation}  
Therefore using purely DM halo dominated spacetime Eq.~\eqref{2a}, the above equations are written as
    \begin{eqnarray}  
        B(r) \left( \frac{1}{r} \frac{B'(r)}{B(r)} + \frac{1}{r^2} \right) - \frac{1}{r^2} =T^{0~(DM)}_0\equiv-\rho~,\non\\
        B(r) \left( \frac{1}{r} \frac{A'(r)}{A(r)} + \frac{1}{r^2} \right) - \frac{1}{r^2} = T^{1~(DM)}_1\equiv p_r~,\label{2j}
    \end{eqnarray}    
with,
    \begin{eqnarray}
        &B(r) \Big\{ \frac{A''(r)A(r) - A'^2(r)}{A^2(r)} + \frac{A'^2(r)}{2A^2(r)} + \frac{A'(r)B'(r)}{2A(r)B(r)}\nonumber\\
        &+ \frac{1}{2} \Big( \frac{A'(r)}{A(r)} + \frac{B'(r)}{B(r)}\Big)\Big\}= 2T^{2~(DM)}_2\equiv2p~,\non\\
        &T^{2~(DM)}_2 = T^{3~(DM)}_3~.\label{2k}
    \end{eqnarray}
Now one can write the solutions for $f_1(r),~f_2(r)$ from Eqs.~\eqref{2h0}, \eqref{2h} as
    \begin{align}
        f_1(r) = &\exp \left[ \int \left(\frac{B(r)}{B(r) - \frac{2M}{r}} \left( \frac{1}{r} + \frac{A'(r)}{A(r)} \right) -\frac{1}{r}\right) dr \right]\non\\
        &- A(r)~,\non\\
        f_2(r) = &-\frac{2M}{r}~.\label{2l}
    \end{align}
Now it is noteworthy to mention that employing the empirically derived DM halo density profile from numerical simulations, one can determine the spacetime metric under the assumption $A(r) = B(r)$. This assumption is justified by two key considerations: First of all, it is well-established that for SSS vacuum BH solutions like the Schwarzschild metric, the equality $A(r)=B(r)$ holds exactly. Given that the gravitational influence of DM is significantly weaker than that of the BH itself, adopting $A(r)=B(r)$ serves as a reasonable approximation in this context. Secondly, as demonstrated in Refs.~\cite{Matos:2003nb,Matos:2004je}, the physical differences between scenarios where $A(r)\neq B(r)$ and those where $A(r)=B(r)$ are substantially smaller than the effects induced by DM halo on the BHs' geometry. Therefore, setting $A(r)=B(r)$ represents a valid simplification. This approach remains applicable to various DM halo density profiles, including both Cold dark matter (CDM) as well as Scalar field dark matter (SFDM) models. Of course the higher order DM halo profile potentially claim for $A(r)\neq B(r)$.\\
Therefore under the assumption of $A(r)=B(r)$, from Eq.~\eqref{2l}, we have $f_1(r)=f_2(r)=-\frac{2M}{r}$ and one can finally write down the Schwarzschild-like BH solution embedded by the Dehnen-$(1,4,5/2)$ type DM halo (i.e., the solution of the metric Eq.~(1)), which is mentioned in Eq.~(2). Before proceeding further, let us mention here that any physically reasonable solution to the Einstein field equations must satisfy all the energy conditions, namely the weak, null, dominant and strong energy conditions \cite{Poisson:2009pwt}. These conditions are also intimately connected to singularity theorems, black hole physics, and the causal structure of general spacetimes. Moreover, the energy conditions play an essential role in excluding non-physical solutions to Einstein's equations. In what follows we have examined all the energy conditions, given at the end of this material and shown all of the energy conditions are satisfied both, inside and outside of the BH-DM halo combined spacetime. \footnote{The authors would like to acknowledge the anonymous Reviewer for suggesting investigation of the validity of all the energy conditions in our considered Schwarzschild-like BH-DM halo combined system}\\
Now in the absence of DM halo (i.e., the parameters $\rhs=\rs=0$), the BH solution Eq.~(2) simplifies to the Schwarzschild BH solution in Einstein gravity, as expected. One can also verify that for $A(r)=B(r)=1$, the metric mentioned in Eq.~\eqref{2b} recovering the Schwarzschild solution. It is also noteworthy to mention that the resulting SSS metric solution (Eq.~(2)) is an exact solution to the Einstein field equations (for more details on the exact solutions of Einstein's equations along with the properties of the curvature tensor see \cite{Al-Badawi:2024asn}).

    \begin{figure}[H] 
        \centering
        \includegraphics[width=0.49\textwidth]{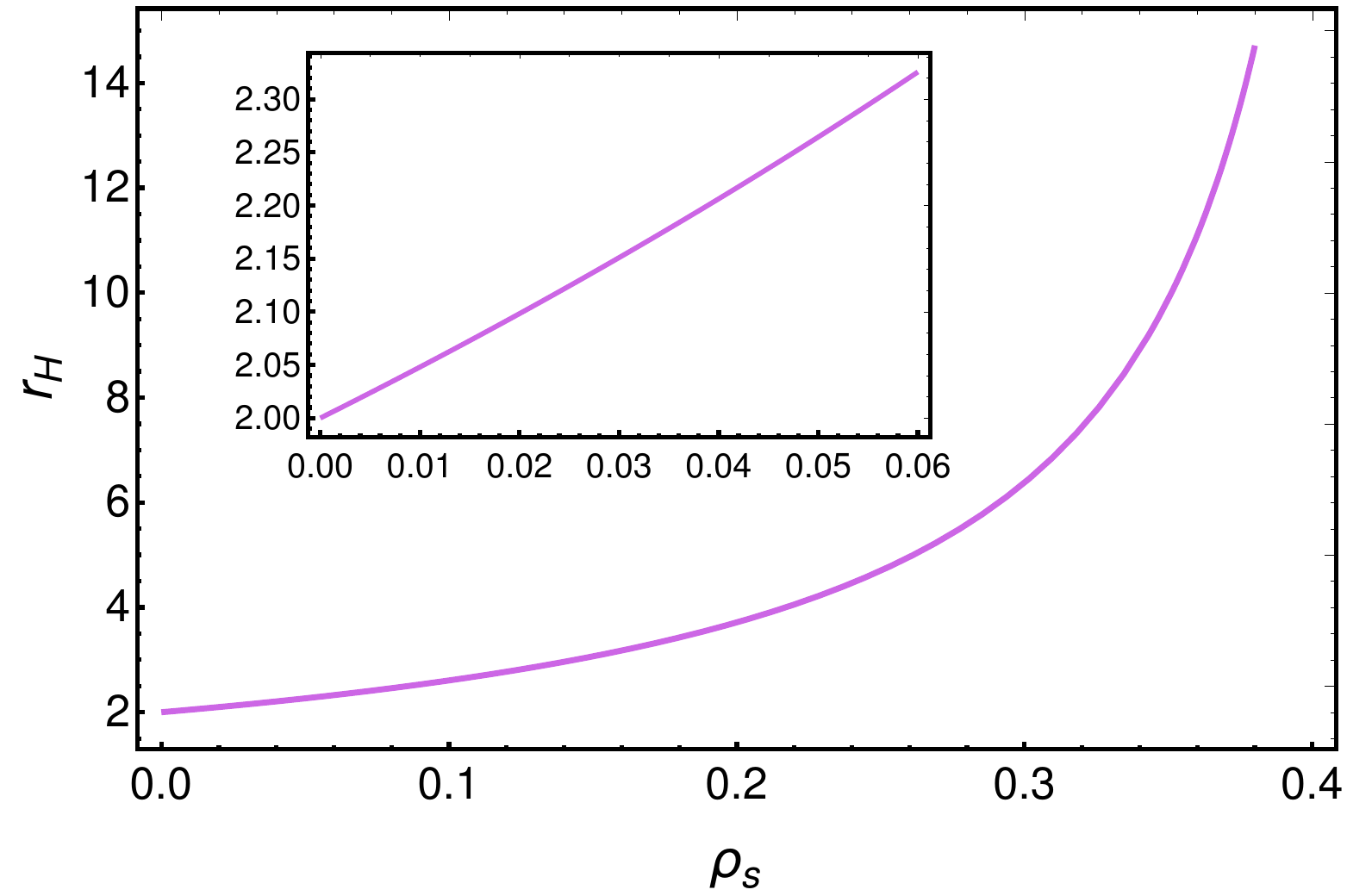}
        \includegraphics[width=0.49\textwidth]{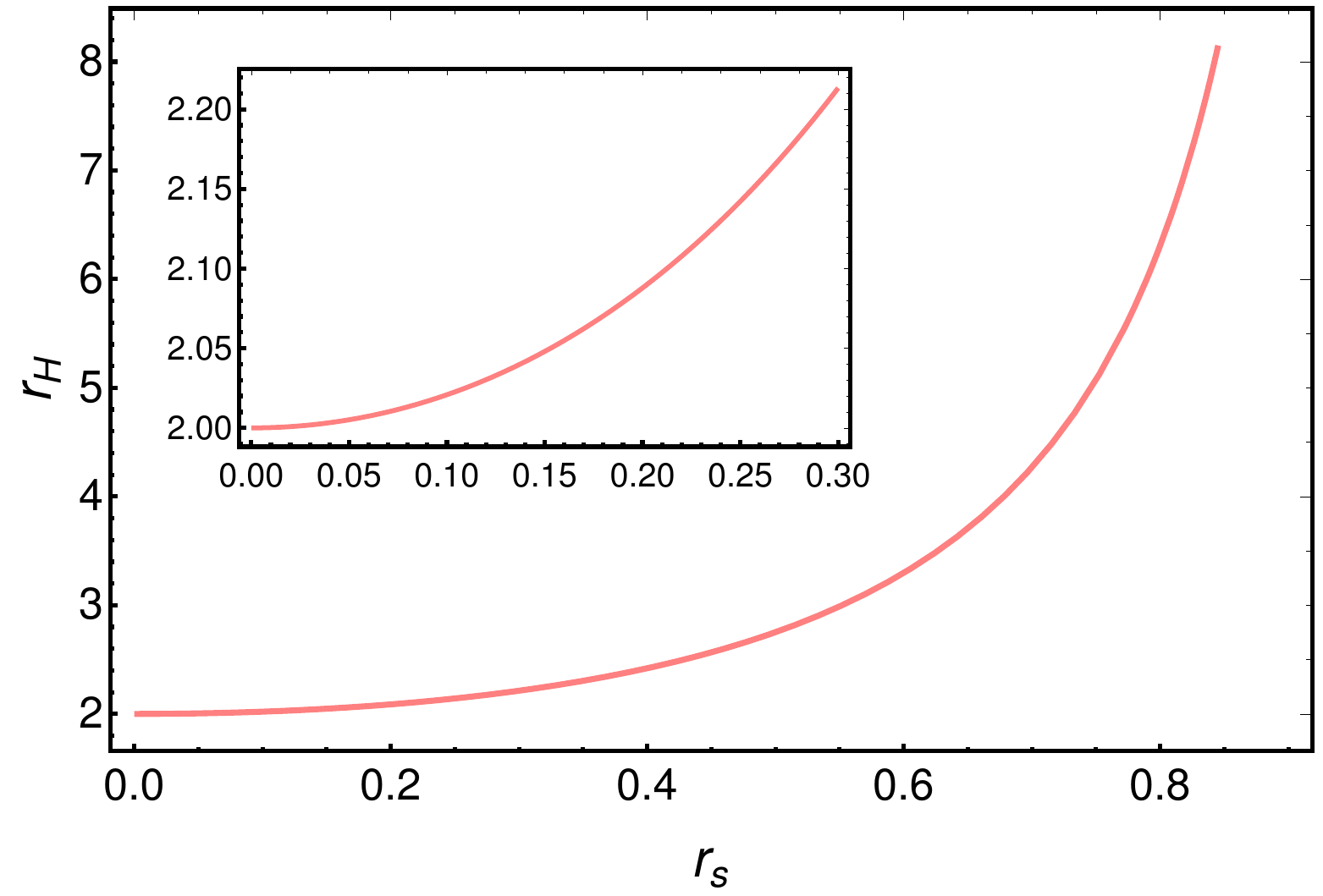}
	    \caption{{\it Upper:} Plot of the horizon radius $r_{H}$ versus core density $\rhs$ of a unit mass Schwarzschild BH-DM halo combined spacetime for fixed scale radius $\rs=0.15$. The inset shows a zoomed version of the horizon variation for central density $\rhs$ in $0\leq \rho_s \leq 0.06$.\\
        {\it Lower:} Plot of the horizon radius $r_{H}$ versus scale radius $\rs$ of a unit mass Schwarzschild BH-DM halo combined spacetime for fixed central density $\rhs=0.01$. The inset shows a zoomed version of the horizon variation in the ranges of scale radius $\rs$ in $0\leq \rs \leq 0.3$.}\label{fhorizon_rhos}
    \end{figure}
\noindent
The horizon of the Schwarzschild-like BH surrounded by a Dehnen-$(1,4,5/2)$ type DM halo are determined by solving $f(r) = 0$ (Eq.~(2)), which gives us
    \begin{eqnarray}  
        &r_H = \frac{2M}{1 - 1024\pi^2 r_s^4 \rho_s^2} + \frac{512\pi^2 r_s^5 \rho_s^2}{1 - 1024\pi^2 r_s^4 \rho_s^2} \non\\
        &+ \frac{32\sqrt{2}\pi r_s^2 \rho_s \sqrt{2M^2 + M r_s + 128\pi^2 r_s^6 \rho_s^2}}{1 -1024\pi^2 r_s^4 \rho_s^2}~.\label{2n}  
    \end{eqnarray}
The horizon radius $r_H$ reduces to the Schwarzschild radius $2M$ in the absence of the DM halo (i.e., $\rhs=\rs=0$). Fig.~\ref{fhorizon_rhos} illustrates the influence of $\rho_s$ and $r_s$ on the horizon radius $r_H$, respectively. As these parameters increase, the horizon radius expands, demonstrating the significant role of the DM halo's core density $\rhs$ and scale radius $\rs$ in determining the horizon.

\section*{A brief discussion of our considered model with others existing in the literature}
\noindent
Recently Shen et al. introduced a novel and generalized framework for deriving a Schwarzschild BH surrounded by DM halos \cite{Shen:2024qbb}. Their approach incorporates three important aspects: (i) the surrounding DM halo is modeled as an Einstein cluster within an anisotropic energy-momentum tensor characterized by vanishing radial pressure i.e., $p_r(r)=0$; (ii) the density profile of the DM halo vanishes at an inner boundary radius given by $r_{\text{in}} = vM$, where $v \geq 2$ and $M$ denotes the mass of the Schwarzschild BH; and (iii) a scaled density distribution $\rho(r) \to \Theta(r - vM) (1 - vM/r)^n \rho(r)$ is adopted, with $\Theta$ representing the Heaviside step function and $n$ a free parameter, where different values of $n$ yield different solutions. As a consequence of this scaled energy distribution and the requirement that all energy conditions be satisfied when the inner boundary radius becomes $r_{\text{in}} \geq 5M/2$ under the Einstein cluster configuration \cite{Shen:2024qbb}. It is also noteworthy that, within the Shen et al. framework \cite{Shen:2024qbb}, a BH surrounded by a DM halo characterizes a Schwarzschild BH with a horizon radius $r_{\text{H}}=2M$.\\
In contrast, our combined BH-DM halo solution, which is adopted from the work of Al-Badawi et al. \cite{Al-Badawi:2024asn} does not employ such a scaled density profile as a starting point. Consequently, the density is monotonically decreasing and non-negative, satisfying $\rho(r)\geq 0$ for all $r$ (so that $d\rho/dr<0$), which specifically ensures the validity of the WEC. Moreover, all three energy conditions hold over the entire domain of the radial coordinate $r > 0$ (see at the end of this material for detail). The same non-zero behavior of the energy density is also observed in another BH-DM halo spacetime, recently presented by Uktamov et al. \cite{UktamjonUktamov:2025emm}, which is surrounded by a Dehnen-$(1,4,2)$ type profile, following the same approach of Xu et al. \cite{Xu:2018wow}. Alternatively, in the work of Al-Badawi et al. \cite{Al-Badawi:2024asn}, understanding the non-vanishing density across all radial distances is as follows: under the assumption of a general SSS line element where $g_{tt} = -g_{rr}^{-1}$, the Einstein field equations automatically lead to the distinctive relation $p_r(r) = -\rho(r)$. This is more naturally associated with dark energy (DE) rather than DM on cosmological scales. However it is the tangential pressure which is positive and associated with the nature of DM. Several studies have investigated fluids satisfying $p_r = -\rho$ or approaching this limit in certain regimes, as possible unified descriptions of DM and DE \cite{Kamenshchik:2001cp,Bento:2002ps,Scherrer:2004au,Dymnikova:2015yma}. Related concepts also arise in models of vacuum-like or self-interacting DM, wherein the dark component effectively behaves as a cosmological-constant term inside overdense regions or compact objects \cite{Balakin:2003tk,Sahni:2002dx}. Therefore, in the work of Al-Badawi et al. \cite{Al-Badawi:2024asn}, the specific relation between radial pressure and density, $p_r = -\rho$, is consistent with the phenomenological freedom inherent in halo modeling. This stands in contrast to the framework of Shen et al. \cite{Shen:2024qbb}, which adopts a different halo modeling approach by setting $p_r = 0$ in the context of the Einstein cluster setup. As a further comparison, it is also noteworthy that, the horizon radius $r_{H}$ depends on the parameters of the DM halo (Eq.~\eqref{2n}), thereby representing a Schwarzschild-like BH with a varying horizon originating from a nontrivial DM halo contribution in our considered BH-DM spacetime surrounded by a Dehnen-$(1,4,5/2)$-type density profile.

\section*{Detailed analysis on homoclinic orbits}
\noindent
Considering a relativistic massive test particle of mass $m$ moving in a curved background described by the spacetime as mentioned in Eq.~(1), we start with the covariant dispersion relation as
    \begin{equation}
        g^{\alpha\beta}p_{\alpha}p_{\beta} = -m^2~,\label{eq:dispersion}
    \end{equation}
where $p_{\alpha}$ denote particle's momenta, canonically conjugate to $x^\a$ by $p_\a=g_{\a\b}p^\b\equiv g_{\a\b}\frac{dx^\b}{d\lambda}$, with $\lambda$ being the affine parameter. The metric (Eq.~(1)) possesses two Killing vectors $\zeta^{\alpha}_{(t)}$ along $t$ and $\eta^{\a}_{(\phi)}$ along $\phi$, leading to the conserved energy $E$ and conserved angular momentum $L$ of the test massive particle respectively, as measured by some observer at infinity as the following way.
    \begin{eqnarray}
        E&\equiv& -\zeta^{\alpha}_{(t)}p_{\alpha}:= -p_t=f(r)\frac{dt}{d\lambda}~,\non\\
        L&\equiv& \zeta^{\alpha}_{(\phi)}p_{\alpha}:= p_{\phi}=r^2\sin^2\theta\frac{d\phi}{d\lambda}~,\label{eq:conserved}
    \end{eqnarray}
where $f(r)$ is given by Eq.~(2).\\
Let us confine our analysis to the trajectories of particles that are constrained to the equatorial plane defined by $\theta = \frac{\pi}{2}$ (a condition for which $p_{\theta} = 0$ remains consistent with the equations of motion within this plane). Under these constraints, the radial coordinate $r$ remains as the sole effective degree of freedom. For a specific values of energy $E$ and angular momentum $L$, a first-order equation of motion governing $r$ can be derived directly from Eq.~\eqref{eq:dispersion}, which becomes
    \begin{eqnarray}
        m^2\Big(\frac{dr}{d\tau}\Big)^2+f(r)\Big(m^2+\frac{L^2}{r^2}\Big)=E^2~,\label{eq:eff1} 
    \end{eqnarray}
where we have used proper time $\tau$ along the world-line instead of affine parameter $\lambda$ and $f(r)$ is given by Eq.~(2).\\
Now to analyze the motion of a massive test particle, it is useful to introduce a change of the dynamical variable from $r$ to $x$ using the relation $x := 2M/r$. Furthermore, it is advantageous to describe in the shape of an orbit using the angle $\phi$ as the parameter instead of some extraterrestrial’s proper time $\tau$ or coordinate time $t$. Employing these new variables transforms the equation of motion given in Eq.~\eqref{eq:eff1} into the following expression:
    \begin{align}
        \Big(\frac{dx}{d\phi}\Big)^2 + \mathcal{V_{\textrm{eff}}}=\mathcal{E_{\textrm{eff}}}~.\label{eq:eff3}
    \end{align}
Here, the newly defined effective potential and effective energy, expressed in terms of the variable $x$, are given by:
    \begin{eqnarray}
        \mathcal{V_{\textrm{eff}}}(x) &=& x^2 - x^3 -32\pi\rhs\rs^2\sqrt{1+\frac{x\rs}{2M}}\Big(x^2+\frac{4M^2m^2}{L^2}\Big)\non\\
        &-&\frac{4M^2m^2x}{L^2} ~,\label{eq:pot2}\\
        \mathcal{E_{\textrm{eff}}}(x) &=& \frac{4M^2 (E^2-m^2)}{L^2}~.\label{eq:en1}
        \end{eqnarray}
When the DM halo parameters are omitted, the expression for the effective potential in Eq.~\eqref{eq:pot2} simplifies to the standard form of the effective potential for Schwarzschild spacetime, now expressed in the variable $x$ \cite{Bombelli:1991eg}. In the provided expression of the effective potential, the initial term represents an attractive $1/r$ potential. This is followed by a repulsive centrifugal term proportional to $1/r^2$ in the second term, which is also present in the analogous Newtonian scenario. The third term introduces an additional attractive contribution as $1/r^3$; this is a consequence of general relativity and is absent in Newtonian gravity. Beyond these, the fourth and fifth terms in Eq.~\eqref{eq:pot2} represent the coupling effects inherent to the combined BH and DM halo system. We will now demonstrate that these coupling contributions from the DM halo parameters $\rhs$ and $\rs$, together with the general relativistic correction, are what facilitate the existence of a homoclinic orbit in this BH-DM halo combined spacetime.\\
In order for a homoclinic orbit to exist, the effective potential $\mathcal{V_{\textrm{eff}}}(x)$ appearing in Eq.~\eqref{eq:pot2} has to have an unstable equilibrium point at which its value is less than zero. However in this context of BH-DM halo combined system, for the expression as mentioned in Eq.~\eqref{eq:pot2}, it is not easy to find an analytical solution for such a real unstable point $x_{\textrm{un}}$ \footnote{It is worthy to mention that the stability characteristics of circular orbits are governed by two constants of geodesic motion associated with infinity: energy and angular momentum, corresponding to the temporal and azimuthal directions, respectively. In contrast, motion in the latitudinal direction is inherently stable due to the symmetry about the equatorial plane. Consequently, the principal direction in which the geodesic flow exhibits divergence is exclusively the radial one.}, for which $\mathcal{V_{\textrm{eff}}}(x)$ has an extrema. Therefore we numerically solve $\frac{d\mathcal{V_{\textrm{eff}}}}{dx}=0$ for such unstable point $x=x_{\textrm{un}}$ for different values of central density $\rhs$, scale radius $\rs$ and angular momentum $L$.\\
When the parameters of the DM halo are not considered, it is straightforward to show that the effective potential $\mathcal{V_{\textrm{eff}}}(x)$ possesses extrema at real values of $x$ if and only if the condition $\sqrt{1-\frac{12m^2M^2}{L^2}}>0$ is satisfied. These extreme points are situated at the positions of 
    \begin{eqnarray}
        x_{\pm}=\frac{1}{3}\left(1\pm\sqrt{1-\frac{12m^2M^2}{L^2}}\right)~\label{eq:xpm},
    \end{eqnarray}
where $x_{-}$ corresponds to stable circular orbits, analogous to the scenario in Newtonian gravity, and $x_{+}\equiv x_{\textrm{un}}$ represents an unstable equilibrium. The effective potential at this unstable location is expressed as
    \begin{eqnarray}
        &\mathcal{V_{\textrm{eff}}}(x_{\textrm{un}})=-\frac{1}{27} \left(1+\sqrt{1-\frac{12m^2M^2}{L^2}}\right)^2\non\\
        &\left(1-2\sqrt{1-\frac{12m^2M^2}{L^2}}\right)~.\label{eq:V_xun}
    \end{eqnarray}
Consequently, to ensure the unstable point has a negative potential value, $\mathcal{V_{\textrm{eff}}}(x=x_{\textrm{un}})<0$, the angular momentum $L$ for a particle around a Schwarzschild BH without a DM halo must be constrained within the range
    \begin{eqnarray}
        2\sqrt{3}M < \frac{L}{m} < 4M~\label{eq:bound_L}.
    \end{eqnarray}
Note that in the bound given by Eq.~\eqref{eq:bound_L}, the lower limit of $L = 2\sqrt{3}mM$ is the angular momentum associated with the ISCO at $r=6M$, while the upper limit of $L = 4mM$ corresponds to the angular momentum of the MBO at $r=4M$ for a Schwarzschild BH in general relativity. This analysis clearly demonstrates that not all values of angular momentum are permissible; instead, $L$ must be bounded between the specific values defining the ISCO and MBO to facilitate the existence of a homoclinic orbit. Accordingly, in the following subsection, we will first establish the framework for determining the ISCO and MBO in order to derive the precise bounds on $L$ for varying central densities and scale radii within our combined Schwarzschild BH-DM halo system, before proceeding to examine particle motion along a homoclinic trajectory.

\subsection{MBO and ISCO around a Schwarzschild black hole with dark matter halo}
\begin{figure}[H]
        \centering
        \includegraphics[width=0.49\textwidth]{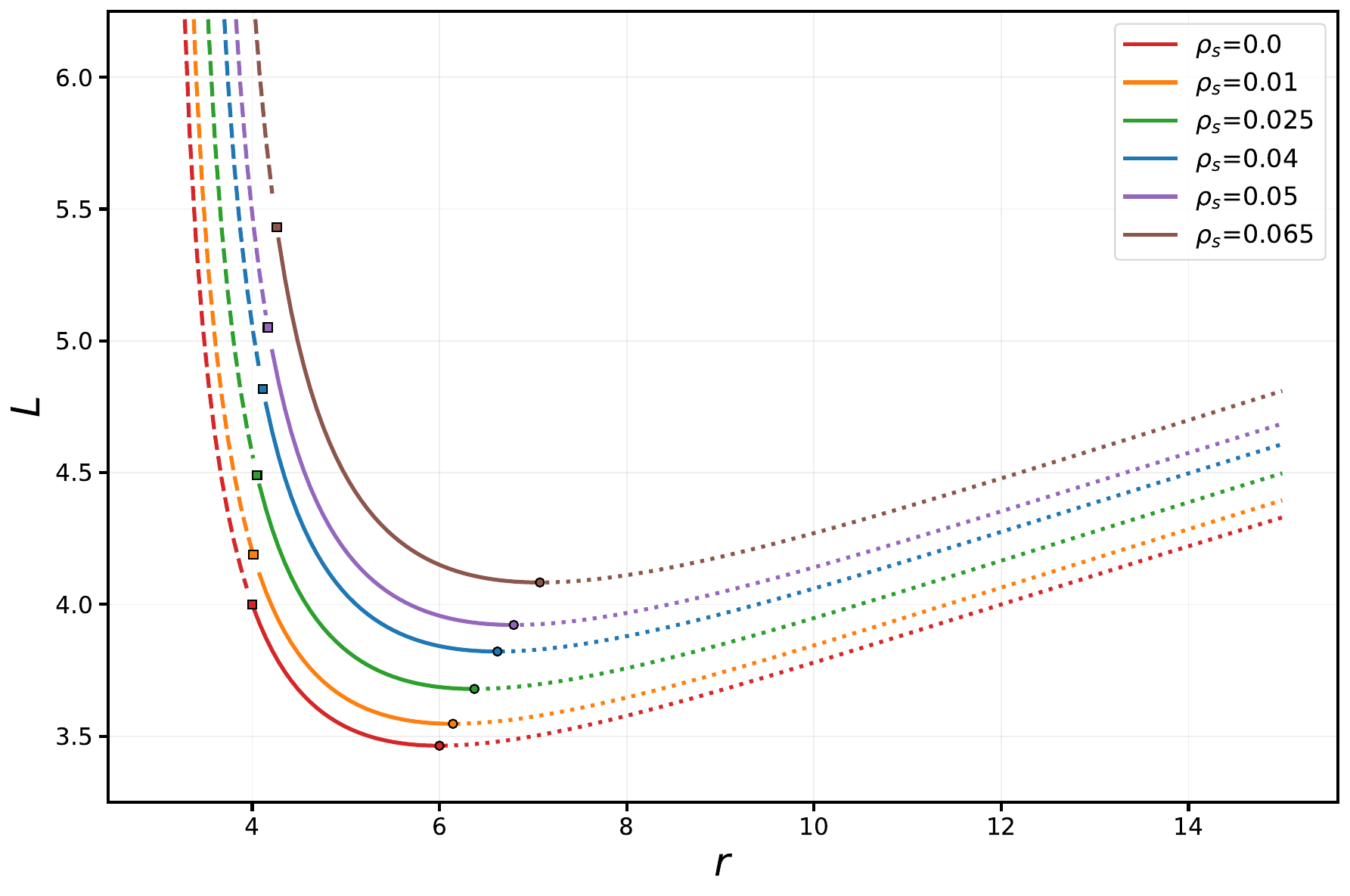}
        \includegraphics[width=0.49\textwidth]{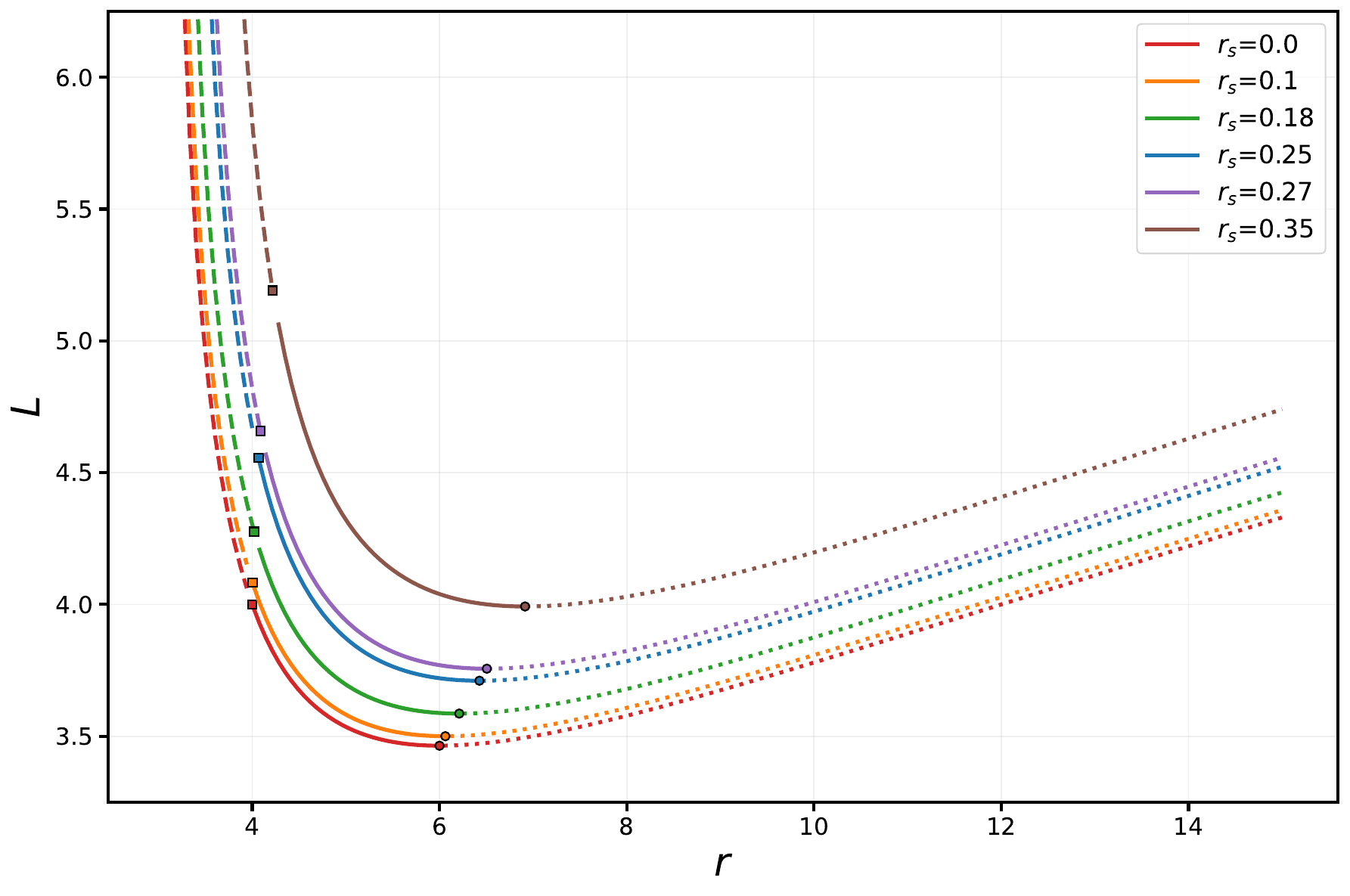}
        \caption{{\it Upper:} The conserved angular momentum $L$ is plotted for circular orbits as a function of their circumferential radius $r$. Each curve corresponds to a different value of the central density $\rhs$ while the scale radius of the DM halo is kept to be constant at $\rs=0.15$. The dotted segments of these curves indicate the regions where the orbits are stable. In contrast, the solid and dashed segments represent regions where the orbits are unstable. The locations of the ISCO for each configuration are marked by circular points, MBO radii are denoted by square points. Note that the displayed range for $L$ corresponds to units of $10^{-5}$. \\
        {\it Lower:} The conserved angular momentum $L$ is plotted for circular orbits as a function of their circumferential radius $r$. Each curve corresponds to a different value of the scale radius $\rs$ while the central density of the DM halo is kept to be constant at $\rhs=0.01$. The dotted segments of these curves indicate the regions where the orbits are stable. In contrast, the solid and dashed segments represent regions where the orbits are unstable. The locations of the ISCO for each configuration are marked by circular points, MBO radii are denoted by square points. Note that the displayed range for $L$ corresponds to units of $10^{-5}$.}
        \label{f:L_diff_rhos}
    \end{figure}

\noindent
First, we examine the MBO for a Schwarzschild BH embedded within a Dehnen-type DM halo. This orbit represents a critical circular trajectory where the particle possesses the maximum possible energy while remaining gravitationally bound. The MBO is characterized by the following conditions:
    \begin{eqnarray}
        \left(\frac{dx}{d\phi}\right)^2 = \mathcal{E_{\textrm{eff}}}(x) - \mathcal{V_{\textrm{eff}}}(x) = 0~, \quad \frac{d\mathcal{V}_{\textrm{eff}}(x)}{dx} = 0~,\label{eq:MBO}
    \end{eqnarray}
\noindent
where the energy parameter is set to unity in units of $10^{-5}$ by considering an EMRI system, i.e. the ratio of $q\equiv m/M=10^{-5}$. Solving this system of equations yields the specific radius $r_{\rm MBO}$ and the corresponding angular momentum $L_{\rm MBO}$ for the orbit.\\
For a Schwarzschild BH influenced by a DM halo in our context, the values for $r_{\rm MBO}$ and $L_{\rm MBO}$ can be determined numerically, and their solutions inherently incorporate the effects of the DM halo parameters, namely the central density $\rhs$ and the scale radius $\rs$.\\
On the other hand as previously stated, the MBO is defined as the bound orbit possessing the maximum possible energy, $E=1$. All other bound orbits with energy $E<1$ are confined to regions beyond the radius of the MBO, i.e., they must satisfy $r > r_{\rm MBO}$. On the other hand, the stability of an orbit is determined by the second derivative of the effective potential; specifically, stable orbits fulfill the condition $d^2 \mathcal{V_{\rm eff}}(x)/dx^2>0$, while unstable orbits are characterized by $d^2 \mathcal{V_{\rm eff}}(x)/dx^2<0$. The radius of the ISCO is consequently defined by the conditions:
    \begin{eqnarray}
        \frac{d\mathcal{V}_{\textrm{eff}}(x)}{dx} = 0~, \quad \frac{d^2 \mathcal{V_{\rm eff}}(x)}{dx^2}=0~.\label{eq:ISCO}
    \end{eqnarray} 
\noindent
As noted earlier, bound orbits are only permissible for energies $E<1$ (in units of $10^{-5}$ in our study). Therefore, to locate the ISCO—which signifies the innermost, stable, bound orbit within the combined Schwarzschild BH and DM halo spacetime—one must guarantee that $E<1$ is satisfied. Employing the conditions specified in Eq.~\eqref{eq:ISCO}, we numerically compute the ISCO parameters. These parameters, namely the ISCO radius, orbital angular momentum, and energy, are evaluated as functions of the DM halo's central density and scale radius.\\ 
In Fig.~\ref{f:L_diff_rhos} (Upper), the functional relationship between angular momentum $L$ and orbital circumferential radius $r$ is illustrated for a range of central densities $\rhs$, while the scale radius is held constant at $\rs=0.15$. Correspondingly, Fig.~\ref{f:L_diff_rhos} (Lower) displays the same relationship, but for various values of the scale radii $\rs$ at a fixed central density $\rhs=0.01$.\\
In Fig.~\ref{f:L_diff_rhos}, the ISCO radius is marked by a circular dot, and the MBO radius of the Schwarzschild BH embedded within a Dehnen-type DM halo is denoted by a square box. The lowest red curve in these figures corresponds to the scenario without any DM halo influence ($\rhs=0,\rs = 0$), for which the ISCO and MBO radii are located at $r=6$ and $r=4$ (both in units of $M=1$), with associated angular momentum values of $L=3.464101$ and $L=4$ (both in units of $M=1,m=10^{-5}$), respectively. The influence of the DM halo parameters is evident, as both pairs of values—$r_{\rm MBO}$, $L_{\rm MBO}$ and $r_{\rm ISCO}$, $L_{\rm ISCO}$—show a positive correlation, increasing with higher values of both the central density and the scale radius of the DM halo.

    \begin{figure}[H]
        \centering
        \includegraphics[width=0.49\textwidth]{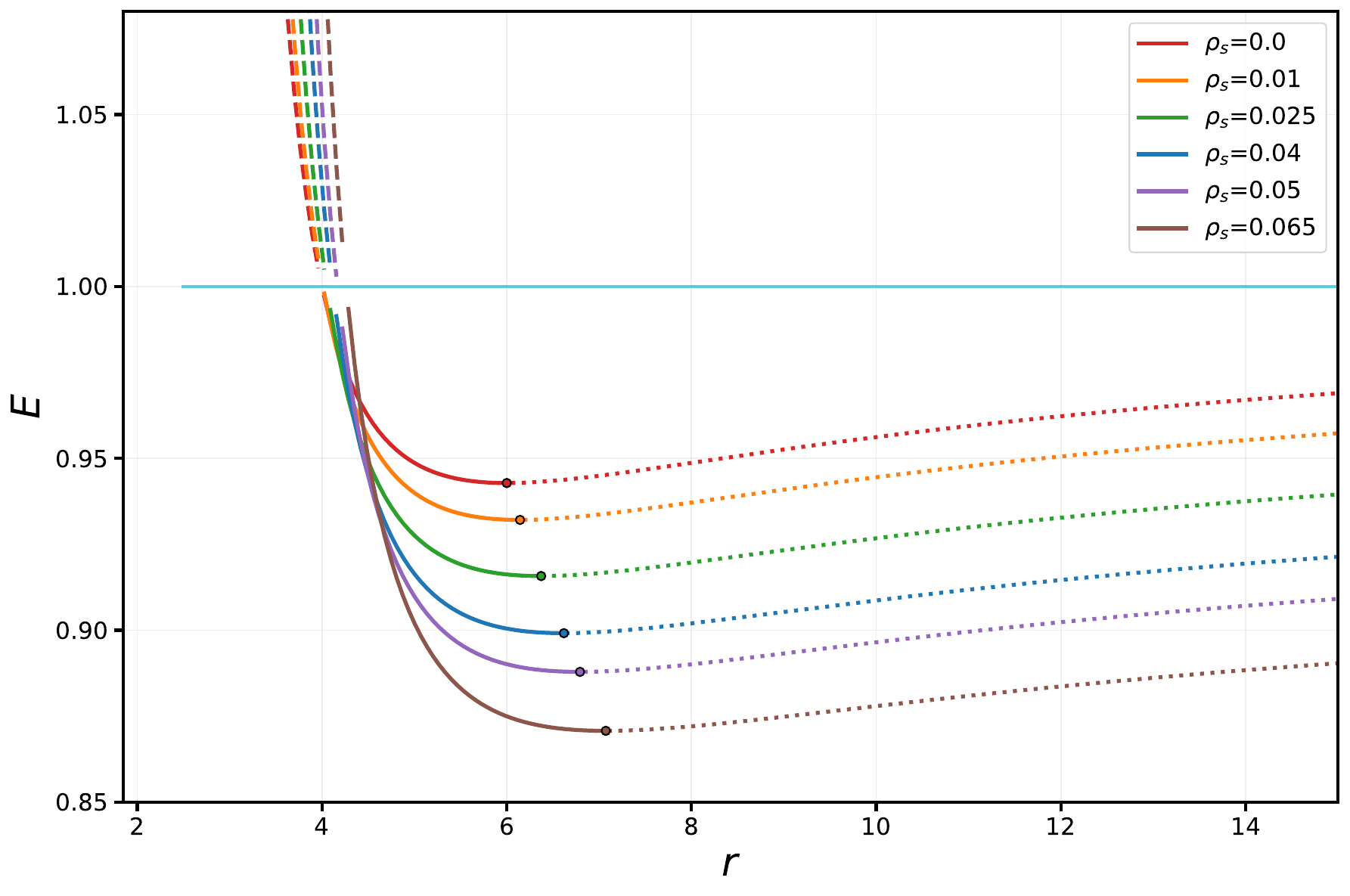}
        \includegraphics[width=0.49\textwidth]{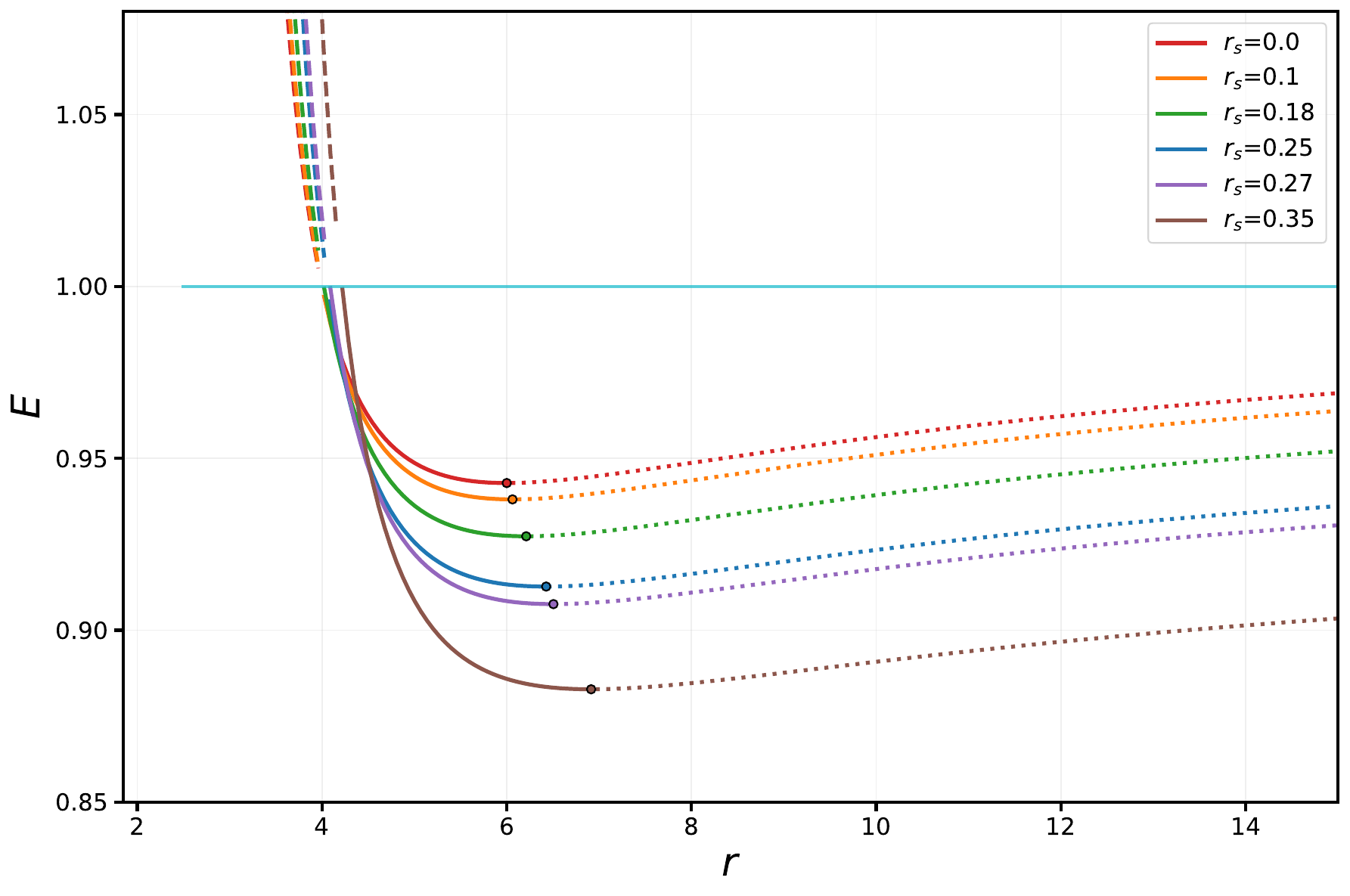}
        \caption{{\it Upper:} The conserved energy $E$ is plotted for circular orbits as a function of their circumferential radius $r$. Each curve corresponds to a different value of the central density $\rhs$ while the scale radius of the DM halo is kept to be constant at $\rs=0.15$. The dotted segments of these curves indicate the regions where the orbits are stable. In contrast, the solid and dashed segments represent regions where the orbits are unstable. The locations of the ISCO for each configuration are marked by circular points. The horizontal line $E=1$ marks the limit of bound orbits. Note that the displayed range for $E$ in vertical axis corresponds to units of $10^{-5}$. \\
        {\it Lower:} The conserved energy $E$ is plotted for circular orbits as a function of their circumferential radius $r$. Each curve corresponds to a different value of the scale radius $\rs$ while the central density of the DM halo is kept to be constant at $\rhs=0.01$. The dotted segments of these curves indicate the regions where the orbits are stable. In contrast, the solid and dashed segments represent regions where the orbits are unstable. The locations of the ISCO for each configuration are marked by circular points. The horizontal line $E=1$ marks the limit of bound orbits. Note that the displayed range for $E$ in vertical axis corresponds to units of $10^{-5}$.}
        \label{f:E_diff_rhos}
    \end{figure}
\noindent
To analyze the influence of the DM halo parameters on the energy of such a BH-DM halo combined system, the conserved energy $E$ is plotted against the radial coordinate $r$ for circular orbits. Fig.~\ref{f:E_diff_rhos} (Upper) displays the change in energy for varying central density values at a constant scale radius $\rs=0.15$, while Fig.~\ref{f:E_diff_rhos} (Lower) illustrates this variation for different scale radii at a fixed central density $\rhs=0.01$.\\
The results clearly indicate that the energy corresponding to the ISCO decreases as both the central density and the scale radius of the DM halo are increased.

\subsection{Homoclinic orbits around a Schwarzschild black hole with dark matter halo}
\begin{figure}[H]
        \centering
        \includegraphics[width=0.49\textwidth]{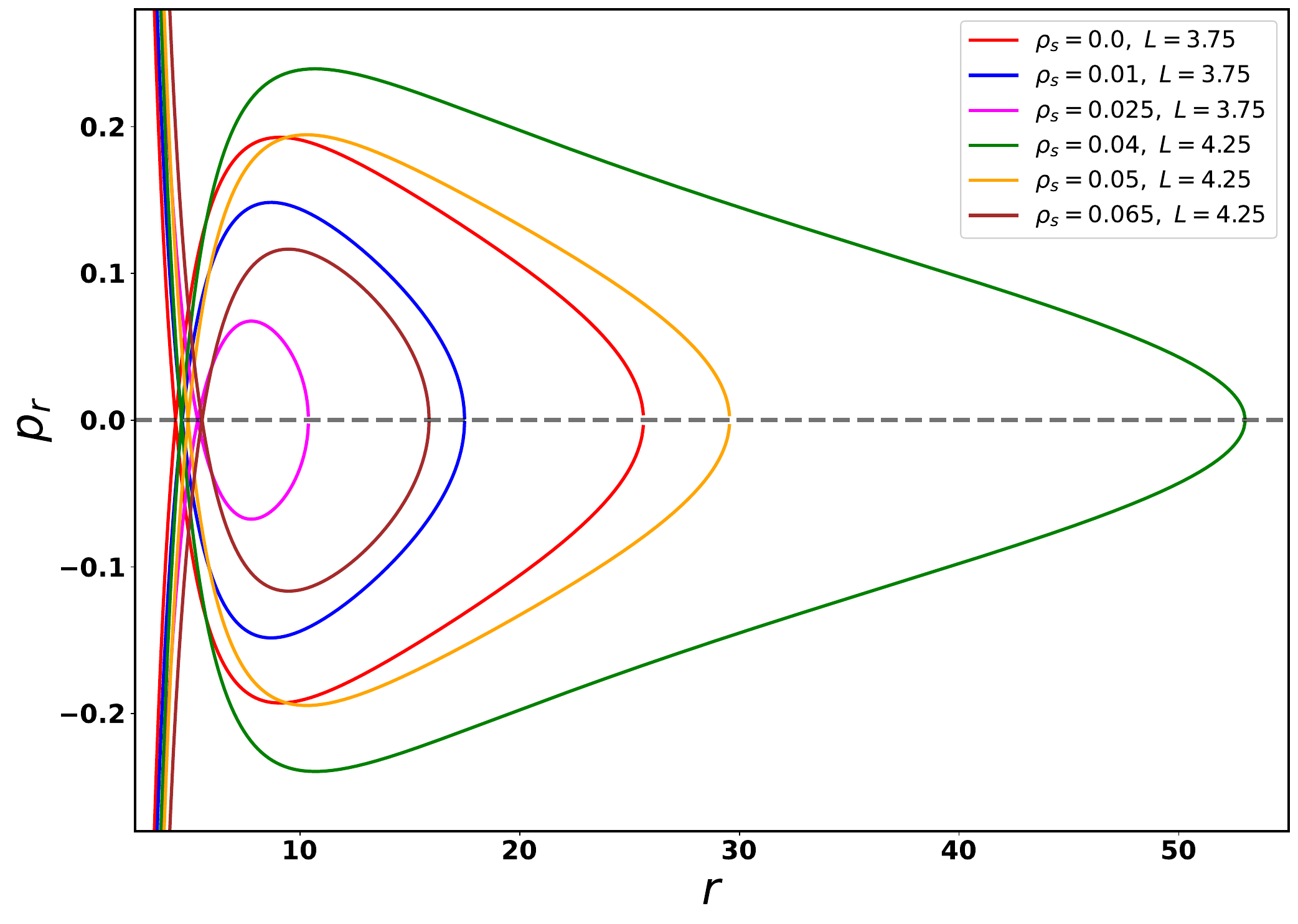}
        \includegraphics[width=0.49\textwidth]{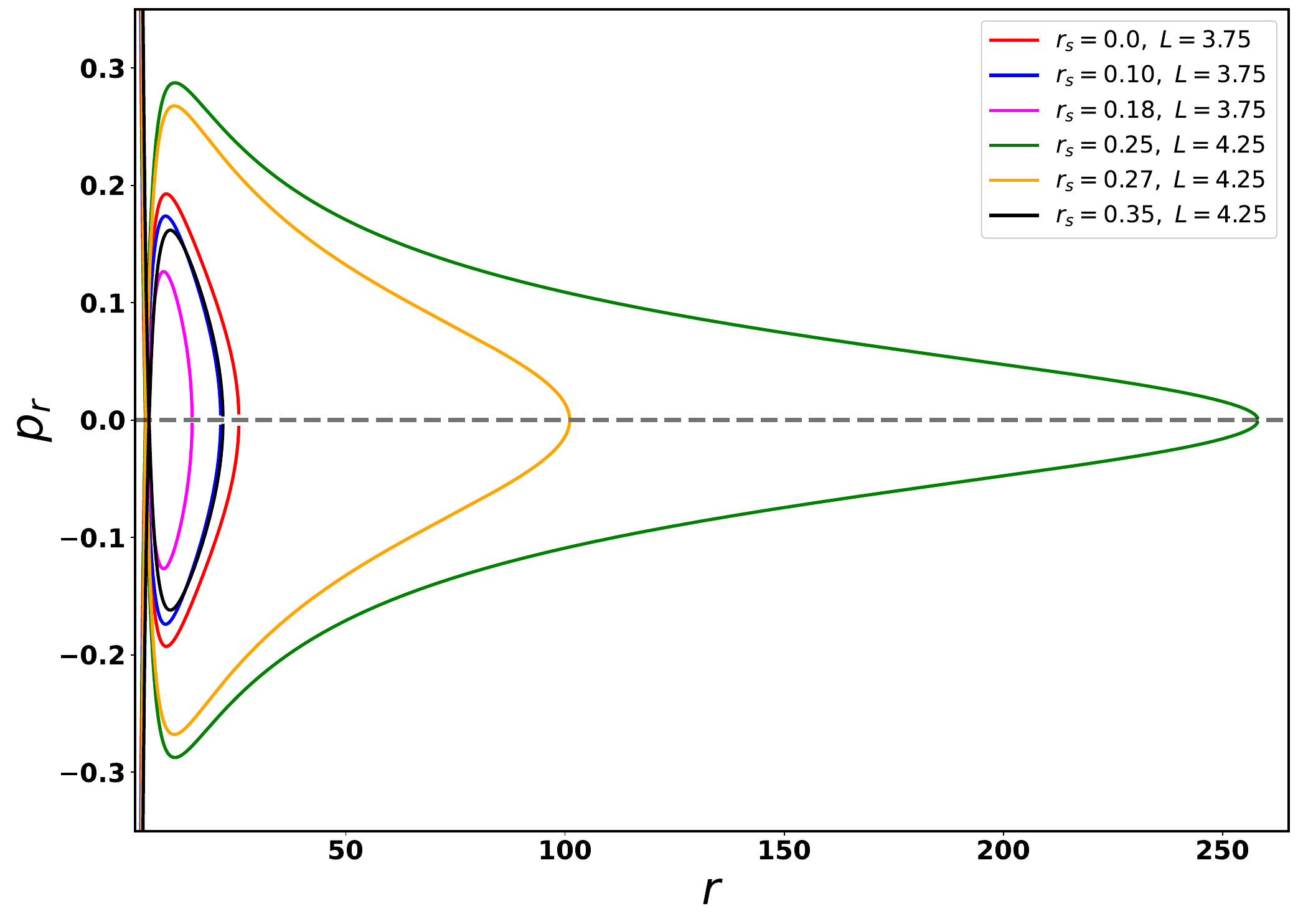}
        \caption{{\it Upper:} The figures illustrate the dual branches of a complete homoclinic orbit of a stellar massive probe particle around a supermassive Schwarzschild BH immeresed in a Dehnen-type DM halo within the $(r, p_r)$ phase space. This is shown for multiple values of the DM halo's central density and the particle's angular momentum. For all cases presented, the scale radius of the DM halo is held constant at $\rs = 0.15$. The trajectory represented by the red curve is a homoclinic orbit occurring in a Schwarzschild spacetime (the specific case corresponds to the absence of DM halo's core density $\rhs = 0$). Note that the displayed range for $p_r$ in vertical axis and the legend labels of $L$ are in units of $10^{-5}$. \\
        {\it Lower:} The figures illustrate the dual branches of a complete homoclinic orbit of a stellar massive probe particle a supermassive Schwarzschild BH immeresed in a Dehnen-type DM halo within the $(r, p_r)$ phase space. This is shown for multiple values of the DM halo's scale radius and the particle's angular momentum. For all cases presented, the central of the DM halo is held constant at $\rhs = 0.01$. The trajectory represented by the red curve is a homoclinic orbit occurring in a Schwarzschild spacetime (the specific case corresponds to the absence DM halo's radius $\rs = 0$). Note that the displayed range for $p_r$ in vertical axis and the legend labels of $L$ are in units of $10^{-5}$.}
        \label{f:pr_diff_rhos}
    \end{figure}

\begin{figure}[htbp]
        \centering
        \includegraphics[width=0.49\textwidth]{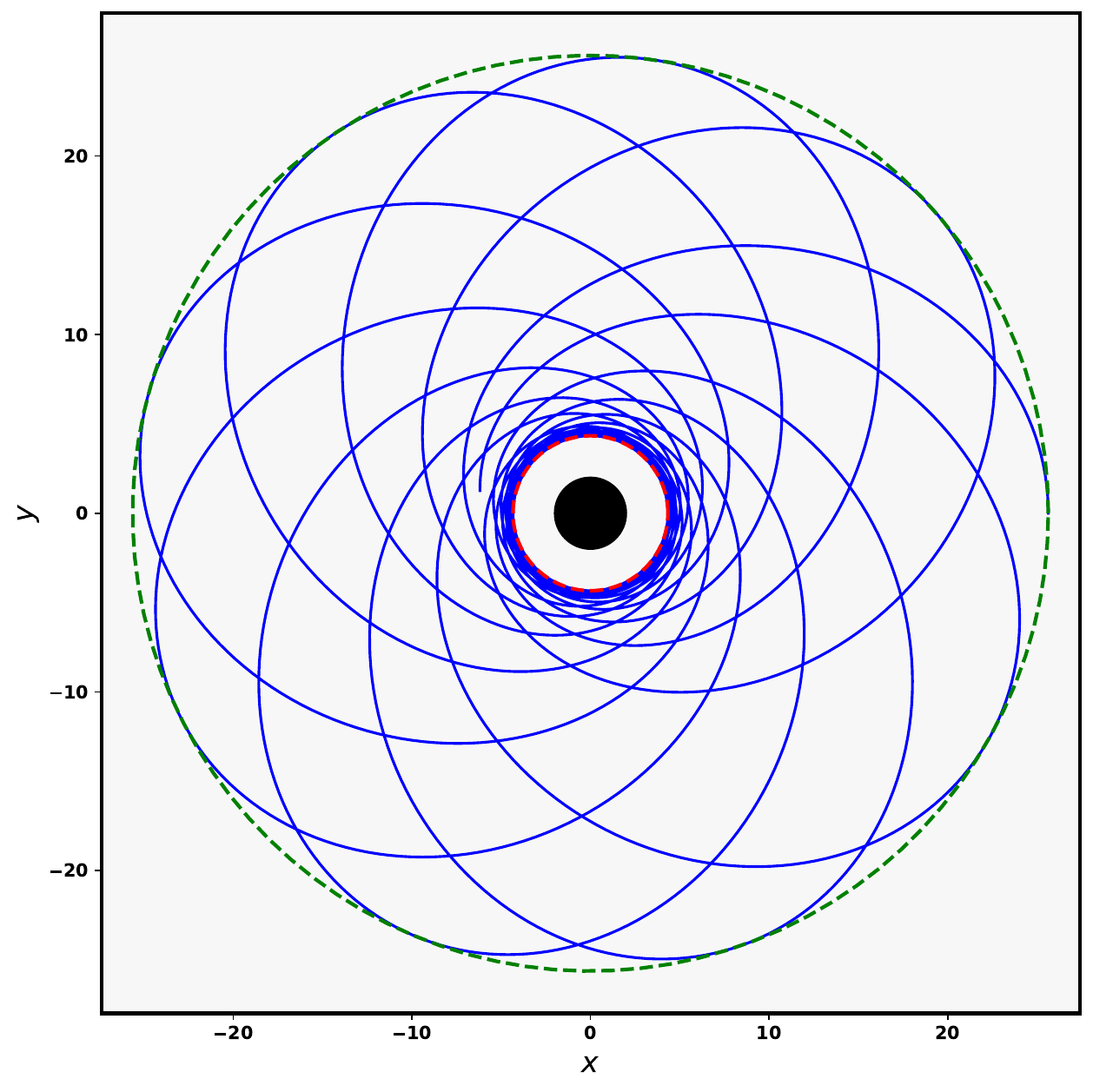}
        \caption{A segment of a homoclinic orbit is plotted around a Schwarzschild BH with $L=3.75\times10^{-5}$. The boundaries of the unstable circular orbits, $r_{\rm un}$, and the maximum attainable radii, $r_{\rm max}$, are indicated by the red and green dotted circles, respectively. The central solid black circle denotes the location of the Schwarzschild BH.}
        \label{f:homo_SC}
    \end{figure}

\noindent
The preceding analysis makes it clear that the combined BH and DM halo spacetime possesses precisely one homoclinic orbit within the $(r, p_r)$ phase space for any combination of the parameters $M$, $m$, $\rhs$, and $\rs$, provided the angular momentum $L$ meets the condition $L_{\rm ISCO} < L < L_{\rm MBO}$. For such a homoclinic orbit, which is associated with a single hyperbolic fixed point located at $(r_{\rm un} \equiv 2M/x_{\rm un}, 0)$, the corresponding energy value is not arbitrary. This specific energy can be determined by identifying a circular orbit solution (i.e., enforcing $dx/d\phi = 0$) at the radial coordinate $x = x_{\rm un}$ for a given value of $L$ and other DM halo parameters, can be derived from Eq.~\eqref{eq:eff3} using Eqs.~\eqref{eq:pot2} and \eqref{eq:en1}.\\
It is important to note that in Figs.~\ref{f:L_diff_rhos} and \ref{f:E_diff_rhos}, the permissible ranges of energy and orbital angular momentum for the existence of an unstable homoclinic orbit around a Schwarzschild BH embedded in a Dehnen-type DM halo are illustrated. These allowed parameter regions are denoted by solid colored lines, which correspond to different values of the DM halo parameters $\rhs$ and $\rs$ (see Figs.~\ref{f:L_diff_rhos} and \ref{f:E_diff_rhos}). An analysis of these figures reveals that for higher values of the DM halo parameters, the corresponding homoclinic orbit is characterized by a reduced energy boundary and an elevated angular momentum boundary.\\
In the context of a Schwarzschild BH with a Dehnen-type DM halo spacetime, a homoclinic orbit originates from an unstable circular orbit located at $r_{\text{MBO}} < r_{\text{un}} = 2M / x_{\text{un}} < r_{\text{ISCO}}$ in the infinite past. This orbit spirals outward to reach a maximum radial distance of $r_{\text{max}} = 2M / x_{\text{max}}$, before asymptotically winding back toward $r_{\text{un}}$. To determine this value of $r_{\text{max}}$, it is necessary to find the third numerical solution to the third-order algebraic equation $\mathcal{V}_{\text{eff}}(x) = \mathcal{V}_{\text{eff}}(x = x_{\text{un}})$ (Eq.~\eqref{eq:pot2}), for which $x = x_{\text{un}}$ is already a two-fold degenerate solution. A crucial point is to note that for a Schwarzschild BH surrounded by a Dehnen-type DM halo, this maximum solution $r_{\text{max}}$ is always guaranteed to exist within the range $r_{\text{ISCO}} < r_{\text{max}} = 2M / x_{\text{max}} < \infty$.\\    
Let us now examine how a homoclinic orbit can be formed in a reduced two-dimensional $(r,p_r)$ phase space corresponding to a single hyperbolic fixed point situated at $(r_{\rm un},0)$ for a given spacetime parameters $M,~m,$ the DM halo parameters $\rhs,~\rs$ along with a suitable choice of angular momentum $L$.\\    
The corresponding energy $E$ is a function of these same parameters. It is straightforward to write Eq.~\eqref{eq:eff1} using Eq.~(2) as
    \begin{eqnarray}
        E^2-m^2&=&m^2\left(\frac{dr}{d\tau}\right)^2-\frac{2Mm^2}{r}-32\pi\rhs\rs^2m^2\sqrt{1+\frac{\rs}{r}}\non\\
        &+&\frac{L^2}{r^2}\left(1-\frac{2M}{r}-32\pi\rhs\rs^2\sqrt{1+\frac{\rs}{r}}\right)~.\label{eq:eff4}
    \end{eqnarray}  
\noindent
From Eq.~\eqref{eq:eff4}, we can derive an expression for the radial momentum $p_r$ of a particle in the spacetime of a Schwarzschild BH surrounded by a Dehnen-type DM halo. We obtain the following expression of $p_r$:
    \begin{align}
        p_r &:= g_{rr}p^r \nonumber \\
        &=  \pm\, \frac{ \sqrt{ E^2 - \left( m^2 + \frac{L^2}{r^2} \right) \left( 1 - \frac{2M}{r} - 32\pi \rhs \rs^2 \sqrt{1 + \frac{\rs}{r}} \right) } }{ \left( 1 - \frac{2M}{r} - 32\pi \rhs \rs^2 \sqrt{1 + \frac{\rs}{r}} \right) }~.\label{eq:pr}
    \end{align}
The $\pm$ signs denote the two distinct branches of the orbit, corresponding to the trajectory before and after it reaches the maximum radial distance, $r_{\rm max}$. These signs are effectively functions of the angular coordinate $\phi$. Utilizing the parameter sets specified in Table~\ref{tab1} and Table~\ref{tab2}, a homoclinic orbit for a stellar massive object in the spacetime of a supermassive Schwarzschild BH surrounded by a Dehnen-type DM halo can be plotted in phase space. This resulting orbit is presented in Fig.~\ref{f:pr_diff_rhos}.\\
As is clearly illustrated by the preceding plots, along with the data in Table~\ref{tab1} and Table~\ref{tab2}, a reduction in the homoclinic energy $E_{\rm homo}$ results in a corresponding decrease of the maximum radial distance $r_{\rm max}$, for a given value of angular momentum and other DM halo parameters. Conversely, the stable radial point $r_{\rm st}$ is situated between the two fixed points $r_{\rm un}$ and $r_{\rm max}$, while consistently remaining at a greater distance than the ISCO radius, fulfilling the condition $r_{\rm st}>r_{\rm ISCO}$.\\
Note that the phase space exhibits distinct instability within the region defined by $r<r_{\rm un}$. In the vicinity of the event horizon, the radial momentum increases significantly, and the phase space resembles the form of a rectangular hyperbola. This characteristic is a fundamental trait of BH spacetime near the event horizon, as described in \cite{Dalui:2018qqv}. In the following sections, we will explore the role of chaotic dynamics for an observer positioned at such an unstable radial point near the event horizon of a combined BH and DM halo spacetime. We will further examine the influence of the DM halo parameters on the behavior of this chaotic dynamics in the proceeding section.\\
A key observation from Fig.~1 is that the structure of a homoclinic orbit segment is distinct for each combination of $(\rhs,L)$ and $(\rs,L)$. For a given values of DM halo parameter ($\rhs$ and $\rs$), no two homoclinic trajectories are identical. As clearly illustrated in Fig.~1, these homoclinic orbits originate from an unstable circular orbit radius (marked by a red dotted circle in Figs.~1 and \ref{f:homo_SC}),

\noindent
expand outward to a maximum radius (indicated by a green dotted circle in Figs.~1 and \ref{f:homo_SC}), and then spiral back into the same unstable circular orbit $r_{\rm un}$.\\    
Additionally, let us mention here that the energy value chosen for these homoclinic orbit is critical. For a specific set of $\rhs$, $\rs$, and $L$, a test particle with energy higher than our selected value would plunge directly into the event horizon. Conversely, with lower energy, the particle would follow a different, non-homoclinic trajectory. Lastly, while the homoclinic orbits in Figs.~1 and \ref{f:homo_SC} are drawn for a sufficiently long but finite duration, a complete homoclinic orbit around a Schwarzschild BH embedded in a Dehnen-type DM halo formally requires an infinite time to approach a complete homoclinic orbit.

    \begin{table*}[htbp]
        \centering
        \caption{The numerical values of horizon radius ($r_H$), unstable point $(r_{\rm un})$, maximum point $(r_{\rm max})$, stable point $(r_{\rm st})$ and the homoclinic energy $(E_{\rm homo})$ of the orbits with a set of different central halo density $\rhs$ and angular momentum $L$ for a Schwarzschild BH immeresed in a Dehnen-type DM halo. The scale radius of the DM halo is fixed at $\rs=0.15$. Note that the first row $\rhs=0.0$ denotes the Schwarzschild BH case.}\label{tab1}
        \begin{tabular}{cccccc}
        \hline
        \hline
        $\left(\rhs,L(10^{-5})\right)$ & $r_{H}$ & $r_{\rm un}=2M/x_{\rm un}$ & $r_{\rm max}=2M/x_{\rm max}$ & $r_{\rm st}=2M/x_{\rm st}$ & $E_{\rm homo}(10^{-5})$\\
        \hline
        (0.0, 3.75)  & 2.0 & 4.338486 & 25.634637 & 9.724013 & 0.970418\\
        \hline
        (0.01, 3.75)  & 2.04799 & 4.639192 & 17.490476 & 9.093595 & 0.950055\\
        \hline
        (0.025, 3.75)  & 2.124295 & 5.341590 & 10.381593 & 7.897674 & 0.921035\\
        \hline
        (0.04, 4.25)  & 2.206295 & 4.604188 & 53.017635 & 11.768592 & 0.936628\\
        \hline
        (0.05, 4.25)  & 2.26445 & 4.904680 & 29.555263 & 11.047455 & 0.914265\\
        \hline
        (0.065, 4.25)  & 2.357467 & 5.537265 & 15.874213 & 9.785232 & 0.882241\\
        \hline
        \hline
        \end{tabular}   
    \end{table*}
    
    \begin{table*}[htbp]
        \centering
        \caption{The numerical values of horizon radius $(r_H)$, unstable point $(r_{\rm un})$, maximum point $(r_{\rm max})$, stable point $(r_{\rm st})$ and the homoclinic energy $(E_{\rm homo})$ of the orbit with a set of different scale radii $\rs$ and angular momentum $L$ for a Schwarzschild BH immeresed in a Dehnen-type DM halo. The central density of the DM halo is fixed at $\rhs=0.01$. Note that the first row $\rs=0.0$ denotes the Schwarzschild BH case.}\label{tab2}
        \begin{tabular}{cccccc}
        \hline
        \hline
        $\left(\rs,L(10^{-5})\right)$ & $r_{H}$ & $r_{\rm un}=2M/x_{\rm un}$ & $r_{\rm max}=2M/x_{\rm max}$ & $r_{\rm st}=2M/x_{\rm st}$ & $E_{\rm homo}(10^{-5})$\\
        \hline
        (0.0, 3.75)  & 2.0 & 4.338486 & 25.634637 & 9.724013 & 0.970418\\
        \hline
        (0.10, 3.75)  & 2.020812 & 4.461267 & 21.483409 & 9.456366 & 0.961329\\
        \hline
        (0.18, 3.75)  & 2.070304 & 4.805105 & 14.970734 & 8.779485 & 0.941247\\
        \hline
        (0.25, 4.25)  & 2.142238 & 4.320333 & 258.063870 & 12.541229 & 0.964178\\
        \hline
        (0.27, 4.25)  & 2.168529 & 4.432143 & 101.069048 & 12.224454 & 0.953116\\
        \hline
        (0.35, 4.25)  & 2.304603 & 5.148448 & 22.001844 & 10521493 & 0.902365\\ 
        \hline
        \hline
        \end{tabular}   
    \end{table*} 

\section*{Lagrangian of a massive probe in curved background}
\noindent
The Lagrangian, subject to an external potential $V(x^\alpha)$ of a massive test particle of mass $m$ is given by
    \begin{equation}
        \mathcal{L} = -m \sqrt{ -g_{\mu\nu} \dot{x}^\mu \dot{x}^\nu }-V(x^\mu)~,\label{eq:Lagrangian}
    \end{equation}
where $g_{\mu\nu}$ denotes the metric tensor. Throughout this analysis, we employ the static gauge condition where $t = \tau$, meaning the over dot denotes a derivative with respect to the proper time $t$. It is noteworthy that an alternative, yet equivalent, formulation of the Lagrangian exists apart from the one, mentioned in Eq.~\eqref{eq:Lagrangian}.\\
This other common form is: $2\mathcal{L} = g_{\mu\nu} \dot{x}^\mu \dot{x}^\nu - V(x^\mu)~$, which is often used to characterize the motion of a massive test particle in a curved spacetime \cite{Wang:2026whj,Paddy}. Now, regarding the external potential, in a realistic astrophysical setting this external potential $V(x^\mu)$ may be interpreted as an effective perturbation arising, for example, from the gravitational influence of nearby stars or from perturbations of the surrounding DM halo. Such effects can break the exact integrability of the idealized background motion. In the present work, we introduce $V(x^\mu)$ phenomenologically in order to keep the particle in a bounded near-horizon region and to probe the resulting horizon-induced nonlinear dynamics in a controlled manner. Similar effective external potentials have also been employed in earlier studies in chaotic dynamics \cite{Hashimoto:2016dfz,Lei:2020clg,Wang:2026whj}, where the origin of such potentials can be considered as the form of electromagnetic, scalar or higher spin potentials.

\section*{A better choice of coordinates in our analysis}
\noindent
The Painlev\'e-Gullstrand coordinates transforms the time coordinate as
    \begin{equation}
        dt \rightarrow dt - \frac{\sqrt{1-f(r)}}{f(r)} dr.~\label{3.1}
    \end{equation}
This transformation yields the modified metric (which is regular at the horizon) as
    \begin{equation}
        ds^2 = -f(r)dt^2 + 2\sqrt{1-f(r)}dt dr + dr^2 + r^2 d\Omega^2~.\label{3.2}
    \end{equation}
\noindent
Before proceeding further into the dynamics, let us pause to understand in brief why we choose to proceed with this particular coordinate transformation. First of all, the Painlev\'e-Gullstrand coordinate system is particularly advantageous because it remains well-defined at the horizon, enabling more robust numerical computations of particle motion in its immediate vicinity (which is a primary focus of our study) \cite{Martel:2000rn,pg2}. Additionally, the time coordinate in this framework describes spacetime foliations that maintain regularity across the horizon. Unlike the Schwarzschild coordinates, where time appears to ``freeze" at the horizon because of extreme gravitational time dilation effect, this formulation avoids such coordinate artifacts. Previous work by two of the present authors has shown that analyzing near-horizon particle dynamics in Painlev\'e-Gullstrand coordinates provides better insights into chaotic phenomena and its quantification compared to the Schwarzschild coordinates, while confirming that the conclusions remain coordinate-independent \cite{Das:2024iuf}. This is because, when approaching the horizon (rather than the asymptotic region), the regular behavior of the Painlev\'e-Gullstrand system at the horizon makes numerical computations more reliable and effective.

\section*{Analytical equations of motion}
\noindent
The primary objective of this work is to analyze how the BH's event horizon as well as the core density and scale radius of the particular Dehnen-type DM halo influence particle's motion. Due to the existence of spherical symmetry, we are considering the particle motion in the equatorial plane, defined by $\theta=\frac{\pi}{2}$ with $p_{\theta}=0$. Therefore from Eq.~\eqref{eq:Lagrangian} by using the metric Eq.~\eqref{3.2}, the Lagrangian can be written as
    \begin{eqnarray}
        \mathcal{L}=-m\sqrt{f(r)-2\sqrt{1-f(r)}\dot{r}-\dot{r}^2-r^2\dot{\phi}^2}\non\\
        -V(r,\phi)~,\label{eq:Lag1}
    \end{eqnarray}
After restricting the dynamics to the equatorial plane, we take the general external potential $V(x^\mu)$ appearing in Eq.~\eqref{eq:Lagrangian} in a separable form, $V(x^\mu)\big|_{\theta=\pi/2}=a(r)+b(\phi)$, where $a(r)$ and $b(\phi)$ represent, the radial and azimuthal contributions to the same effective external perturbation, respectively. Therefore, the two potentials introduced in Eq.~\eqref{eq:Lag1} should be understood as a decomposition of $V(x^\mu)$ in the reduced phase-space description, rather than as two unrelated external interactions. For the numerical analysis, in the latter part of the paper, we adopt the simplest phenomenological choice, i.e. the harmonic potentials which provides a controlled confinement of the particle near the horizon and allows us to isolate the resulting nonlinear dynamics.\\
Now from the above equation the generalized momenta along $r$ and $\phi$ directions can be written as
    \begin{eqnarray}
        p_r=\frac{m\left(\dot{r}+\sqrt{1-f(r)}\right)}{\sqrt{f(r)-2\sqrt{1-f(r)}\dot{r}-\dot{r}^2-r^2\dot{\phi}^2}}~,\non\\
        p_\phi=\frac{mr^2\dot{\phi}}{\sqrt{f(r)-2\sqrt{1-f(r)}\dot{r}-\dot{r}^2-r^2\dot{\phi}^2}}~.\label{eq:momenta}
    \end{eqnarray}
Therefore we can write the conserved energy of massive test particle as the following.
    \begin{align}
        E&=p_r\dot{r}+p_\phi\dot{\phi}-\mathcal{L}\non\\
        &=-\sqrt{1-f(r)}~p_r \pm \sqrt{p^2_r + \frac{p^2_\phi}{r^2}+m^2}+\left(a(r)+b(\phi)\right)~,\label{3.4}
    \end{align}    
\noindent
where the negative sign corresponds to ingoing particle and the positive sign is for outgoing particle by general convention. Since we are interested to investigate the dynamics of a probe particle when it is very near to the horizon, our analysis focuses on outgoing particle.

\section*{The choice of the external potentials}
\noindent
As we have mentioned earlier that it is instructive to consider a phenomenological choice that describes the near-horizon geometry of the BH-DM halo combined spacetime with external potentials as a harmonic oscillator, as the simplest case, since a particle motion near the event horizon of a Schwarzschild BH-DM halo spacetime does not refer to any concrete form of the external potentials. Importantly, these harmonic potentials represent an externally imposed constraint rather than an inherent feature of such Schwarzschild BH-DM halo spacetime. At sufficiently large distances from the horizon, the system remains integrable and regular, exhibiting periodic orbital behavior with preserved invariant tori. However, close to the horizon introduces strong nonlinear gravitational effects that disrupt these regular tori (in the sense of a Kolmogorov-Arnold-Moser tori) and thus the breakdown of integrability facilitates the detection of the horizon's distinctive influence.\\
It is also noteworthy to mention that in the background (Painlev\'e-Gullstrand coordinates) spacetime of a Schwarzschild BH-DM halo, we have also identified a local maximum, located very close to the horizon in the effective potential, resembles an ``inverted harmonic oscillator" (see the detailed analysis regarding such unstable maxima at the end of this Supplemental Material) \footnote{Note that one can also find similar kind of unstable maxima for the expression of the one dimensional effective potential of the BH-DM halo combined system in the Schwarzschild coordinates \cite{Al-Badawi:2024asn}, as discussed in Sec.~3}. Therefore existing of such instability, close to the horizon indicates that if we perturb the probe particle, it may either fall into the black hole or escape to infinity, resulting in an unbound phase space trajectory. To determine whether this instability leads to chaos within a certain parameter range of the existing geometry, we need to confine the particle in the phase space.\\
Therefore to probe the horizon, we consider the form of the external potentials $a(r)=\frac{1}{2}K_r(r-r_c)^2$ and $b(\phi)=\frac{1}{2}K_\phi r_H^2\phi^2$. Here $K_r$ and $K_{\phi}$ are two strength constants of $a(r)$ and $b(\phi)$, respectively and $r_c$ denotes the center position of $a(r)$. It is reasonable to mention that this framework follows established approaches (see \cite{Hashimoto:2016dfz,Lei:2020clg,Wang:2026whj} for massive particle and \cite{Dalui:2018qqv,Bera:2021lgw,Das:2024iuf} for massless cases), while alternative potential choices would modify the trajectory dynamics.

\section*{Investigation on Poincar$\Acute{E}$ sections}
\noindent
The Poincar$\acute{e}$ map serves as a fundamental technique for analyzing nonlinear dynamical systems. It is constructed by examining the points where periodic (regular or non-chaotic) or aperiodic (chaotic) orbits intersect a cross-sectional subspace that is transverse to the flow in the complete state space. The core principle involves projecting higher-dimensional phase space trajectories onto a lower-dimensional subspace using this mapping technique \cite{Strogartz}. In our analysis, we select $\phi=0$ plane as the Poincar$\acute{e}$ section. Within this framework, we record the coordinates on the $(r-p_r)$ phase plane whenever the particle crosses this section, adhering to the conditions of constant energy $E$ and $p_{\phi}>0$. For periodic orbits, the resulting points form a toroidal structure in phase space, whereas chaotic dynamics manifest through the disintegration of these tori. The presence of fragmented tori distributed across various regions of the phase space serves as a distinctive indicator of chaotic behavior \cite{Strogartz,Stock}.\\
Next, we proceed with the numerical integration of the equations of motion [Eqs.~(3),~(4),~(5) and~(6)] corresponding to a Schwarzschild BH embedded within a DM halo. These computational solutions enable us to generate the Poincar$\acute{e}$ sections. Our study provides numerical verification by investigating these sections for a SSS BH solutions, with particular attention to how the DM halo parameters $\rhs$ and $\rs$ influence the dynamics. Furthermore, it is possible to gain a comprehensive insight into how the parameters of the DM halo affect the chaotic dynamics exhibited within such a BH-DM halo spacetime.\\
To investigate the influence of the DM halo parameters $\rhs,~\rs$ and energy $E$ on the dynamics of particle, which is located close to the event horizon, we first construct the Poincar\'{e} section by examining four distinct scenarios. These cases are outlined as follows:
    \begin{itemize}
        \item \underline{Case-I:} Fixed scale radius $\rs = 0.15$ and central density $\rhs = 0.02$, with varying energy levels.
        \item \underline{Case-II:} Fixed central density $\rhs = 0.01$ and scale radius $\rs = 0.25$, with varying energy levels.
        \item \underline{Case-III:} Fixed scale radius $\rs = 0.15$ and energy $E = 90$, with varying central density.
        \item \underline{Case-IV:} Fixed central density $\rhs = 0.01$ and energy $E = 115$, with varying scale radius.
    \end{itemize}
When $\rhs$ and $\rs$ are set to zero, the background resembles that of a Schwarzschild BH in Einstein's general relativity. By introducing non-zero values for $\rhs$ and $\rs$ as the contributions of a Dehnen-type DM halo, we can explore how DM halo affects the chaotic behavior of particle near the event horizon. Additionally, we note that increasing $\rhs$ and $\rs$ lead to an expansion in the BH's size, a crucial factor in our subsequent analysis. The condition $\sqrt{1 - f(r)} > 0$ in Eq.~(4) restricts the radial coordinate $r$ to the range $r > r_H$, where $r_H$ is defined by Eq.~\eqref{2n}. We numerically solve the equations of motion [Eqs.~(3),~(4),~(5) and~(6)] using the fourth-order Runge-Kutta method, treating as a mass ratio $q=10^{-5}$ for a large set of initial values with a fixed step size of $h = 0.01$. For the Poincar\'e-section analysis considered across all four cases, we use the parameter values $K_r=100$, $K_\phi=25$, and $r_c=3.65$ (to ensure particle remains close to the event horizon). The initial values of the remaining phase-space variables are chosen randomly within the ranges
\[3.0<r<3.8~,~~~ -0.5<p_r<0.5~,~~~ -0.05<\phi<0.05~,\]
while the initial conjugate momentum $p_\phi$ is determined from the conserved-energy relation in Eq.~\eqref{3.4}, with the additional condition $p_\phi>0$ at the section $\phi=0$ for determining the Poincare plots. Different colors in the Poincar\'e plots correspond to trajectories generated from different initial conditions chosen within these ranges.\\
In Figs.~2(a) and 2(b), we present the Poincar\'{e} sections of a massive particle around a Schwarzschild BH with DM halo, projected onto the $(r-p_r)$ phase plane, with $\phi = 0$ and $p_{\phi} > 0$, for two distinct energy configurations. The first case (Fig.~2(a)) considers different values of energy as $E = 108$, $130$, and $132.5$ for a fixed DM halo parameters $\rs = 0.15$ and $\rhs = 0.02$, while the second case (Fig.~2(b)) examines energies $E = 95$, $118$, and $122$ with another set of DM halo parameters $\rhs = 0.01$ and $\rs = 0.25$. At the lowest energies ($E = 108$ and $E = 95$ for the respective cases), the Poincar\'{e} sections exhibit regular Kolmogorov-Arnold-Moser (KAM) tori \cite{KAM}, suggesting that the particle's motion remains largely confined near the center of the external potential, with only a single dominant frequency present. However, as the total energy rises—consistent with the Hamiltonian's conservation—the momentum increases, driving the trajectory closer to the BH's event horizon. Therefore for intermediate energies ($E = 130$ and $E = 118$ in the two cases), the Poincar\'{e} sections reveal distorted tori, signaling a departure from regularity. At the highest energies ($E = 132.5$ and $E = 122$ in the two cases), as illustrated in Figs.~2(a) (Right) and 2(b) (Right), the tori disintegrate entirely, giving way to scattered points in the phase plane. This disordered distribution of points signifies the chaotic dynamics, with the system exhibiting fully chaotic behavior.\\
The emergence of chaotic dynamics in this system exhibits notable variations depending on the choice of DM halo parameters $\rhs$ and $\rs$, as clearly illustrated by the Poincaré maps (see Figs.~2(c) and 2(d)) for a given constant energy. When the core radius is fixed at $\rs = 0.15$ and the system's energy is set to $E = 90$, the toroidal structures start breaking down at a comparatively higher central density, as shown in Fig.~2(c) (Right). This behavior can be explained as follows: at large distances from the horizon, the BH-DM halo combined system constitutes an integrable system characterized by regular or non-chaotic trajectories. However, when the particle approaches the horizon, the nonlinear effects induced by the BH-DM halo spacetime geometry dominate, leading to chaotic motion. Therefore, particle's distance from the event horizon significantly influences the development of chaos, a phenomenon previously noted in \cite{Dalui:2018qqv} for Einstein gravity and in \cite{Das:2024iuf} for $f(R)$ gravity (both, in the case of massless particle). A similar trend occurs for a fixed central density of the DM halo ($\rhs = 0.01$) with a different energy ($E = 115$). In this case, as seen in Fig.~2(d), a fully chaotic regime arises at a relatively larger scale radius of the DM halo ($\rs = 0.27$).\\
When the energy of the particle is increased while keeping the DM halo parameters fixed, the momentum rises, and the associated orbits shift closer to the event horizon, leading to a disruption in the structure of the regular tori. Additionally, as discussed in the previous section, higher values of the DM halo parameters $\rhs$ and $\rs$ cause the event horizon to expand, thereby amplifying its influence on the particle's trajectory. Consequently, even with a constant energy, the effect of the DM halo parameters on the particle's dynamics becomes evident—almost as if the horizon moves toward the particle to interact with it (for more clear visualization, see Fig.~\ref{T1}). As a result, chaotic behavior is expected to emerge at higher energy levels for fixed DM halo parameters or for larger values of DM halo parameters at a fixed energy. It is worthy to mention that arbitrarily high energy values cannot be selected in this framework due to the onset of numerical instabilities, as anticipated.

\section*{Trajectory analysis}
\begin{figure*}[htbp]
        \centering   
        \begin{minipage}[b]{0.3\textwidth}
        \centering
        \includegraphics[width=\textwidth, height=4.5cm]{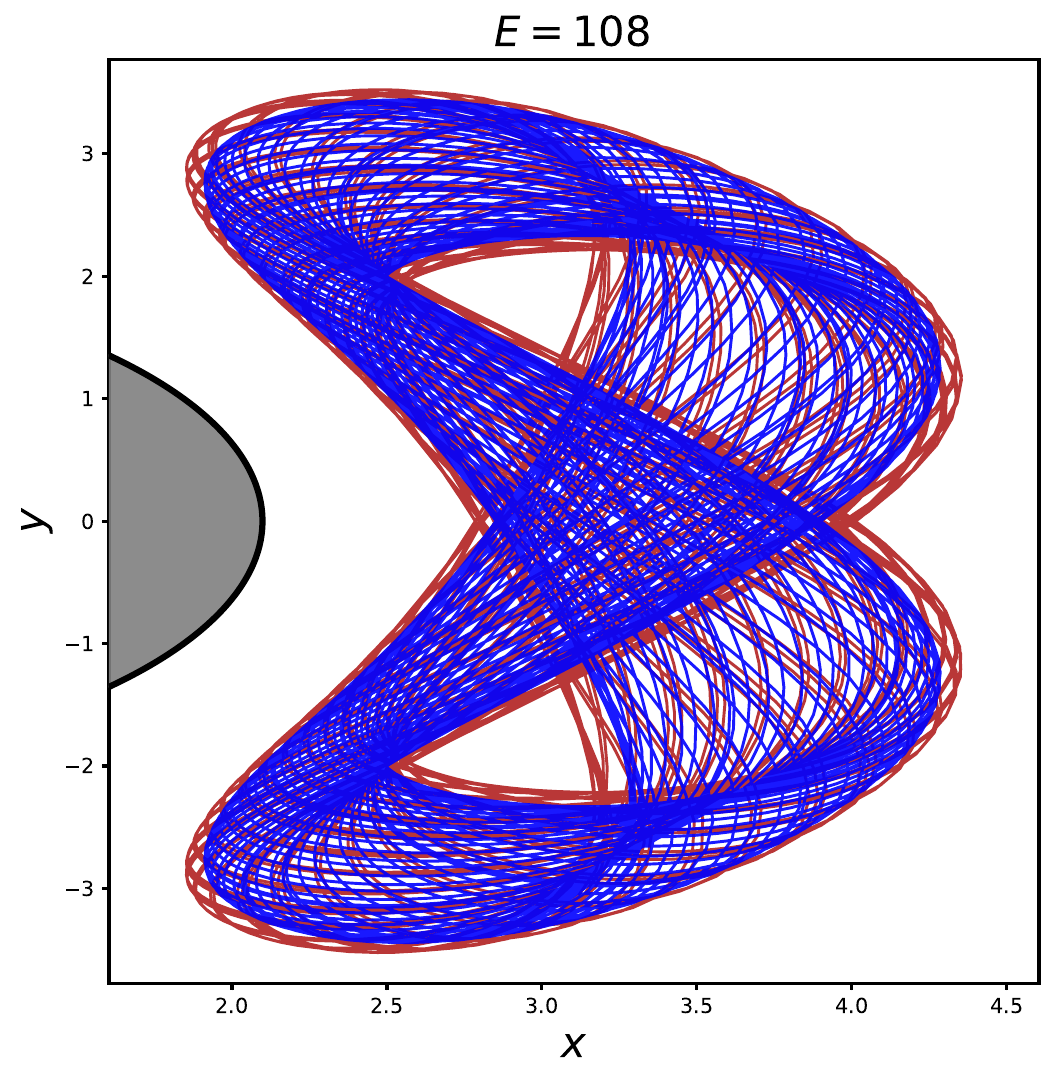}
        {\footnotesize I(a) $\rhs=0.02,\rs=0.15$}
        \end{minipage}
        \begin{minipage}[b]{0.3\textwidth}
        \centering
        \includegraphics[width=\textwidth, height=4.5cm]{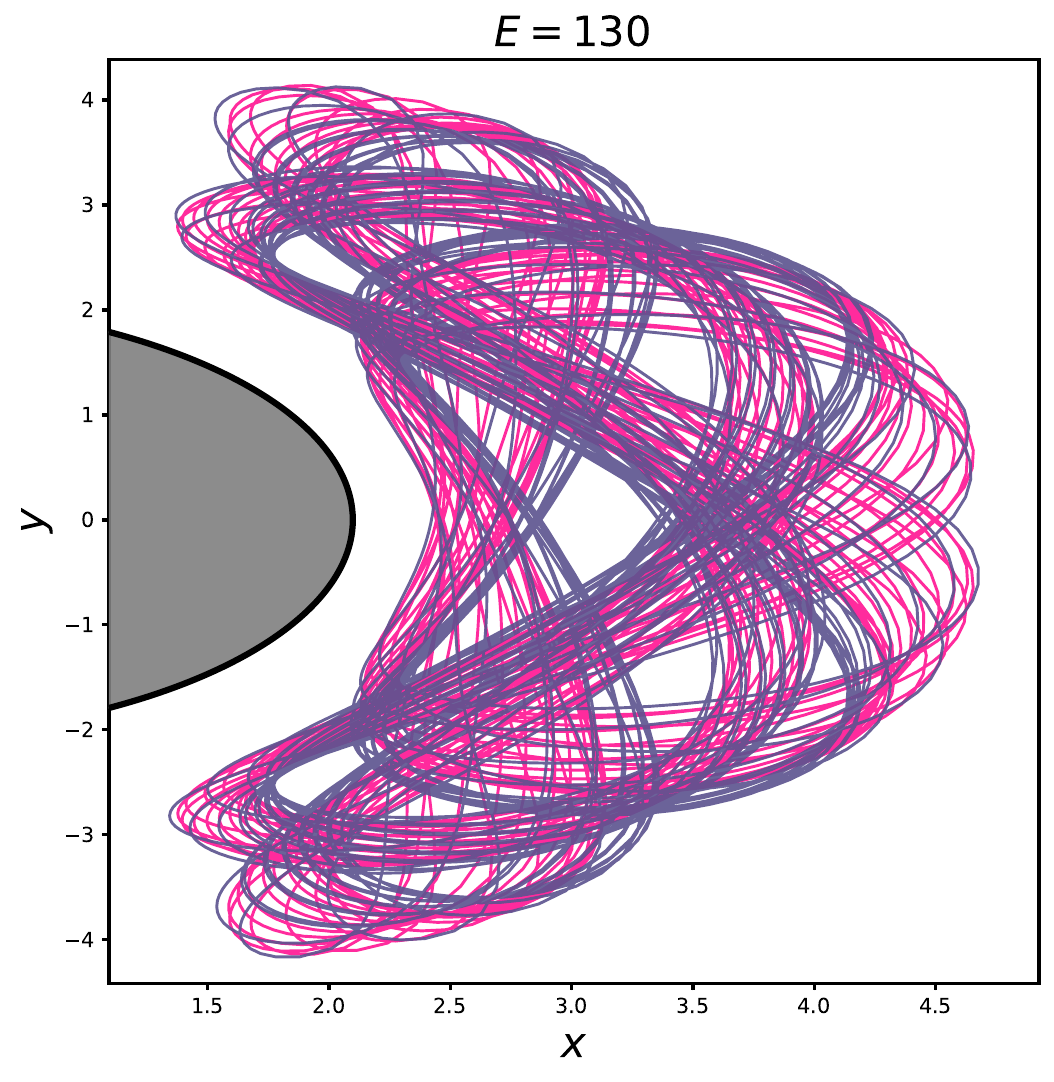}
        {\footnotesize I(b) $\rhs=0.02,\rs=0.15$}
        \end{minipage}
        \begin{minipage}[b]{0.3\textwidth}
        \centering
        \includegraphics[width=\textwidth, height=4.5cm]{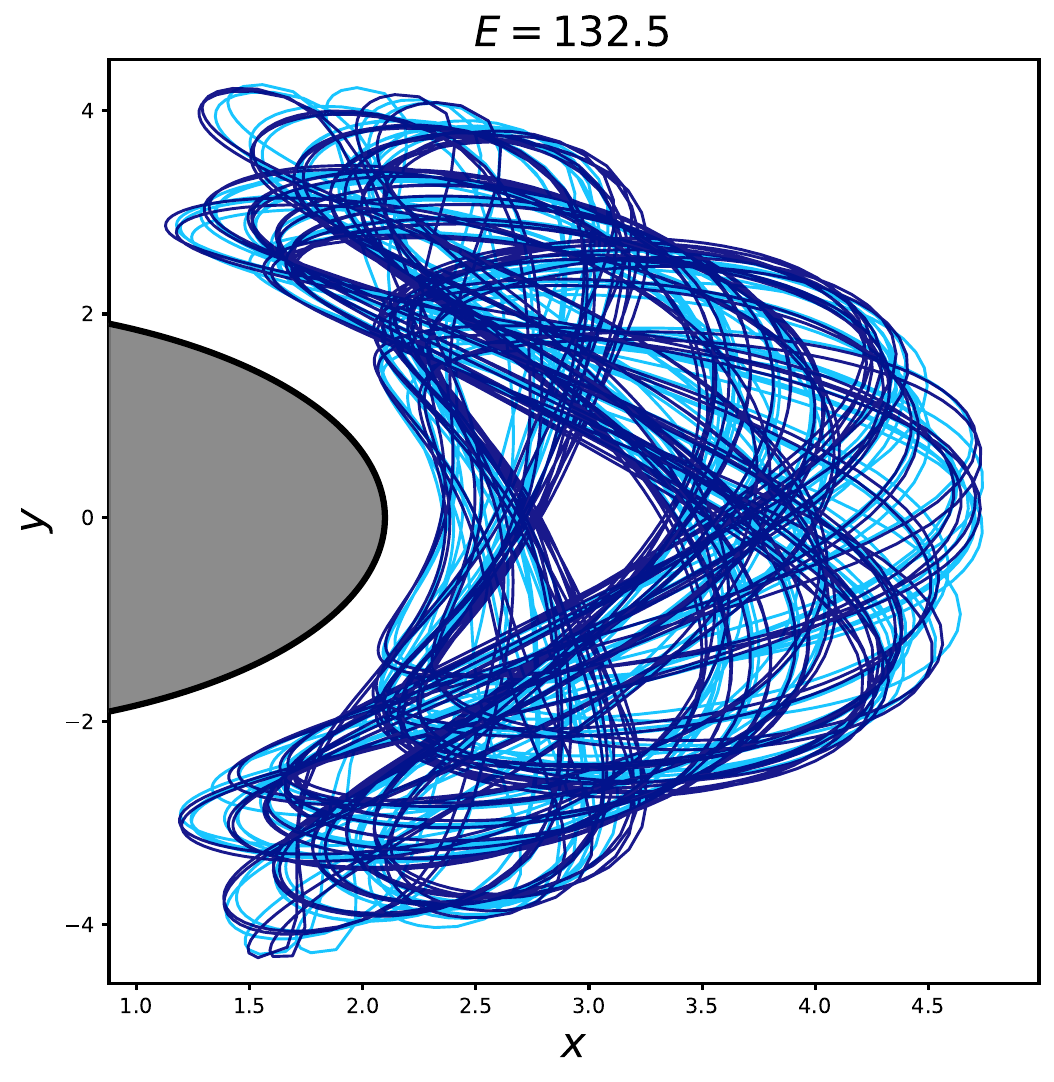}
        {\footnotesize I(c) $\rhs=0.02,\rs=0.15$}
        \end{minipage}
        \vspace{0.25cm}
        \centering
        \begin{minipage}[b]{0.3\textwidth}
        \centering
        \includegraphics[width=\textwidth, height=4.5cm]{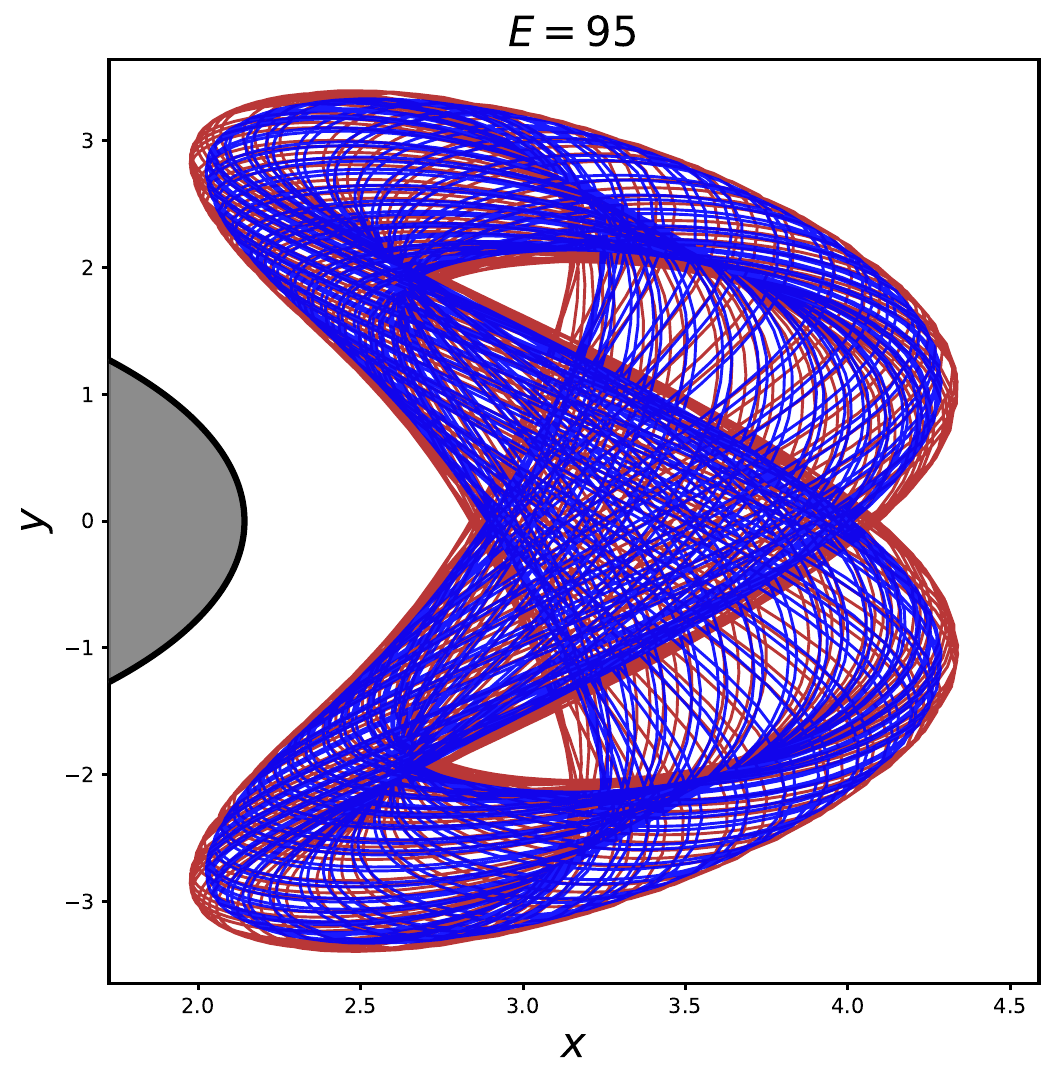}
        {\footnotesize II(a) $\rhs=0.01,\rs=0.25$}
        \end{minipage}
        \begin{minipage}[b]{0.3\textwidth}
        \centering
        \includegraphics[width=\textwidth, height=4.5cm]{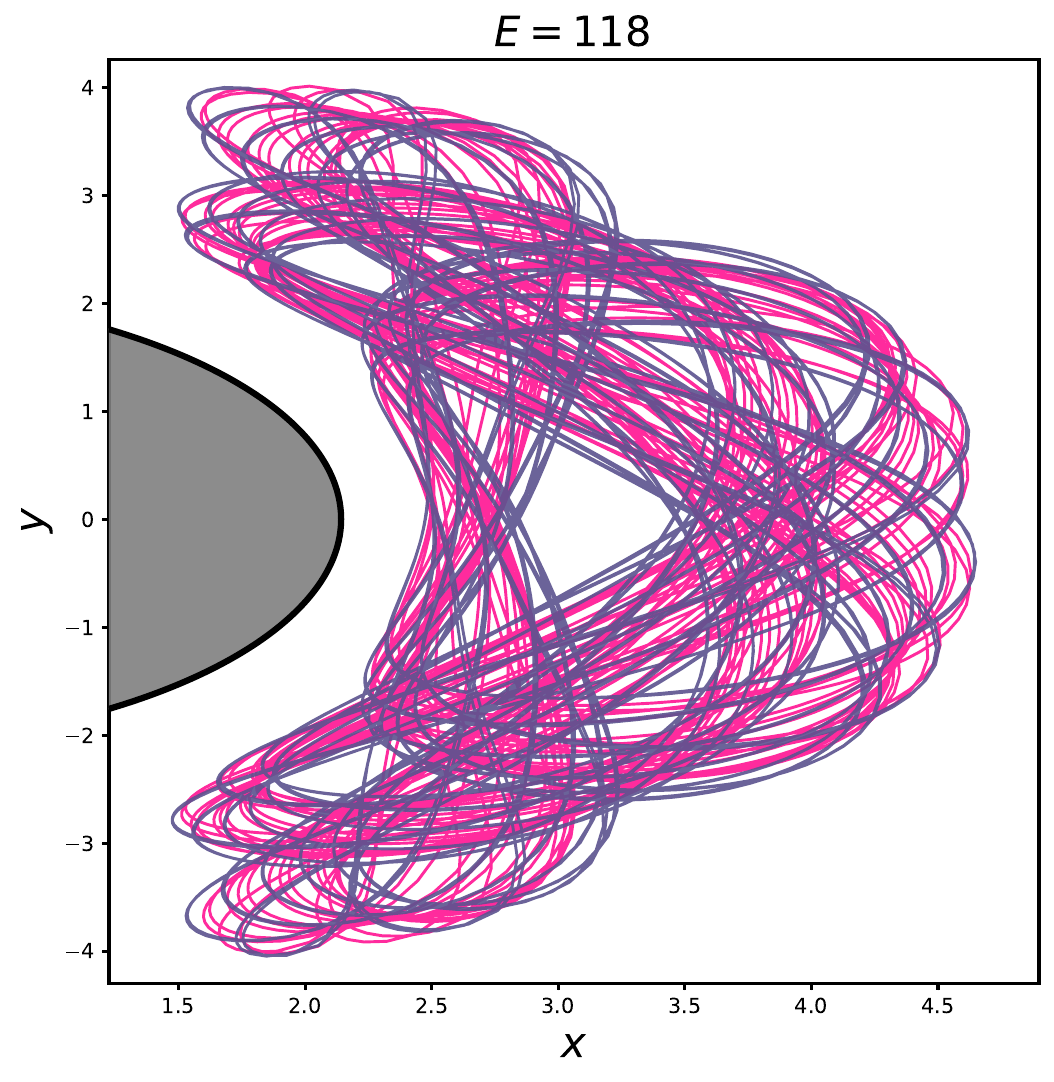}
        {\footnotesize II(b) $\rhs=0.01,\rs=0.25$}
        \end{minipage}
        \begin{minipage}[b]{0.3\textwidth}
        \centering
        \includegraphics[width=\textwidth, height=4.5cm]{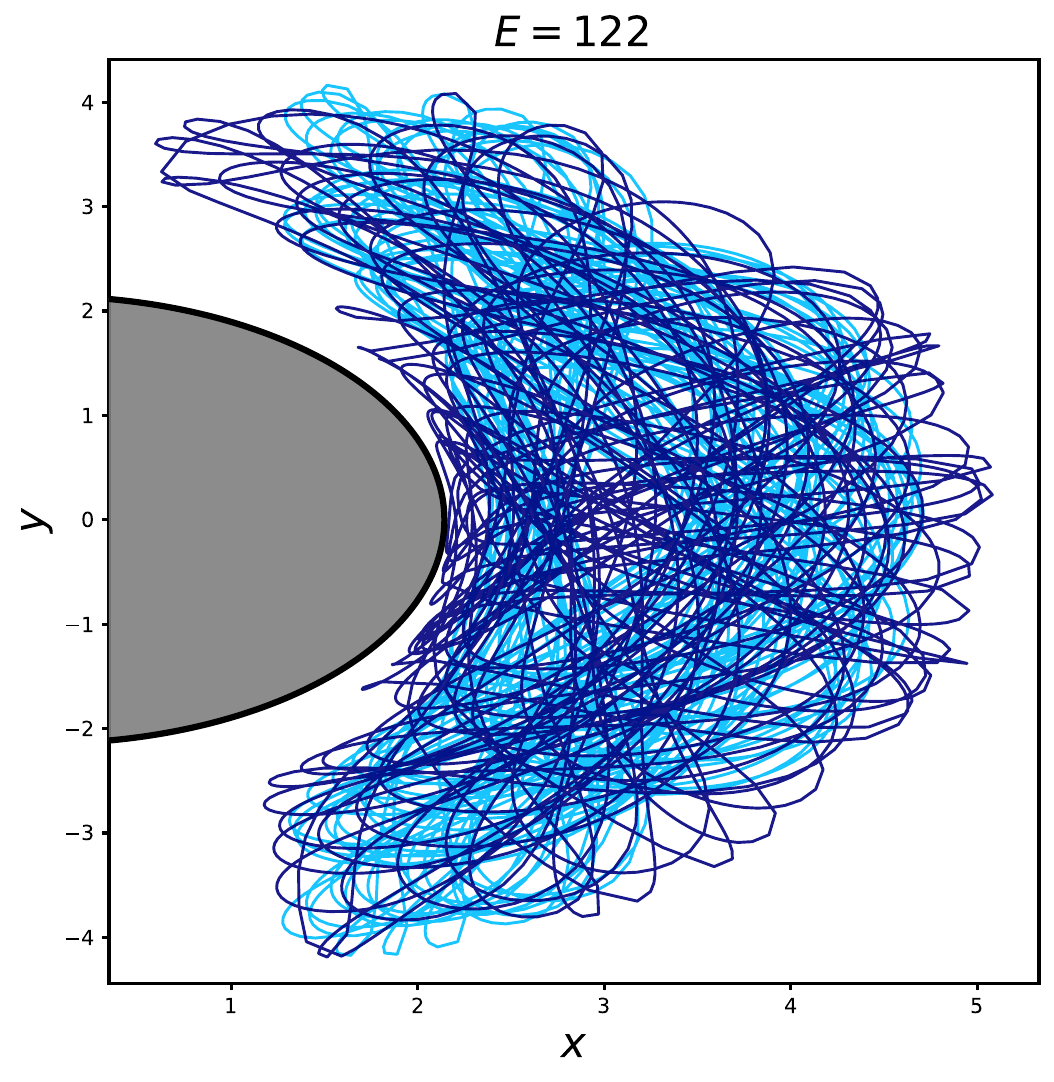}
        {\footnotesize II(c) $\rhs=0.01,\rs=0.25$}
        \end{minipage}
        \vspace{0.25cm}
        \centering
        \begin{minipage}[b]{0.3\textwidth}
        \centering
        \includegraphics[width=\textwidth, height=4.5cm]{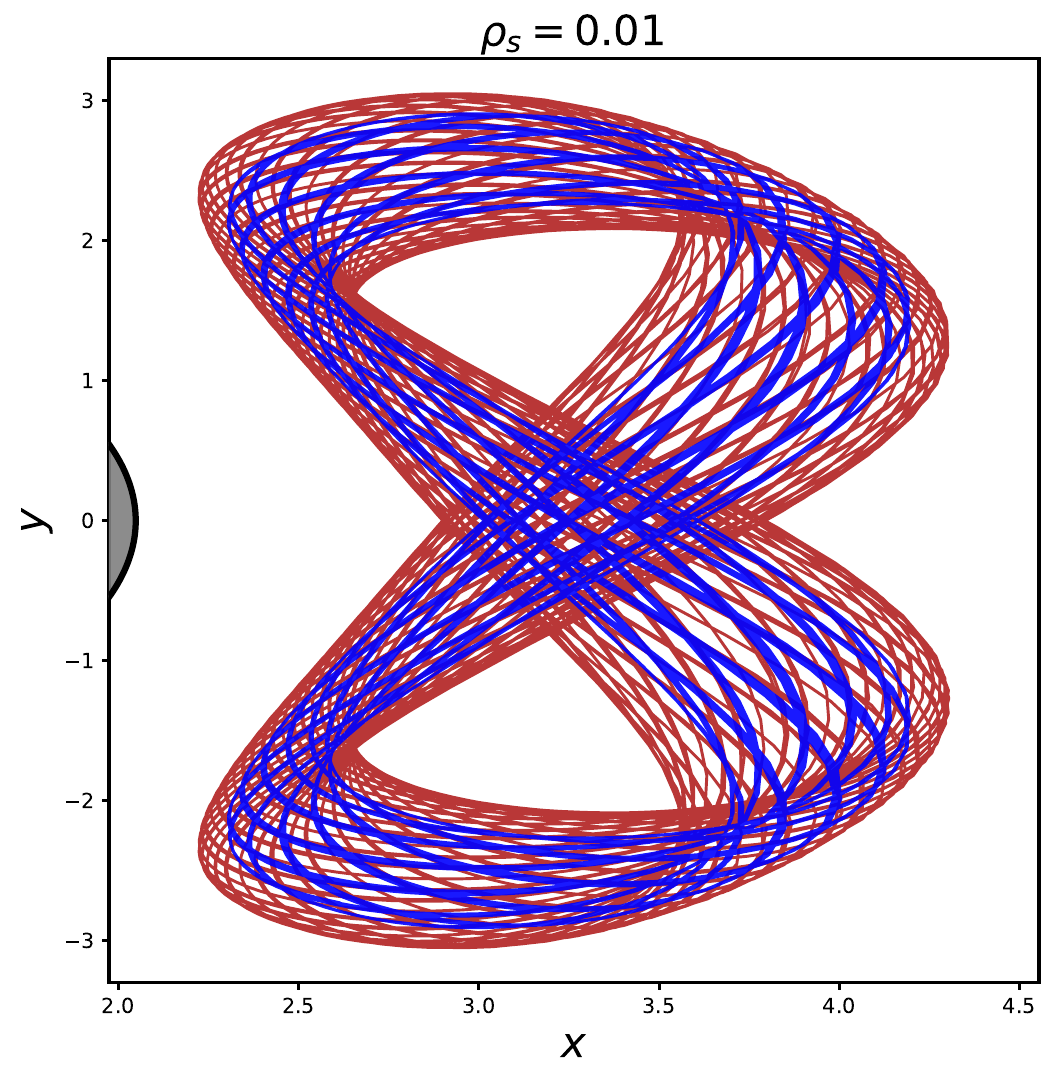}
        {\footnotesize III(a) $\rs=0.15,E=90$}
        \end{minipage}
        \begin{minipage}[b]{0.3\textwidth}
        \centering
        \includegraphics[width=\textwidth, height=4.5cm]{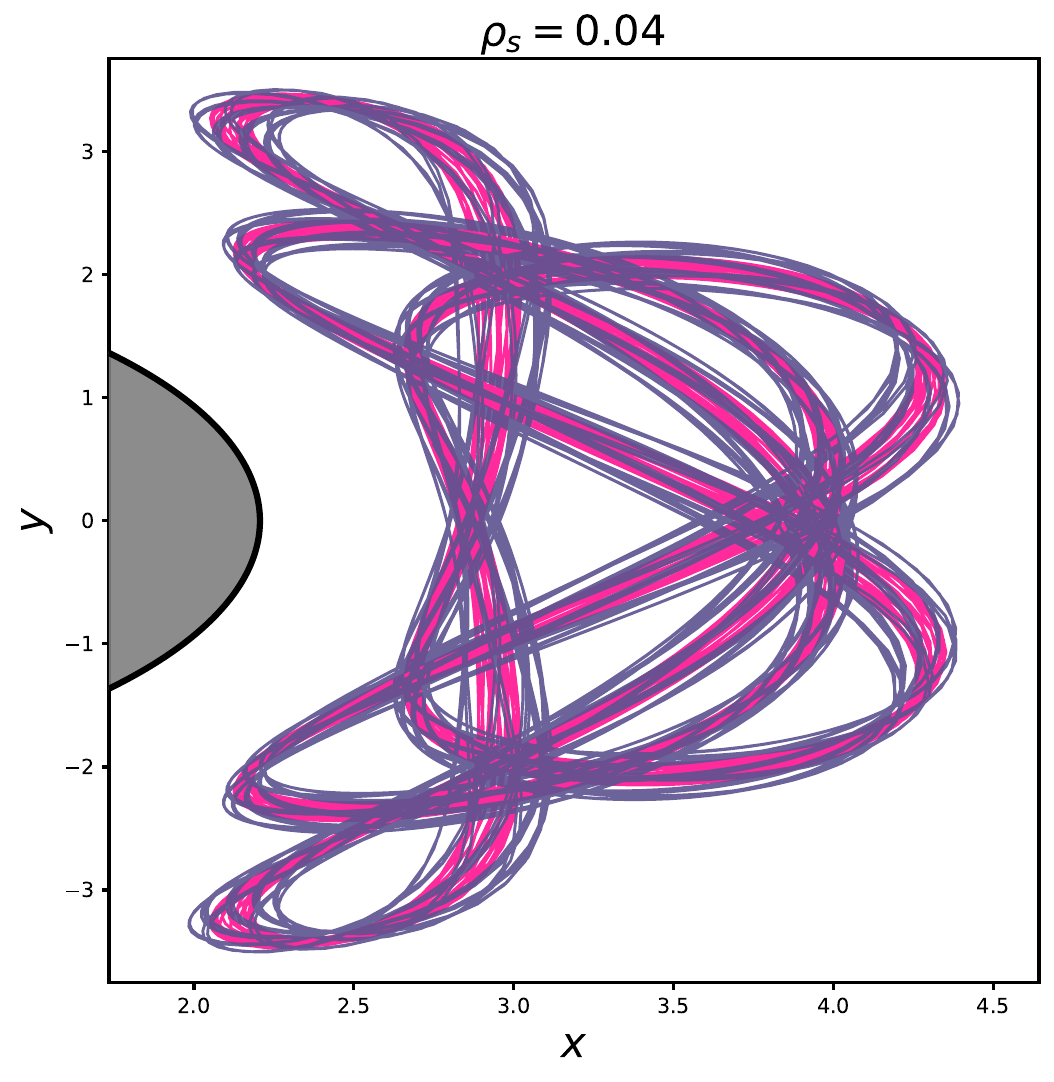}
        {\footnotesize III(b) $\rs=0.15,E=90$}
        \end{minipage}
        \begin{minipage}[b]{0.3\textwidth}
        \centering
        \includegraphics[width=\textwidth, height=4.5cm]{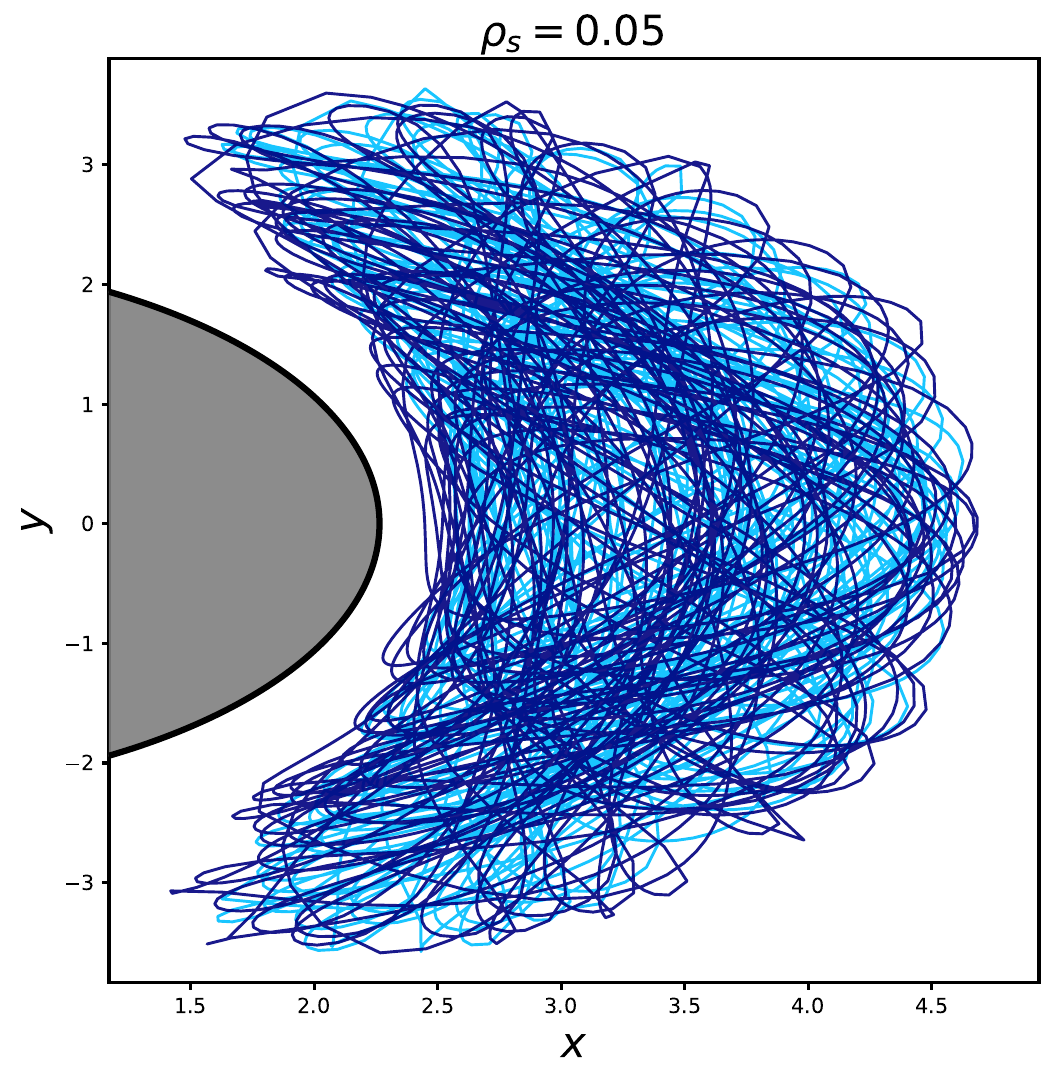}
        {\footnotesize III(c) $\rs=0.15,E=90$}
        \end{minipage}
        \vspace{0.25cm}
        \centering
        \begin{minipage}[b]{0.3\textwidth}
        \centering
        \includegraphics[width=\textwidth, height=4.5cm]{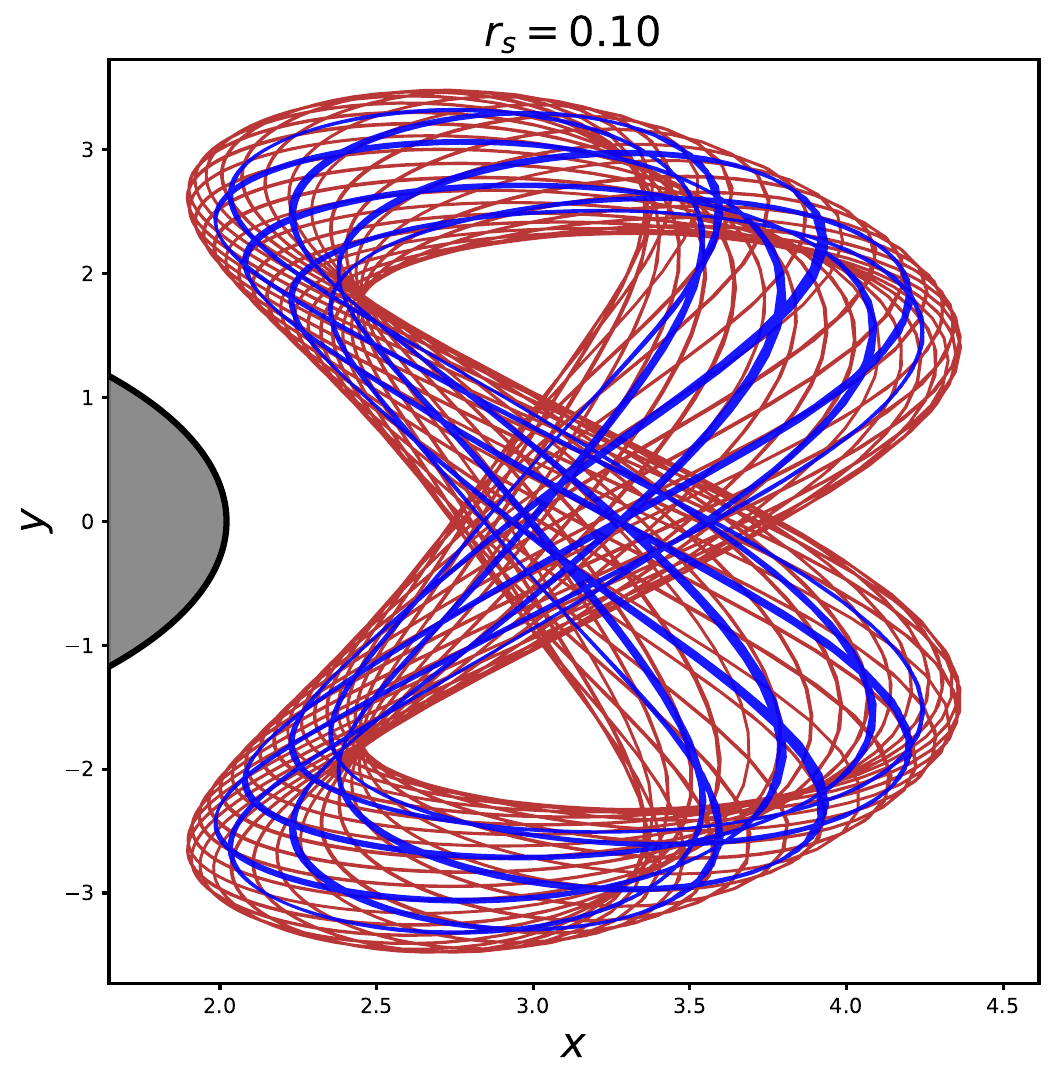}
        {\footnotesize IV(a) $\rhs=0.01,E=115$}
        \end{minipage}
        \begin{minipage}[b]{0.3\textwidth}
        \centering
        \includegraphics[width=\textwidth, height=4.5cm]{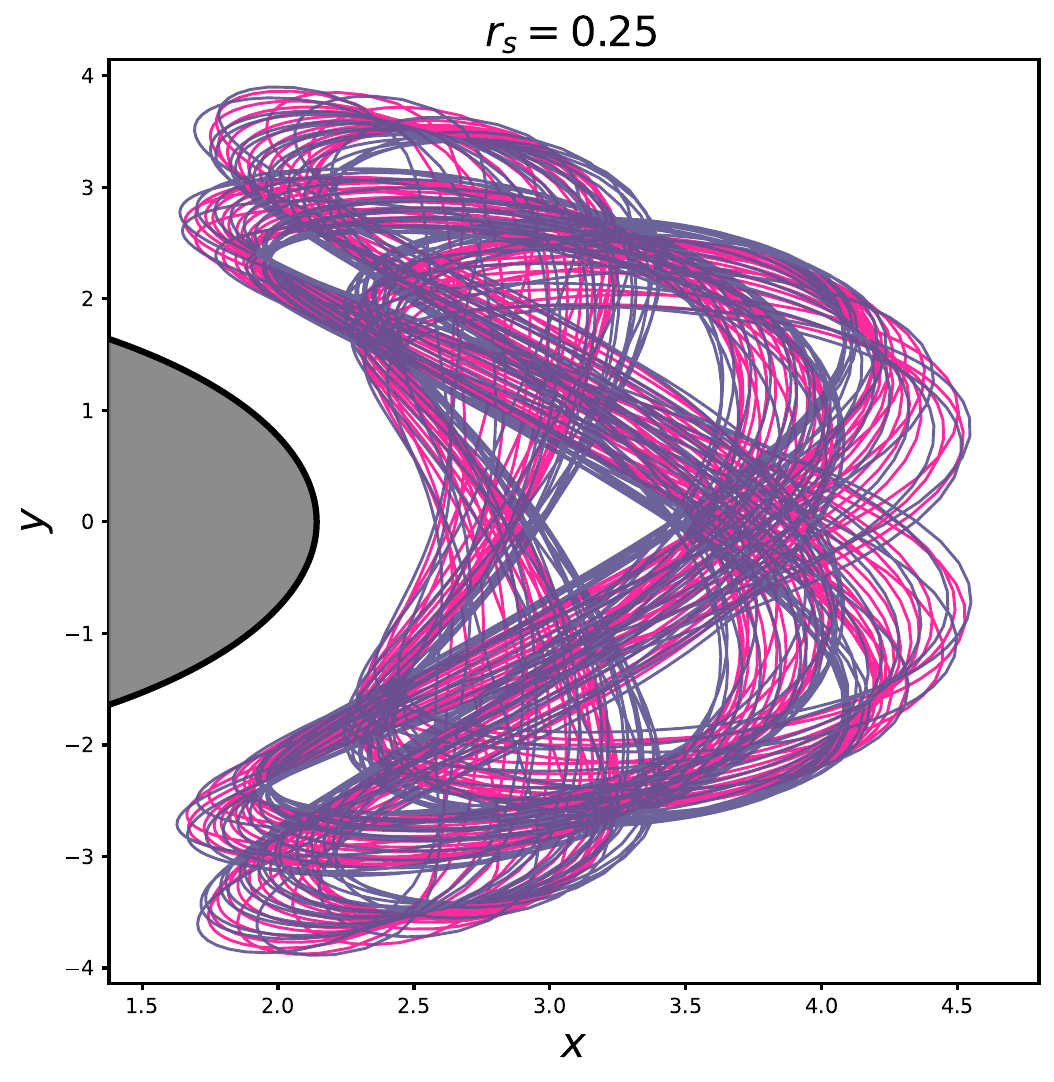}
        {\footnotesize IV(b) $\rhs=0.01,E=115$}
        \end{minipage}
        \begin{minipage}[b]{0.3\textwidth}
        \centering
        \includegraphics[width=\textwidth, height=4.5cm]{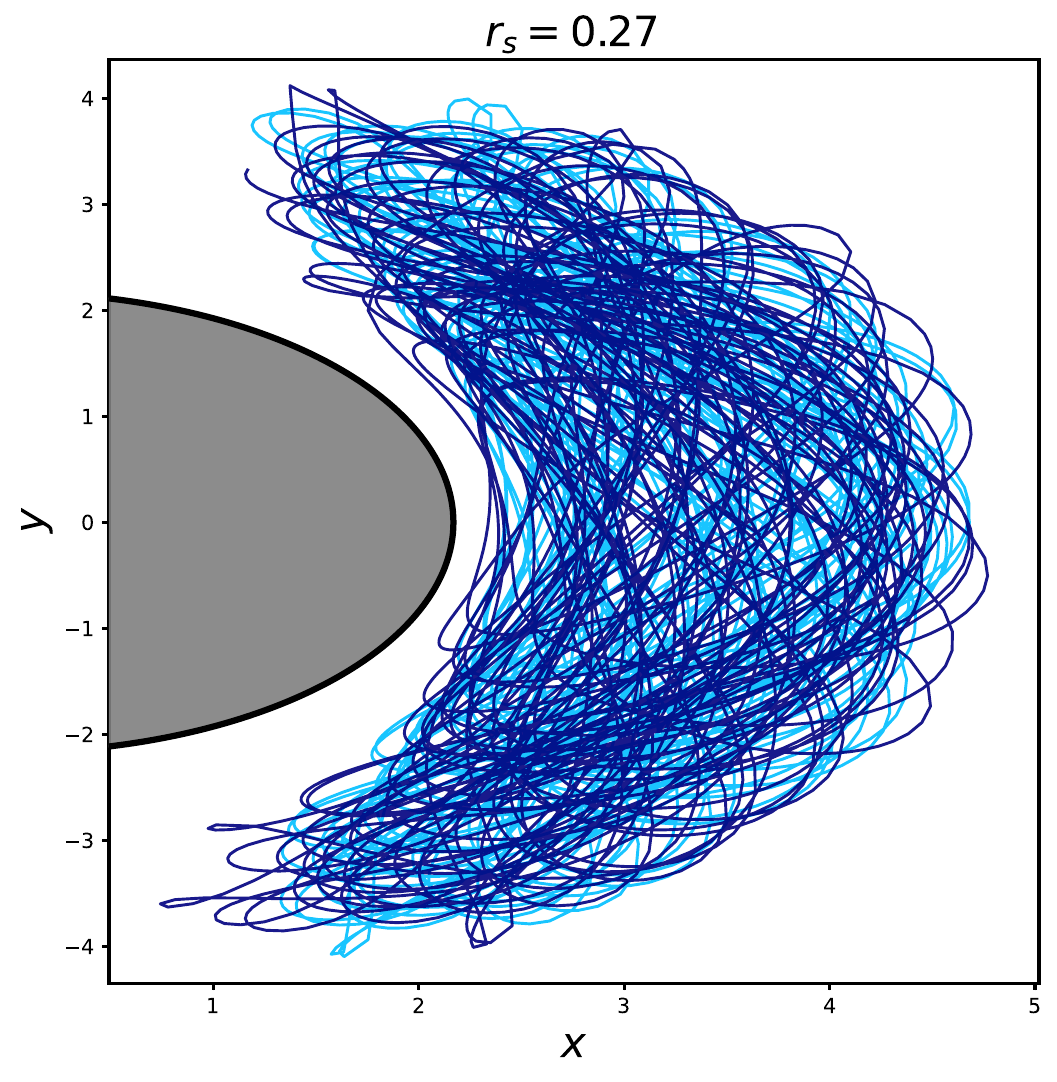}
        {\footnotesize IV(c) $\rhs=0.01,E=115$}
        \end{minipage}
        \caption{These figures depict two closely-starting stellar objects orbiting a supermassive Schwarzschild BH embedded in a Dehnen-type DM halo, modeled as an EMRI system ($q=10^{-5}$). The gray area is the interior of the event horizon, whose boundary (black curve) is shaped by the DM halo parameters. Each row shows three different orbits: Left: A ``non-chaotic" orbit at low energy/central density/and scale radius. Middle: The ``onset of chaotic" orbit at intermediate values of energy/central density/scale radius. Right: A fully ``chaotic orbit" at high values of energy/central density/scale radius. As these parameters increase from left to right in this figure, the trajectories move closer to the horizon. The decreasing overlap between two trajectories indicate the emergence and strengthening of chaotic behavior in the background geometry.}
        \label{T1}
    \end{figure*}

\noindent
Before examining the Lyapunov exponents, let us present a brief overview of representative particle trajectories for two distinct, but very close initial conditions in the case of an EMRI system. Such EMRI system constitute a stellar mass compact object orbiting a supermassive Schwarzschild BH surrounded by a Dehnen-$(1,4,5/2)$ type DM halo. We explore the dependence of energy $E$ and the DM halo parameters $\rhs$ and $\rs$ on such EMRI system. This preliminary analysis aids in developing an intuitive understanding and clearer visualization of the Poincar\'e sections, as illustrated in Fig.~\ref{T1}. In Fig.~\ref{T1} the trajectories are plotted in the $(x-y)$ plane, where $x=r \cos \phi$ and $y=r \sin \phi$, with $\phi \in (-\pi/2, \pi/2)$.\\
Two key observations emerge from these plots. First, the confining effect of the external potentials $a(r),~b(\phi)$ along the $r$ and $\phi$ directions is evident. Without the BH, these potentials would constitute an integrable and regular system. Second, in the presence of horizon, the system's integrability is disrupted, leading to the emergence of chaotic behavior at higher energies or with larger values of the central density $\rhs$ and scale radius $\rs$ of the DM halo (causing particle trajectories to approach the horizon). Thus, both energy and the DM halo parameters significantly influence the onset of chaotic dynamics. Here, the term ``onset-of-chaos" refers to the first appearance of broken tori in the corresponding Poincar\'e section (the middle column of each figures in Fig.~2(a), (b), (c) and (d)). In all the cases, mentioned in Fig.~\ref{T1}, we compare two initially close trajectories with the following initial values: $r_1=3.688131,~r_2=3.658908;~p_{r1}=-0.105617,~p_{r2}=0.411647;~\phi_1=0.028310,~\phi_2=-0.030245~;$ and $p_{\phi1},~p_{\phi2}$ are obtained from Eq.~\eqref{3.4} within the permissible parameter space (as discussed in the text). Additionally, the horizon radius of each BH is represented with its respective horizon contour in Fig.~\ref{T1}.

\null 
\vfill

\section*{Investigation on Lyapunov exponents}
\noindent
Let us now define one another important tool to probe chaotic dynamics, known as the Lyapunov exponent (LE). In dynamical systems, the LE represents a fundamental quantity that describes the exponential divergence rate of initially nearby trajectories \cite{Sandri}. The total Lyapunov exponent is defined as \cite{Sandri,Strogartz},
    \begin{equation}
        \lambda_{T} = \lim_{t\to\infty}\frac{1}{t}\ln\left(\frac{\Delta(t)}{\Delta (0)}\right)~,\label{3.5}
    \end{equation}
where $\Delta(t)$ measures the separation between two initially close trajectories at time $t$ in the full phase space $(r,p_r,\phi,p_{\phi})$ and $\Delta(0)$ is the same at the initial time. In this context, one can also define the radial Lyapunov exponent with the same definition as mentioned in Eq.~\eqref{3.5} but for which, $\Delta(t):=\delta r(t)$ and $\Delta(0):=\delta r(0)$ measure the separation between two initially close trajectories at time $t$ and at the initial time, respectively, along the radial direction ($r$) only. Therefore, the radial Lyapunov exponent $\lambda_r$ is defined as
    \begin{equation}
        \lambda_{r} = \lim_{t\to\infty}\frac{1}{t}\ln\left(\frac{\delta r(t)}{\delta r(0)}\right)~.\label{eq:Lyp_sep}
    \end{equation}
The Lyapunov exponent is bounded by the surface gravity ($\kappa$) of a BH spacetime through the relation \cite{Maldacena:2015waa,Hashimoto:2016dfz}:
    \begin{equation}
        \lambda_{T} \leq \kappa~,\label{3.6}
    \end{equation} 
\noindent
We now turn our attention in analyzing the LE as a quantitative measure of the chaotic behavior evident in the Poincar\'{e} sections examined earlier. Our investigation focuses on two distinct types of LE in this framework by considering an EMRI system $(q=10^{-5})$. The first is the total Lyapunov exponent ($\lambda_{T}$), which is related with the logarithmic separation ratio (Eq.~\eqref{3.5}), characterizing the divergence rate between trajectories across the entire phase space. The second quantity ($\lambda_r$), related only with the radial logarithmic separation ratio (Eq.~\eqref{eq:Lyp_sep}), specifically measures the divergence rate along the radial ($r$) direction between phase space trajectories.\\
It is worth noting that in this classical scenario, chaos exhibits an upper limit determined by a BH's surface gravity ($\kappa$) \cite{Hashimoto:2016dfz}, which emerges from universal features of particle motion near the event horizon. Recent studies about the violations of Lyapunov exponents have been reported in certain scenarios, including charged probes in Kerr-Newman-AdS spacetimes \cite{Gwak:2022xje}, charged particles balanced by Lorentz forces \cite{Zhao:2018wkl}, BH with anisotropic matter fields \cite{Jeong:2023hom}, in Einstein's gravity \cite{Lei:2021koj,Kan:2021blg} and various modified theories of gravity such as $f(R)$ and $f(T)$ \cite{Addazi:2021pty, Addazi:2023pfx}. This observation prompts us to numerically evaluate the LEs for this specific galactic BH-DM halo system and compare them with the established bound. Such analysis proves particularly significant in exploring DM halo characteristics, as we begin to investigate how various DM halo properties influence the combined BH-DM halo system.

\noindent
The onset of chaotic behavior in particle dynamics near the horizon or any deviations from expected LE values in Einstein's gravity might serve as indirect indicators of interactions between the DM halo and BH geometry. With this objective, we now advance to the numerical computation of the LEs.\\
In Figs.~3(a) (Left) and 3(b) (Left), we illustrate the evolution of the total LE $(\lambda_T)$ over time $t$ for various energy values while keeping the DM halo parameters fixed at $\rs=0.15,~\rhs=0.02$ (Case-I) and $\rhs=0.01,~\rs=0.25$ (Case-II), respectively. Additionally, we depict $\lambda_T$ as functions of time for different DM halo central densities ($\rhs$) with fixed scale radius $\rs=0.15$ and energy $E=90$ (case-III) in Fig.~3(c) (Left), and for varying DM halo scale radii ($\rs$) with fixed central density $\rhs=0.01$ and energy $E=115$ (Case-IV) in Fig.~3(d) (Left).\\
It is significant to highlight that across all four cases investigated here, the numerical values for both the total Lyapunov exponent $\lambda_T$ and the radial Lyapunov exponent $\lambda_r$ are evaluated up to a time of $t = 2500$. This is because the separation of two points in phase space does not grow anymore after time $t^*$ (say), where $e^{\lambda t^*}$ is comparable to the size of phase space and our numerical analysis suggests that $t^*\sim 2500$. By this point, every category of orbit studied in this work—namely, non-chaotic, onset-of-chaotic, and chaotic—exhibits initial fluctuations and an exponential decay before stabilizing into a nearly constant average, as depicted in the inset figures of the corresponding main plots. For the sake of clarity, the evolution of $\lambda_T$ and $\lambda_r$ is plotted over an extensive time range. This extended visualization genuinely reveals the saturated convergence of the Lyapunov exponents in the large-time limit, effectively representing the mathematical infinite-time limit.\\
The presented figures demonstrate that the saturation levels of both the total and the radial LEs exhibit a characteristic dependence that scales as $1/t$. Furthermore, for a given set of DM halo parameters, an increase in energy corresponds to a higher saturation value for the total and radial logarithmic separation ratios. This observed trend signifies that the system's dynamics display more pronounced chaotic behavior at higher energies [see Figs.~3(a) and (b)]. A similar trend is observed when varying the DM halo central density and scale radius while maintaining a constant energy, suggesting enhanced chaos for larger values of the DM halo parameters $\rhs,\rs$ [see Figs.~3(c) and (d)]. Thus, we can infer that the DM halo parameters $\rhs$ and $\rs$ significantly influence the chaotic dynamics of an EMRI system where a massive compact object inspirals around a supermassive BH in a galactic environments.\\
A comparable pattern is also noticeable from the behavior of the radial LE $(\lambda_r)$ with time $t$, where the peak values are highest at energies $E=132.5$ and $E=122$ in Figs.~3(a) (Right) and 3(b) (Right), corresponding to Case-I and Case-II, respectively. Furthermore, $\lambda_r$ reaches its maximum for the largest values of the DM halo parameters $\rhs=0.05$ and $\rs=0.27$, highlighting chaotic behavior under the consideration of the radial dynamics, as seen in Figs.~3(c) (Right) and 3(d) (Right), respectively. Additionally, we present the numerically computed total and radial Lyapunov exponents in Table \ref{T2} for each of the four cases within our combined BH and DM halo spacetime. The final row of Table \ref{T2} displays the respective values for the total and radial Lyapunov exponents in a Schwarzschild spacetime, devoid of any DM halo, corresponding to the predictions of Einstein gravity. It is important to note that for chaotic orbital dynamics, we find that the values of both the total and radial Lyapunov exponents in the Schwarzschild spacetime are marginally greater than their counterparts in all four cases involving the galactic BH and DM halo combined spacetime. Conversely, another significant observation is that the radial Lyapunov exponents consistently exhibit values somewhat lower than the corresponding total Lyapunov exponents. This trend holds for the non-chaotic, onset-of-chaotic, and fully chaotic regimes of particle motion across all five cases as mentioned in Table \ref{T2}.

\section*{Investigation on the MSS chaos bound}\label{sec:bound}
\noindent
We now examine whether the violation of the chaos bound of the total Lyapunov exponent ($\lambda_{T}$) and the radial Lyapunov exponent ($\lambda_{r}$) in relation to the DM halo parameters $\rhs$ and $\rs$ and energy $E$ happens or not. The surface gravity of a BH in SSS spacetime can be obtained by the following definition:
    \begin{eqnarray}
        \kappa_{\rm BH}&=& \frac{1}{2} \left( \frac{\partial f(r)}{\partial r} \right) \Bigg\rvert_{r = r_H}~.\label{eq:kappa_def}
    \end{eqnarray}
By using Eq.~(2), from Eq.~\eqref{eq:kappa_def} we have the following expression for the surface gravity of the BH-DM halo spacetime in terms of the DM halo parameters $\rhs,\rs$.
    \begin{eqnarray}
        \kappa:=\kappa_{\textrm{BHDM}}=\frac{\left(1 - 1024 \pi^2 r_s^4 \rho_s^2\right)^2 8\sqrt{2} \pi r_s^3 \rho_s + M \sqrt{\frac{B}{A}}}{4 \sqrt{\frac{B}{A}} \cdot A^2}~,\nonumber\\
        \label{eq:kappa}
    \end{eqnarray}
where,
    \begin{eqnarray}
        &A =  16 \pi r_s^2 \rho_s \left(16 \pi r_s^3 \rho_s + \sqrt{4M^2 + 2M r_s + 256 \pi^2 r_s^6 \rho_s^2}\right)\non\\
        &+M~,\nonumber\\
        &B  =  32 \pi r_s^2 \rho_s \sqrt{4M^2 + 2M r_s + 256 \pi^2 r_s^6 \rho_s^2} + 2M + r_s \non\\
        &-512 \pi^2 r_s^5 \rho_s^2~.\nonumber
    \end{eqnarray} 

    \begin{table*}[htbp]
    \centering
    \caption{Comparison of the numerically computed total and radial Lyapunov exponents in the BH-DM halo combined spacetime with Schwarzschild spacetime in Einstein gravity. Here in the case of Einstein's GR (last row), the NC (Non-chaotic) trajectory appears at energy value $E=140$, whereas OC (Onset-of-chaotic) and C (Chaotic) orbits happen at energy values $E=154$ and $E=156$, respectively.}
    \label{T2}
    \renewcommand{\arraystretch}{1.5}
    \begin{tabular}{|c|c|c|}
    \hline
    \hline
    \textbf{Case Description} & \textbf{Total LE $\lambda_T$} & \textbf{Radial LE $\lambda_r$} \\
    \hline
    \textbf{Case-I:} & \textbf{NC:} $3.914403\times10^{-4}$ & \textbf{NC:} $2.79629\times10^{-5}$ \\[2pt]
    Several energies with fixed & \textbf{OC:} $1.3799717\times10^{-3}$ & \textbf{OC:} $1.0446739\times10^{-3}$ \\[2pt]
    $\rho_s=0.02$, $r_s=0.15$ & \textbf{C:} $1.6140374\times10^{-3}$ & \textbf{C:} $1.3953516\times10^{-3}$ \\
    \hline
    \textbf{Case-II:} & \textbf{NC:} $3.579065\times10^{-4}$ & \textbf{NC:} $-1.91346\times10^{-5}$ \\[2pt]
    Several energies with fixed & \textbf{OC:} $8.155527\times10^{-4}$ & \textbf{OC:} $1.2348984\times10^{-3}$ \\[2pt]
    $\rho_s=0.01$, $r_s=0.25$ & \textbf{C:} $2.3538360\times10^{-3}$ & \textbf{C:} $1.7021501\times10^{-3}$ \\
    \hline
    \textbf{Case-III:} & \textbf{NC:} $1.0163\times10^{-3}$ & \textbf{NC:} $4.825199\times10^{-4}$ \\[2pt]
    Several central densities with fixed & \textbf{OC:} $1.4944646\times10^{-3}$ & \textbf{OC:} $5.719151\times10^{-4}$ \\[2pt]
    $E=90$, $r_s=0.15$ & \textbf{C:} $1.8617405\times10^{-3}$ & \textbf{C:} $1.7505470\times10^{-3}$ \\
    \hline
    \textbf{Case-IV:} & \textbf{NC:} $1.4888789\times10^{-3}$ & \textbf{NC:} $1.2031799\times10^{-3}$ \\[2pt]
    Several scale radii with fixed & \textbf{OC:} $1.5429047\times10^{-3}$ & \textbf{OC:} $1.3952989\times10^{-3}$ \\[2pt]
    $\rho_s=0.01$, $E=115$ & \textbf{C:} $1.7227998\times10^{-3}$ & \textbf{C:} $1.6558557\times10^{-3}$ \\
    \hline
    \textbf{Einstein's GR:} & \textbf{NC:} $1.4171994\times10^{-3}$ & \textbf{NC:} $5.083468\times10^{-4}$ \\[2pt]
    Absence of DM halo parameters & \textbf{OC:} $2.2051817\times10^{-3}$ & \textbf{OC:} $1.4732053\times10^{-3}$ \\[2pt]
    $\rho_s=r_s=0$ & \textbf{C:} $3.4929236\times10^{-3}$ & \textbf{C:} $3.2824898\times10^{-3}$ \\
    \hline
    \hline
    \end{tabular}   
    \end{table*}
\noindent
It is interesting to see the leading order terms of the DM halo parameters $\rhs$ and $\rs$ for the expression of $\kappa_{BHDM}$ in Eq.~\eqref{eq:kappa}. Therefore expanding it as a series in $\rhs$ and $\rs$, we obtain:
    \begin{eqnarray}
        &\kappa& = \kappa_{EG} + \frac{8\sqrt{2}\pi\rs^2\rhs\sqrt{2M+\rs}}{M\sqrt{M}} + \mathcal{O}(\rhs^2)~,\label{eq:kappa_rho}\\
        &\kappa& = \kappa_{EG} - \frac{16 \pi \rhs \rs^2}{M} + \mathcal{O}(\rs^3)\label{eq:kappa_r}~.
    \end{eqnarray}
It is worthy to note that there is no term presents linear in $\rs$ in the above expansion. In the limit where $\rhs,\rs \rightarrow 0$, Eqs.~\eqref{eq:kappa_rho}, \eqref{eq:kappa_r} simplifies to $\kappa=\kappa_{EG} = \frac{1}{4M}$, which represents the surface gravity of a Schwarzschild BH in the context of Einstein gravity.

    \begin{figure*}[t]
        \centering   
        \begin{minipage}[b]{0.45\textwidth}
        \centering
        \includegraphics[width=\textwidth, height=4.0cm]{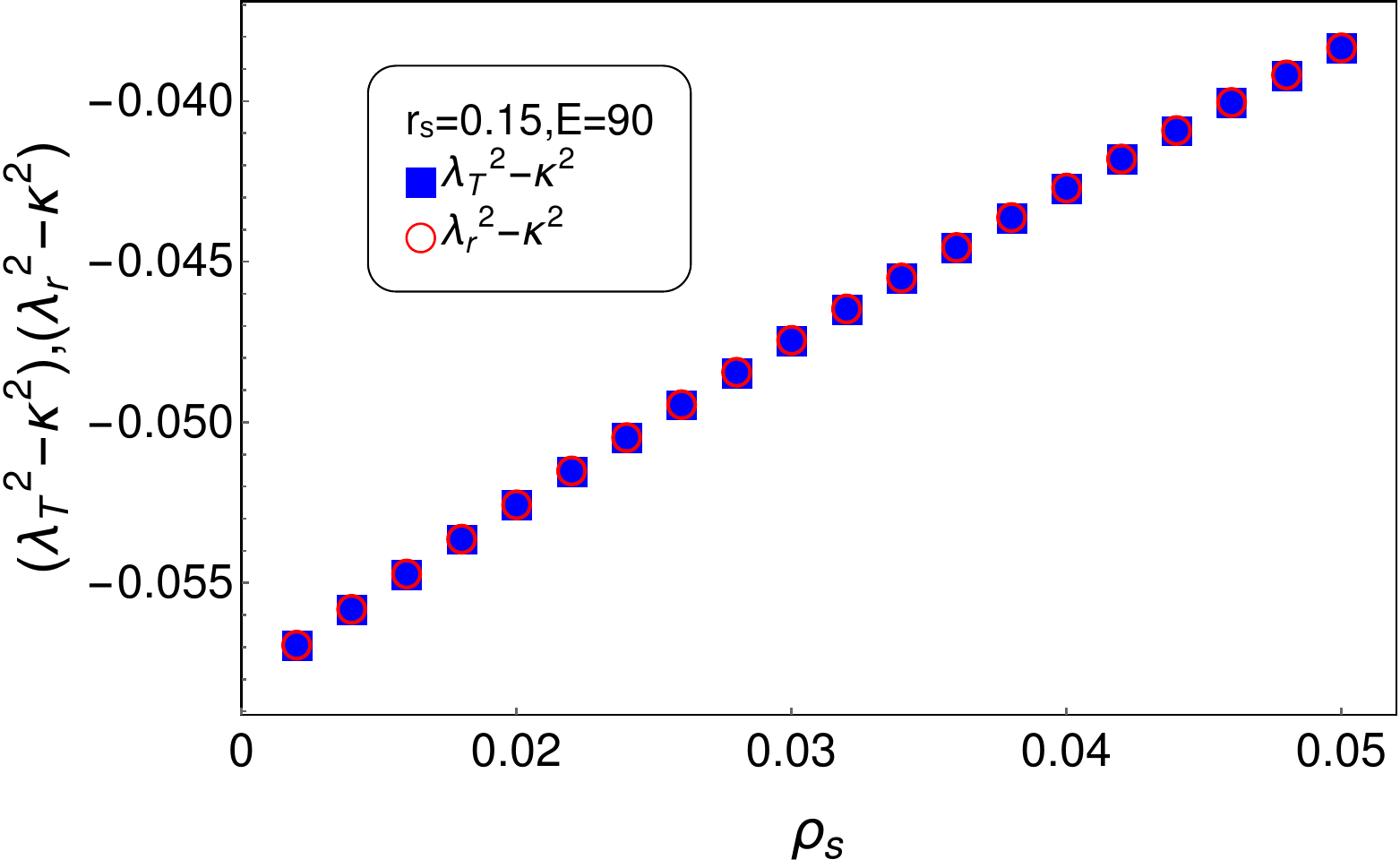}
        \end{minipage}
        \begin{minipage}[b]{0.45\textwidth}
        \centering
        \includegraphics[width=\textwidth, height=4.0cm]{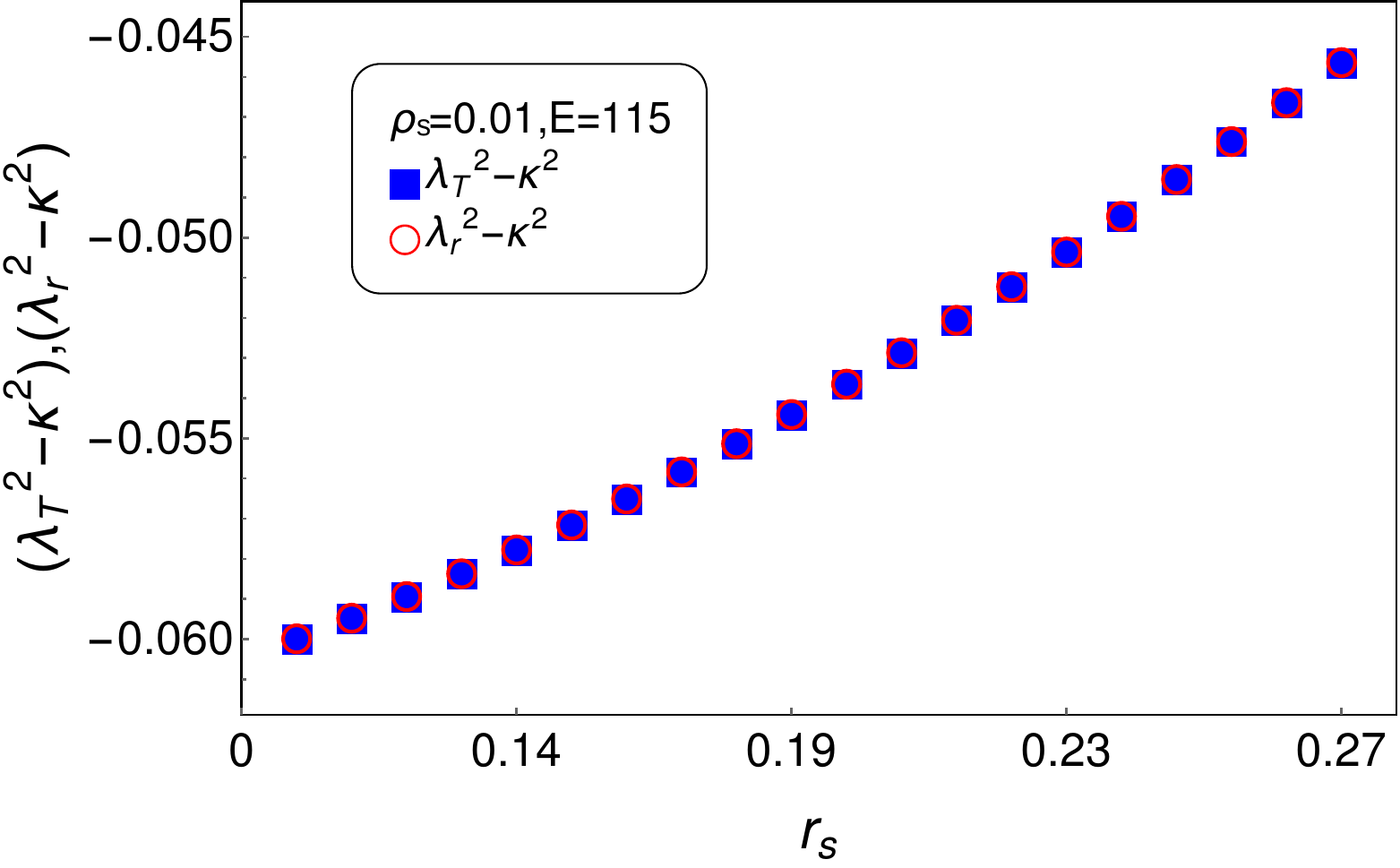}
        \end{minipage}
        \caption{{\it Left:} The plot represents the variation of the total and radial Lyapunov exponents via the relation $(\lambda_T^2-\kappa^2)$ and $(\lambda_r^2-\kappa^2)$ with halo's central density $\rhs$ for fixed energy, $E=90$ and halo's core radius $\rs=0.15$. Due to a tiny numerical differences between $\lambda_T$ and $\lambda_r$, as mentioned in Table~\ref{T2}, the numerical differences between the values of $(\lambda_T^2-\kappa^2)$ and $(\lambda_r^2-\kappa^2)$ are very very small (by overlapping two values, as depicted here). {\it Right:} The plot represents the variation of the total and radial Lyapunov exponents via the relation $(\lambda_T^2-\kappa^2)$ and $(\lambda_r^2-\kappa^2)$ by changing halo's scale radius $\rs$ for fixed energy, $E=115$ and central density $\rhs=0.01$. Due to a tiny numerical differences between $\lambda_T$ and $\lambda_r$, as mentioned in Table~\ref{T2}, the numerical differences between the values of $(\lambda_T^2-\kappa^2)$ and $(\lambda_r^2-\kappa^2)$ are very very small (by overlapping two values, as depicted here).}
        \label{f:b_rho}
        \vspace{0.25cm}
        \centering
        \begin{minipage}[b]{0.45\textwidth}
        \centering
        \includegraphics[width=\textwidth, height=4.0cm]{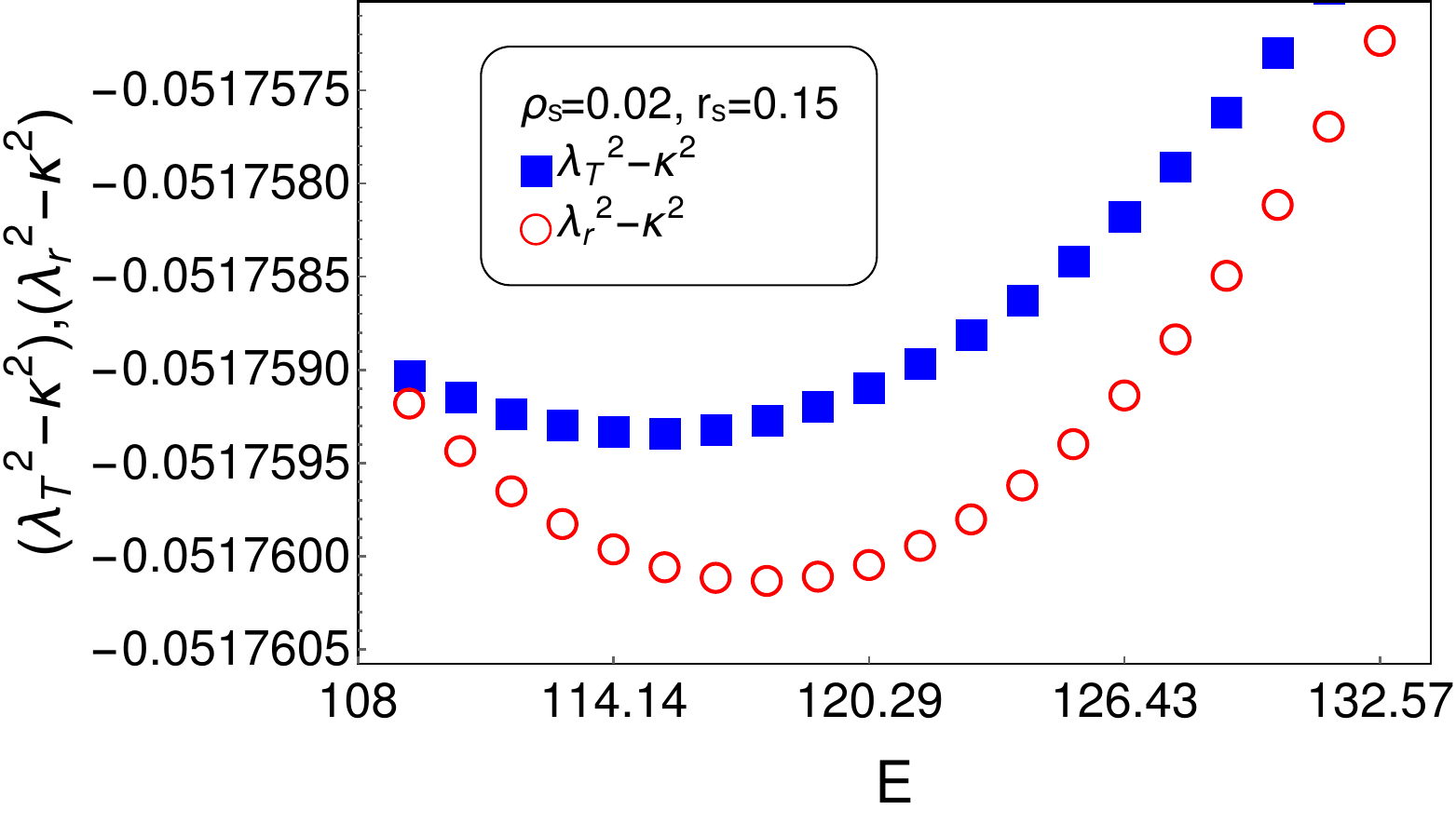}
        \end{minipage}
        \begin{minipage}[b]{0.45\textwidth}
        \centering
        \includegraphics[width=\textwidth, height=4.0cm]{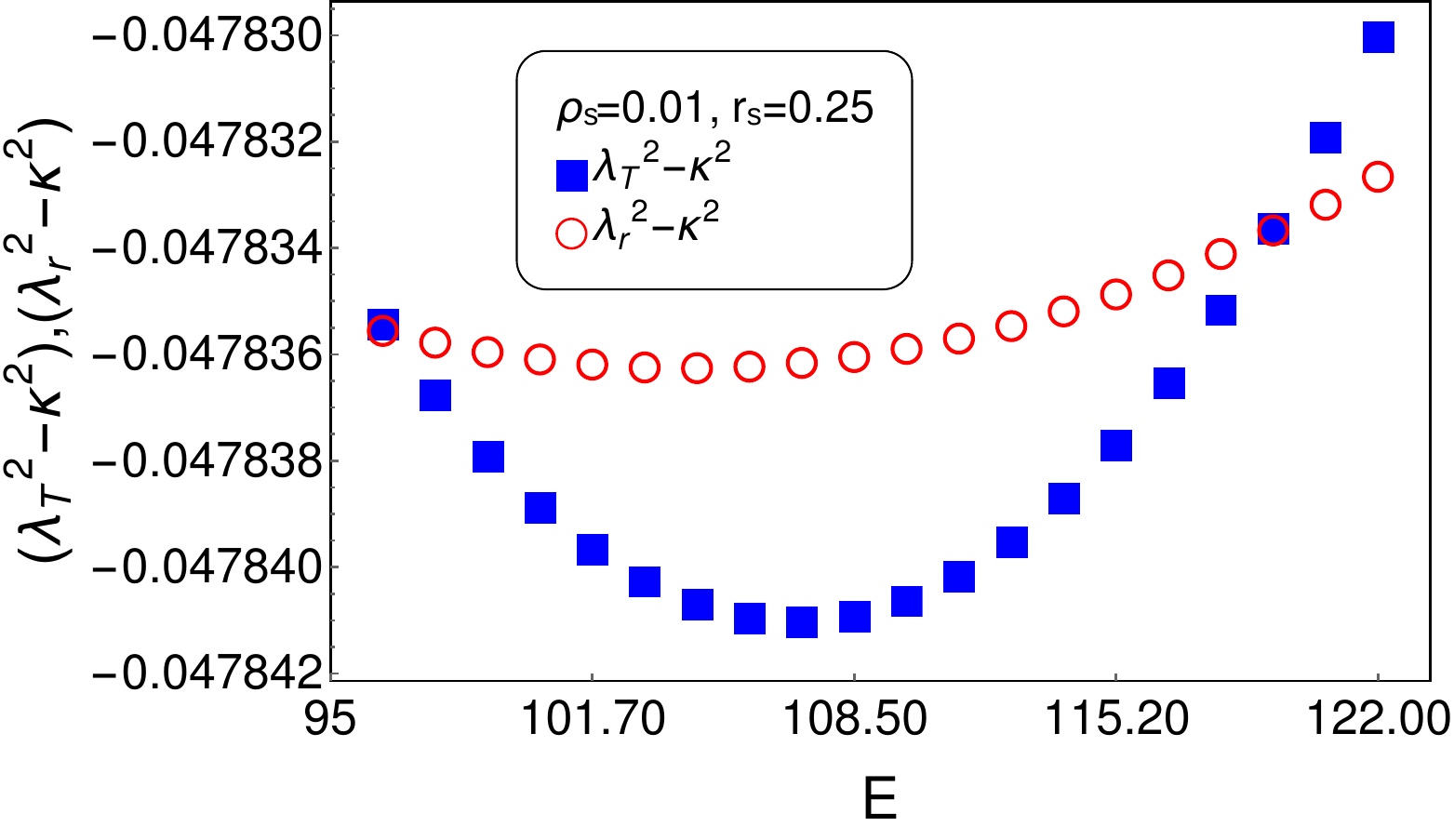}
        \end{minipage}
        \caption{{\it Left:} The plot represents the variation of the total and radial Lyapunov exponents via the relation $(\lambda_T^2-\kappa^2)$ and $(\lambda_r^2-\kappa^2)$ with halo's central density $\rhs$ for fixed energy, $E=90$ and halo's core radius $\rs=0.15$. {\it Right:} The plot represents the variation of the total and radial Lyapunov exponents via the relation $(\lambda_T^2-\kappa^2)$ and $(\lambda_r^2-\kappa^2)$ by changing halo's scale radius $\rs$ for fixed energy, $E=115$ and central density $\rhs=0.01$.}
        \label{f:b_en1}
    \end{figure*}    
\noindent
Let us now briefly discuss on the violation of the chaos bound for a galactic BH with DM halo. Here we address how the DM halo parameters, the central density $\rhs$ and the scale radius $\rs$, along with energy $E$ affect the total and radial Lyapunov exponents to the corresponding non-chaotic, onset-of-chaotic and chaotic dynamics of a particle motion. To do so, we plot the numerically computed values of both, $(\lambda_T^2-\kappa^2)$ and $(\lambda_r^2-\kappa^2)$ with varying the central density $\rhs$, scale radius $\rs$ and energy $E$. Therefore to adhere the chaos bound, both the values of $(\lambda_T^2-\kappa^2)$ and $(\lambda_r^2-\kappa^2)$ must be negative. Fig.~\ref{f:b_rho} (Left) illustrates the variation of both the values of $(\lambda_T^2-\kappa^2)$ and $(\lambda_r^2-\kappa^2)$ with central density $\rhs$, while keeping $\rs = 0.15$ and $E = 90$ constant i.e., Case-III. Similarly, Fig.~\ref{f:b_rho} (Right) shows the dependence of these values on the core radius $\rs$, with fixed values of $\rhs = 0.01$ and $E = 115$ as Case-IV. In both of the cases, we observe that the total as well as the radial Lyapunov exponents remain significantly below the chaos bound followed by the surface gravity of the combined BH-DM halo solution.\\
On the other hand to see the energy effect of the chaos bound violation, we plot the values of $(\lambda_T^2-\kappa^2)$ and $(\lambda_r^2-\kappa^2)$ for different values of energy by keeping fixed DM halo parameters. In Fig.~\ref{f:b_en1} (Left), the differences between $(\lambda_T^2-\kappa^2)$ and $(\lambda_r^2-\kappa^2)$ are clearly visible and it is also important to observe that $(\lambda_T^2-\kappa^2)>(\lambda_r^2-\kappa^2)$ for any values of energy in the range, starting from $E=108$ to $E=132.5$, as Case-I.\\
On a contrary, for Case-II the inequality of $(\lambda_T^2-\kappa^2)>(\lambda_r^2-\kappa^2)$ is only valid for the case of non-chaotic $(E=95)$, onset-of-chaotic $(E=118)$ and chaotic $(E=122)$ scenarios, as presented in Fig.~\ref{f:b_en1} (Right). However, in both of Cases-I and II, the bound of chaos is not violated. Therefore our analysis reveals that both $\lambda_{T}$ and $\lambda_{r}$ adhere to the chaos bounds. This leads us to conclude that while the DM halo parameters $\rhs$ and $\rs$ significantly influence the chaotic dynamics of a massive test particle, they do not violate the surface gravity bounds—whether in the framework of general relativity or in the environment of a galactic BH, surrounded by a Dehnen-$(1,4,5/2)$ type DM halo. This observation aligns with the findings of Hashimoto \textit{et al.} \cite{Hashimoto:2016dfz}.


\section*{Energy conditions for a Schwarzschild-like black hole immersed in a dark matter halo}\label{a11}
\noindent
Let us briefly present all of the energy conditions for our considered BH-DM halo combined spacetime (given by Eqs.~(1), (2)). One can write the metric function as $f(r)=1-\frac{2m(r)}{r}$, where following Eq.~(2), the mass function $m(r)$ for a Schwrzschild-like BH immersed in a Dehnen-$(1,4,5/2)$ type DM halo can be achieved. Now recall the Einstein field equations, mentioned in Eq.~\eqref{2}, where the energy-momentum tensor for the DM halo can be defined in terms of an anisotropic fluid as the following.
    \begin{equation}
        T_{\mu\nu}=(\rho+p_t)u_\mu u_\nu+p_t g_{\mu\nu}+(p_r-p_t)\chi_\mu \chi_\nu~.\label{aemt}
    \end{equation}
Here $\rho$ is the energy density, $p_r$ is the radial pressure along the $\chi_\mu$ direction, whereas in the direction orthogonal to $\chi_\mu$ the pressure along the tangential axis refers to $p_t$ . We note that $\chi_\mu$ represents the unit vector in the radial direction orthogonal to the 4-velocity $u^\mu$ , satisfying the condition $u^\mu u_\mu=-1,~\chi_\mu \chi^\mu=1$.\\
Now the choice of the line element in Eq.~(1), in the Schwarzschild gauge, automatically leads to the equation $G_{tt}=-G_{rr}$, where $G_{\mu\nu}:=R_{\mu\nu}-\frac{1}{2}R g_{\mu\nu}$ is the Einstein tensor. Therefore, using Eq.~\eqref{aemt}, from Eq.~\eqref{2} one may write
    \begin{equation}
        \rho=\frac{2m'}{r^2}=-p_r~,~~~p_t=-\frac{m''}{r}~,\label{eq:ecs}
    \end{equation}
where a prime denotes the derivative with respect to radial coordinate $r$. Thus, for the given mass function $m(r)$ (Eq.~(2)), one can easily compute all the necessary parameters from Eq.~\eqref{eq:ecs} to calculate all the energy conditions.\\
Let us now examine the different energy conditions (see \cite{Poisson:2009pwt} for details on energy conditions) for the matter required to support such a Schwarzschild-like BH-DM halo combined solution (given by Eqs.~(1), (2)). 
    
\begin{figure}[H]
        \centering
        \includegraphics[width=0.49\textwidth]{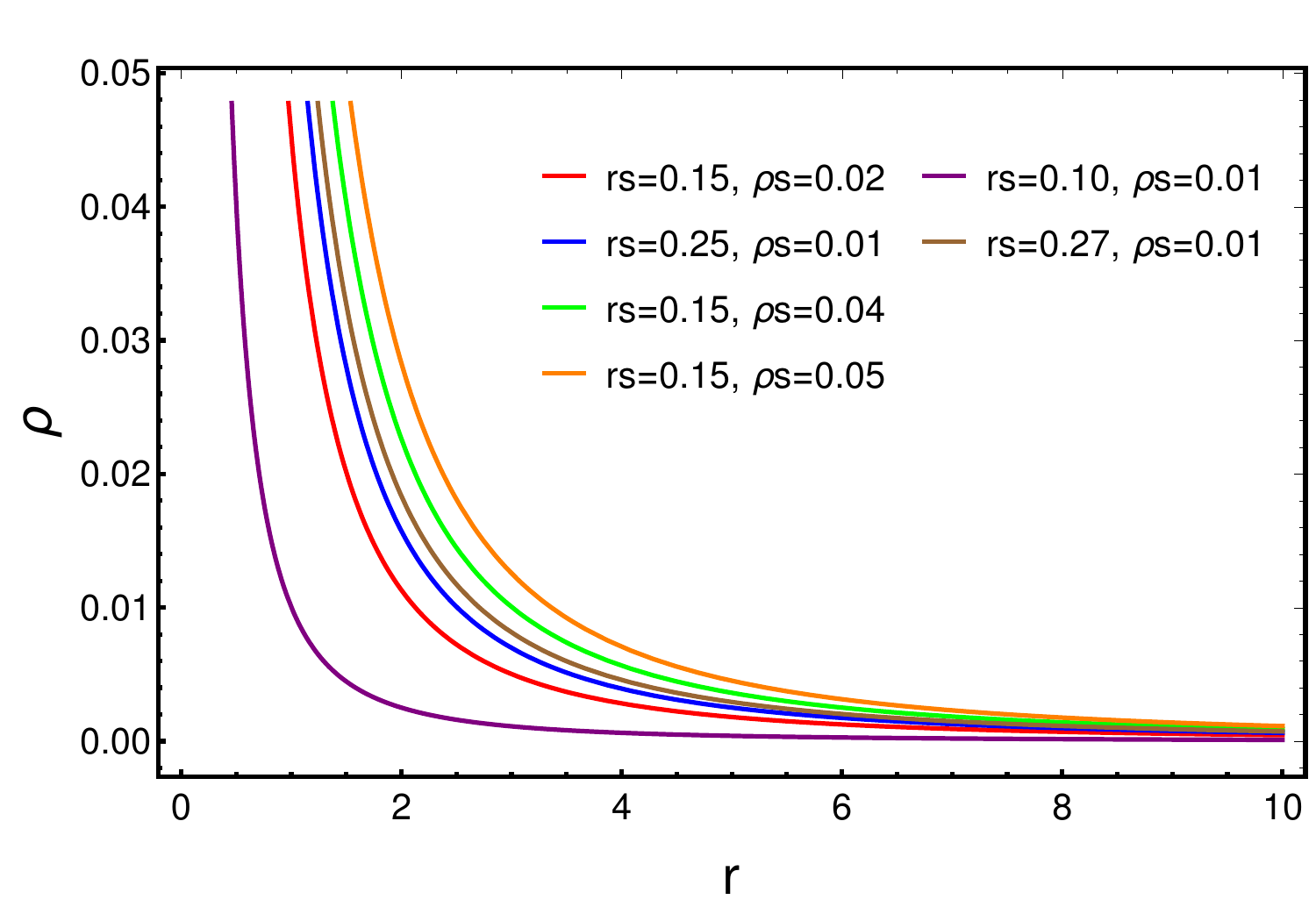}
        \includegraphics[width=0.49\textwidth]{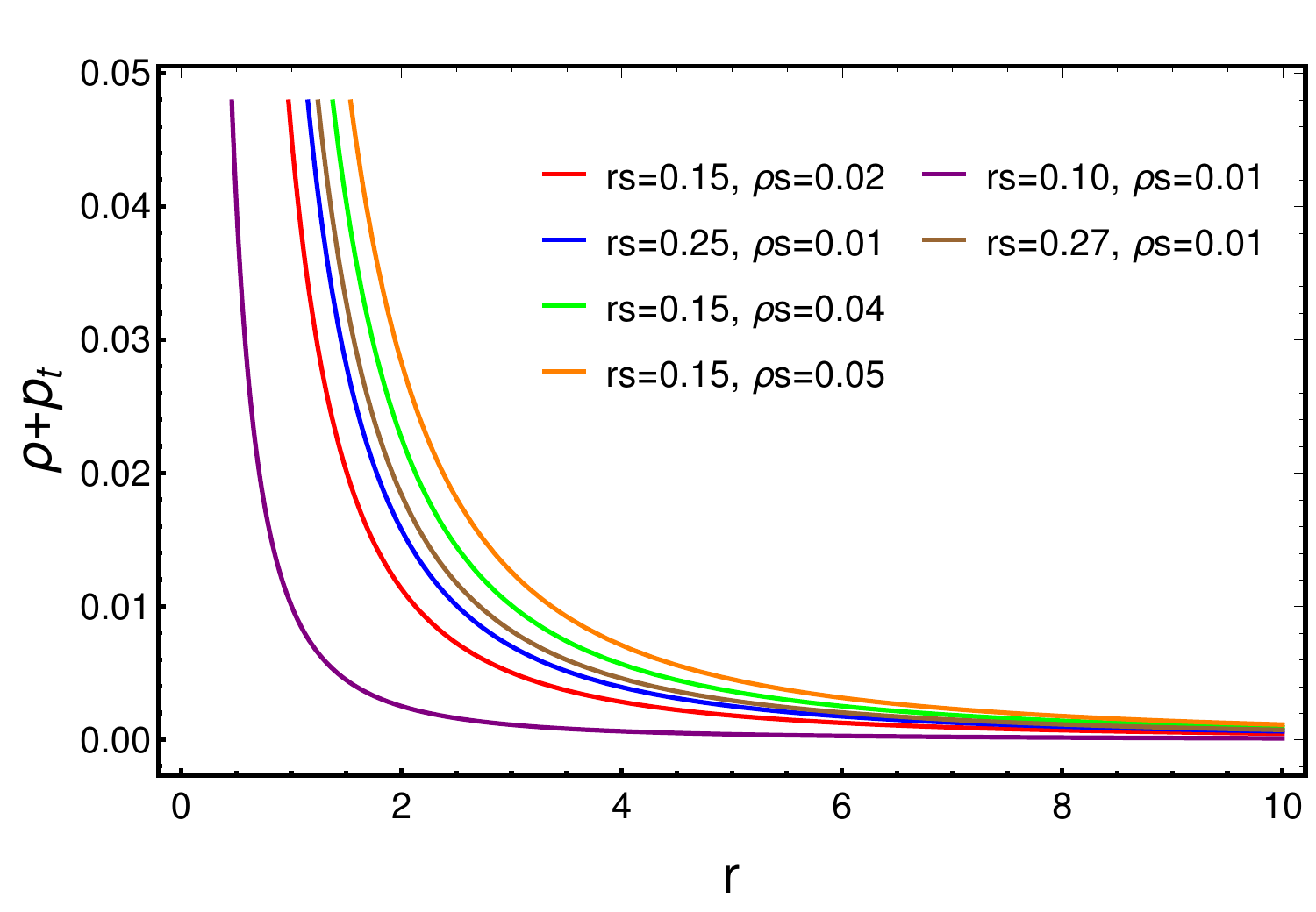}
        \caption{The variation of $\rho$ and $\rho+p_t$ with radial coordinate $r$ for different pair of $(r_s,\rho_s)$. This is clear that $\rho\geq0~,~~\rho+p_t\geq0$ in the entire ranges of $r$.}\label{fwec}
    \end{figure}

\begin{figure}[H]
        \centering
        \includegraphics[width=0.49\textwidth]{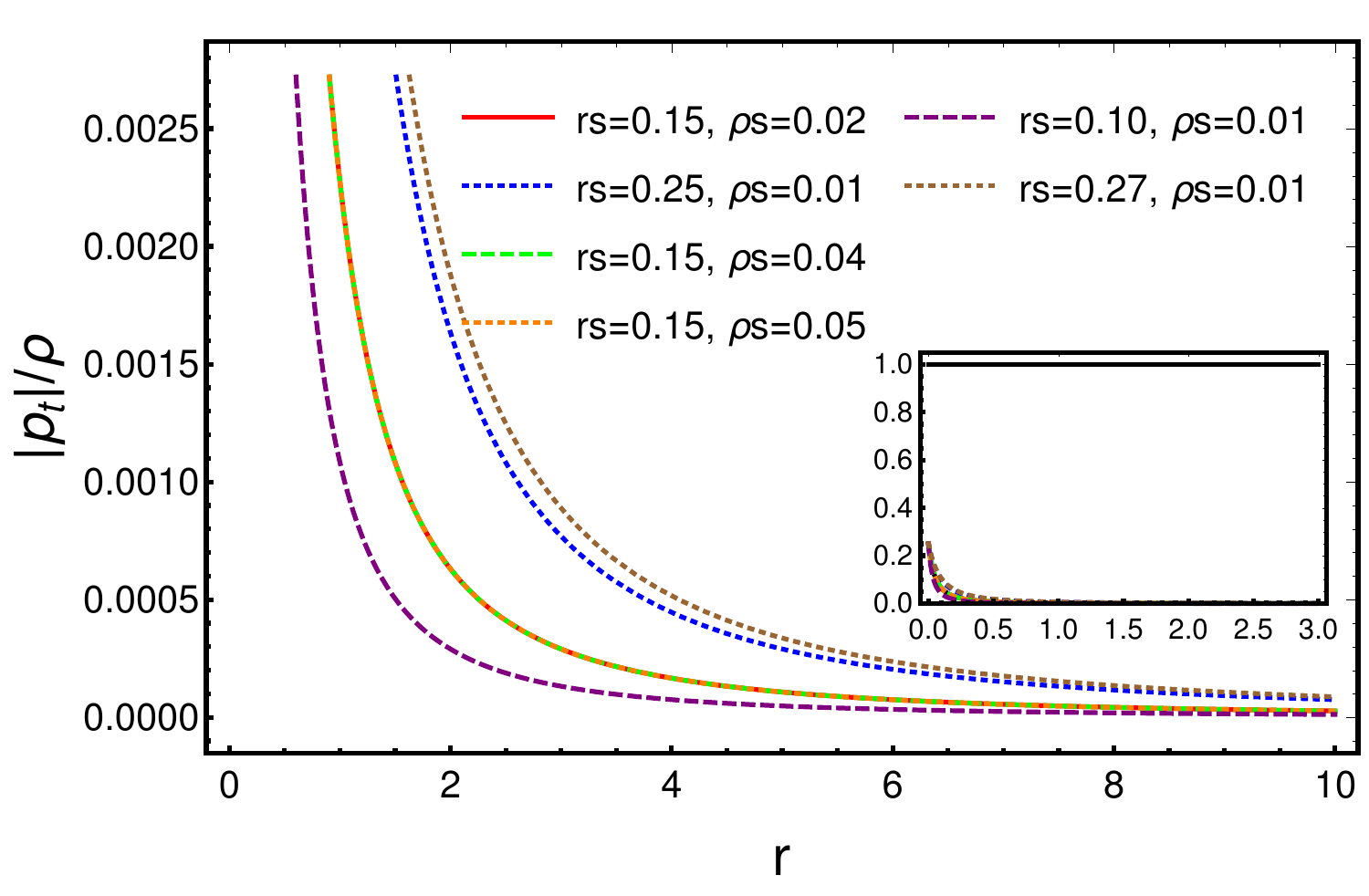}
        \caption{The variation of the ratio of energy density $\rho$ and tangential pressure $p_t$ with radial coordinate $r$ for different pair of $(r_s,\rho_s)$. The inset plot shows that $\frac{|p_t|}{\rho}$ is always less than 1 (denoted as a solid black color in the inset plot) in the entire ranges of $r$ for the considered values of $(r_s,\rho_s)$, clearly states that DEC is always satisfied.}\label{fdec}
    \end{figure}

\begin{figure}[H]
        \centering
        \includegraphics[width=0.49\textwidth]{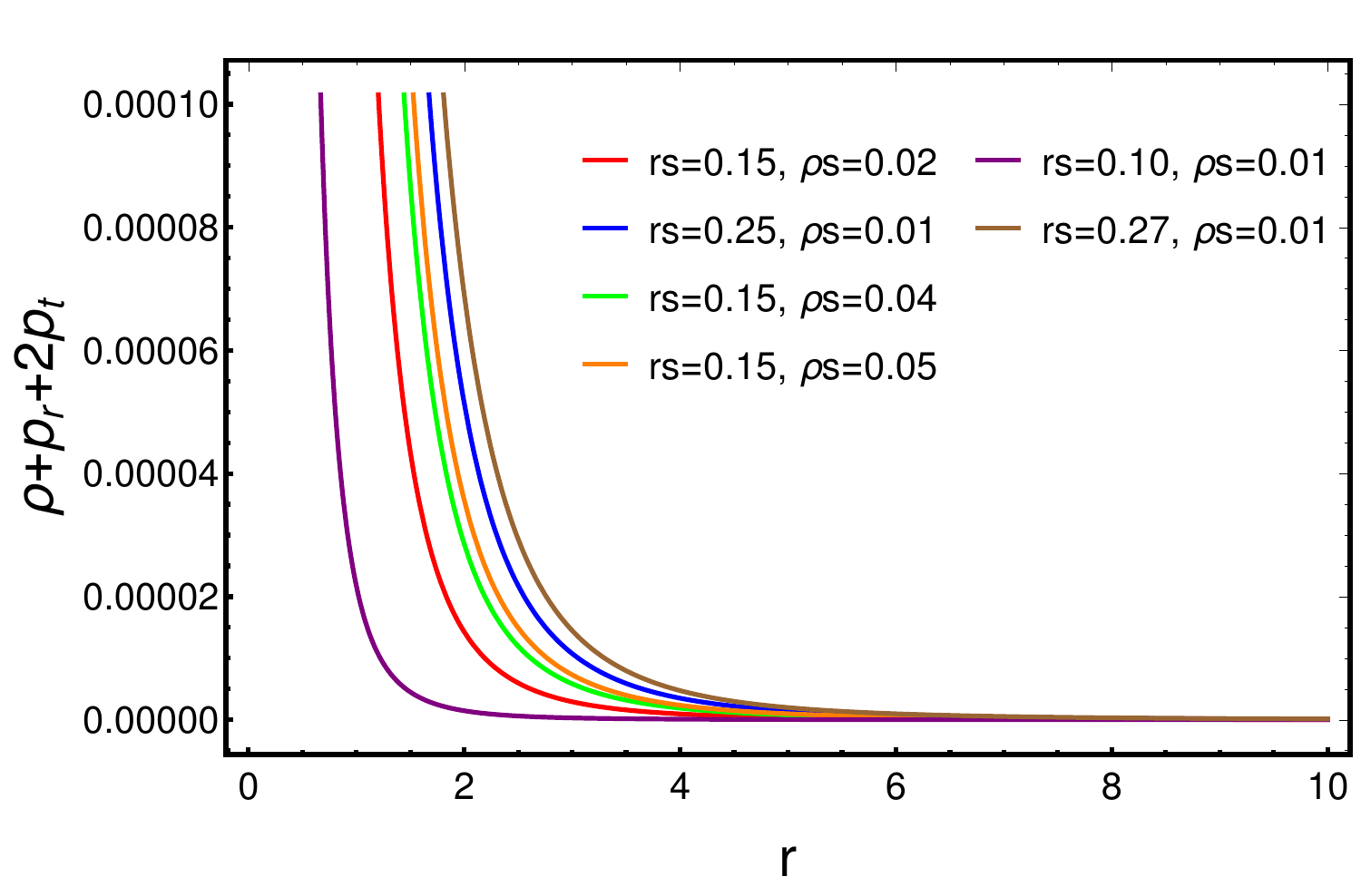}
        \caption{The variation of $\rho+p_r+2p_t$ with radial coordinate $r$ for different pair of $(r_s,\rho_s)$. This is clear that $\rho+p_r+2p_t$ is always non-negative, which obeys the SEC in the entire ranges of $r$.}\label{fsec}
    \end{figure}

\begin{itemize}

    \item[1.] {\bf \textit{WEC:}}\\
    The WEC states that $T_{\mu\nu}v^\mu v^\nu\geq0$ anywhere, for a future-directed timelike vector $v^\mu$. Therefore with a SSS line element (Eq.~(1)), it implies
    \begin{equation}
        \rho\geq0~,~~~\rho+p_i\geq0~,~~~(i=r,\theta,\phi)~.
    \end{equation}
    Therefore in our considered spacetime, from Eq.~\eqref{eq:ecs} we have
    \begin{eqnarray}
        &\rho>0~,~~~\rho+p_r=0~,\non\\
        &\rho+p_t=\frac{4\pi\rho_s r_s^2(8r^2+12r r_s+5 r_s^2)}{r^\frac{5}{2}(r+r_s)^\frac{3}{2}}\geq0~.
    \end{eqnarray}

    \item[2.] {\bf \textit{NEC:}}\\
    The NEC makes the same statement as the WEC except that $v^\mu$ is replaced by a future-directed null vector $k^\mu$. Thus,
    \begin{eqnarray}
        &T_{\mu\nu}k^\mu k^\nu\geq0~,&\non\\
        &\implies \rho+p_i\geq0~,~~~(i=r,\theta,\phi)&~.
    \end{eqnarray}
    Therefore, the required matter obeys the WEC and NEC over the entire domain of the radial coordinate. This is also confirmed by Fig.~\ref{fwec}.\\

    \item[3.] {\bf \textit{DEC:}}\\
    The DEC states the notion that matter should ﬂow along timelike or null world lines. Hence for a future-directed timelike vector ﬁeld $v^\mu$, the quantity $-T^\mu_{\nu} v^\nu$ is a future-directed, timelike or null, vector ﬁeld and therefore requires that
    \begin{equation}
        \rho\geq |p_t|~,\implies~~\frac{|p_t|}{\rho}\leq1~.\label{eq:dec}
    \end{equation}
    In our case, the above relation turns out to be
    \begin{equation}
        \frac{|p_t|}{\rho}=\frac{r_s^2}{4(r+r_s) (2r+r_s)}< 1~.
    \end{equation}
    In Fig.~\ref{fdec}, we have plotted the ratio of tangential pressure $|p_t|$ with density $\rho$, which ensures complete fulfillment of this energy condition.\\

    \item[4.] {\bf \textit{SEC:}}\\
    The strong energy condition is 
    \begin{eqnarray}
        &\big(T_{\mu\nu}-\frac{1}{2} T g_{\mu\nu}\big)v^\mu v^\nu\geq0~,\non\\
        &\implies\rho+\sum_{i} p_i\geq0~.
    \end{eqnarray}
    Here $T$ is the trace of the energy-momentum tensor. Thus in our considered BH-DM halo solution, the expression of SEC gives
    \begin{equation}
        \rho+p_r+2p_t=\frac{8\pi\rho_s r_s^4}{r^\frac{5}{2}(r+r_s)^\frac{3}{2}}\geq0~.
    \end{equation}
    From Fig.~\ref{fsec}, it is clear that SEC is also satisfied over the whole range of radial coordinate.

\end{itemize}
\noindent
Therefore the analytical examinations of all the energy conditions allow us to presume that all the energy conditions are well-satisfied in our considered BH-DM halo combined spacetime. It can also be clearly seen from the graphical analysis, as presented in Figs.~\ref{fwec}, \ref{fdec} and \ref{fsec}, where we have considered to plot all those pair values of $(r_s,\rho_s)$, for which all the analysis presented in our work is taken into account. The validation of all energy conditions has also been examined by Al-Badawi et al. in the considered BH-DM halo spacetime for a Dehnen-$(1,4,5/2)$ type DM halo \cite{Al-Badawi:2024asn}, which mirrors the same results of ours. Additionally, we also note here that Uktamov et al. recently proposed a new combined Schwarzschild-like BH-DM halo solution surrounded by a Dehnen-$(1,4,2)$ type density distribution, in which they systematically analyzed all the energy conditions and concluded that all of them are well satisfied over the entire domain of $r>0$ \cite{UktamjonUktamov:2025emm}.

\section*{An unstable maxima: resemblance to an inverted harmonic oscillator in SSS Painlevé-Gullstrand coordinate}\label{a1}
\noindent
We start with the Lagrangian of a massive test particle with conserved energy $E$ and conserved angular momentum $L$ around a Schwarzschild BH-DM halo spacetime, considering the equatorial plane symmetry $(\theta=\pi/2)$ in Painlev$\Acute{e}$-Gullstrand coordinate, which can be written as
    \begin{eqnarray}
        &2\mathcal{L}=-f(r)E^2+2\sqrt{1-f(r)}Em\left(\frac{dr}{d\tau}\right)\non\\
        &+\left(\frac{dr}{d\tau}\right)^2+\frac{L^2}{r^2}~,\label{eq:a1}
    \end{eqnarray}
where $\tau$ being the proper time along the world-line of the massive test particle. From the conserved Hamiltonian associated with the invariance of the Lagrangian $\mathcal{L}$ under translations of the proper time ($\tau$), one may have
    \begin{equation}
        \frac{d}{d\tau}(2\mathcal{L})=0\implies 2\mathcal{L}=\epsilon\label{A1}~,
    \end{equation}

    \begin{figure}[H]
        \centering
        \includegraphics[width=0.49\textwidth]{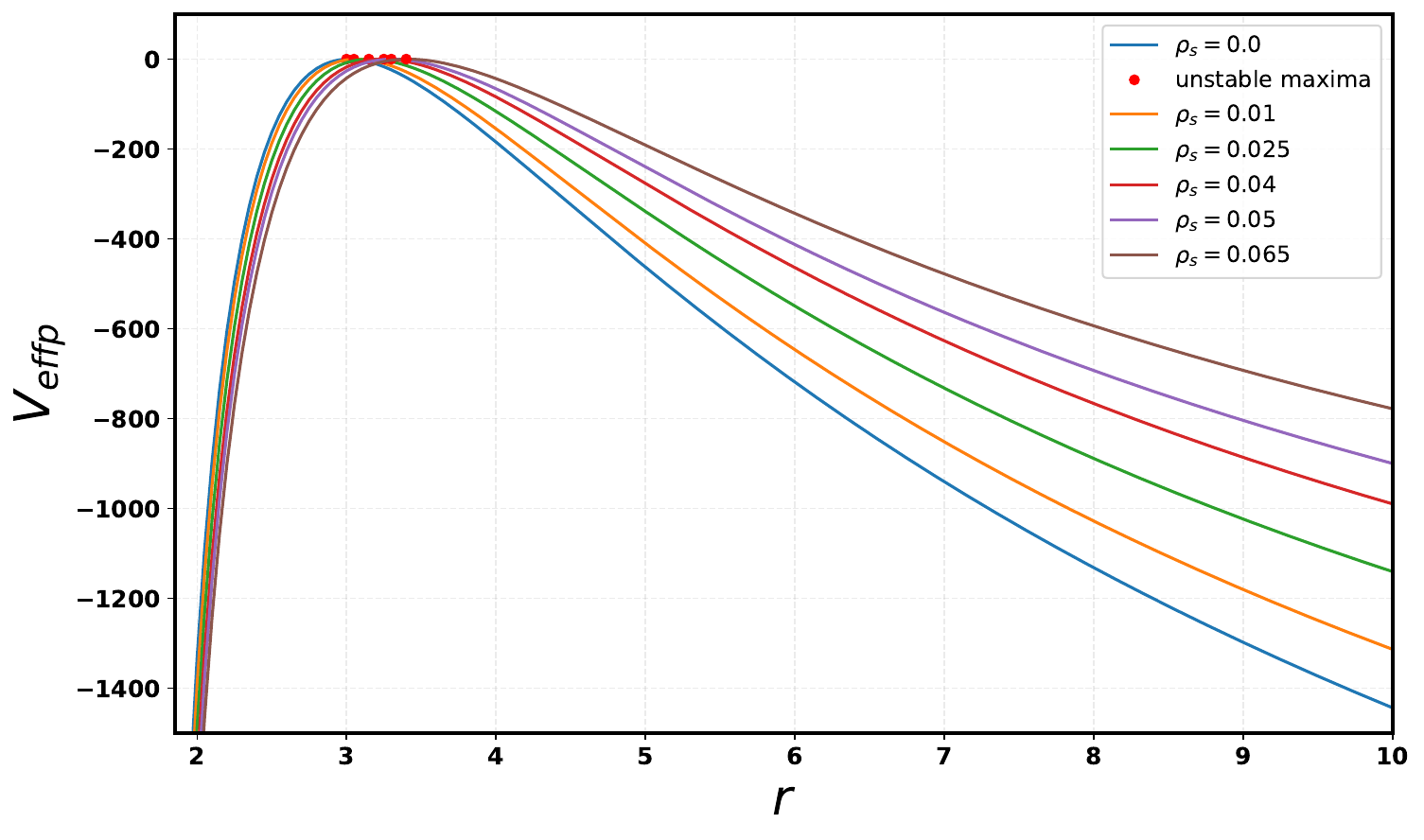}
        \includegraphics[width=0.49\textwidth]{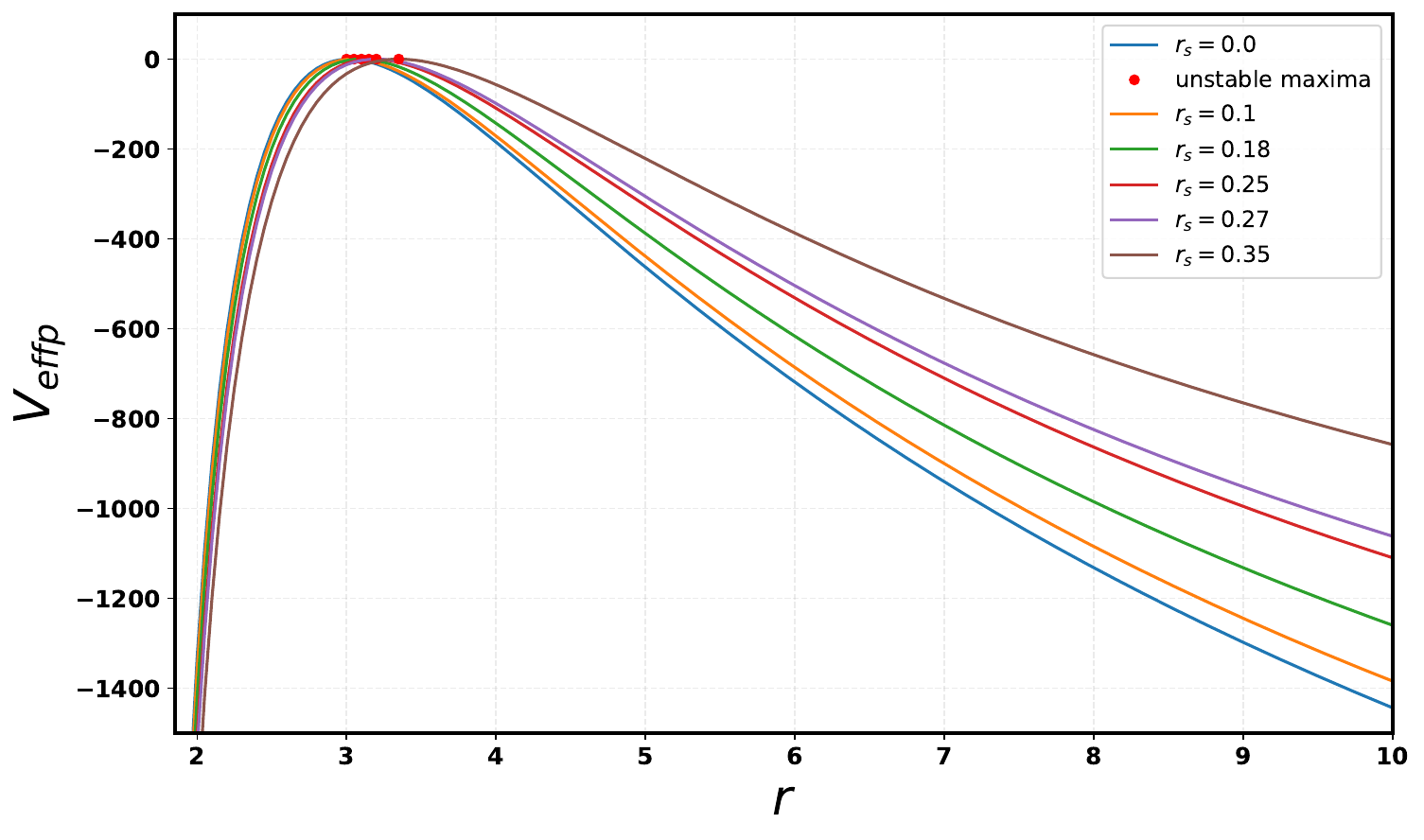}
        \caption{{\it Upper:} Plot of $V_{\rm effp}$, defined in the Painlev$\Acute{e}$-Gullstrand coordinate with radial coordinate $r$ by considering an EMRI system $(q=10^{-5})$ for different values of the DM halo central density $\rhs$ with fixed $L=175, E=100$ and the DM halo scale radius $\rs=0.15$.\\
        {\it Lower:} Plot of $V_{\rm effp}$, defined in the Painlev$\Acute{e}$-Gullstrand coordinate with radial coordinate $r$ by considering an EMRI system $(q=10^{-5})$ for different values of the DM halo scale radius $\rs$ with fixed $L=3.75, E=1.65$ and the DM halo central density $\rhs=0.01$.}\label{f1A}
    \end{figure}
\noindent
where $\epsilon=-m^2$. Therefore using Eq.~\eqref{A1} from Eq.~\eqref{eq:a1}, we obtain the radial equation for the geodesic motion,
    \begin{eqnarray}
        \frac{1}{2}m\left(\frac{dr}{d\tau}\right)^2+V_{\rm effp}(r)=0~,
    \end{eqnarray}
where the expression of the effective potential in the Painlev$\Acute{e}$-Gullstrand coordinate is given by
    \begin{eqnarray}
        &V_{\rm effp}(r)=\frac{1}{2m}\Bigg(2E\sqrt{\left(1-f(r)\right)\left(E^2-m^2-\frac{L^2}{r^2}\right)}-2E^2\nonumber\\
        &+f(r)E^2+m^2+\frac{L^2}{r^2}\Bigg)~,\nonumber\\
        \label{A4}
    \end{eqnarray}
The solution of the Schwarzschild BH-DM halo spacetime is given by Eq.~(2) and in the above expression, the effective potential depends on $E,L,M,m$ and the DM halo parameters $\rhs,\rs$. One can explore that an unstable local maxima arises in the effective potential, mentioned in Eq.~\eqref{A4}. For those radii of unstable circular orbits, we have
    \begin{eqnarray}
         V_{\rm effp}(r)=0~, \quad \frac{dV_{\rm effp}(r)}{dr}=0~.
    \end{eqnarray}
\noindent
The position of the unstable local maximum, indicated by a red point in Fig.~\ref{f1A}, varies for different values of the DM halo parameters: the central density $\rho_s$ and the scale radius $r_s$. It is noteworthy to mention that all points situated to the left of this identified maximum are inherently unstable. Consequently, the inclusion of a harmonic oscillator as an external potential become essential to probe the event horizon by placing a test particle within this unstable region (which is also very close to the location of the event horizon).


\bibliography{references}

@article{LIGOScientific:2016emj,
    author = "Abbott, B. P. and others",
    collaboration = "LIGO Scientific, Virgo",
    title = "{GW150914: The Advanced LIGO Detectors in the Era of First Discoveries}",
    eprint = "1602.03838",
    archivePrefix = "arXiv",
    primaryClass = "gr-qc",
    reportNumber = "LIGO-P1500237",
    doi = "10.1103/PhysRevLett.116.131103",
    journal = "Phys. Rev. Lett.",
    volume = "116",
    number = "13",
    pages = "131103",
    year = "2016"
}

@article{TheLIGOScientific:2017qsa,
    author = "Abbott, B. P. and others",
    collaboration = "LIGO Scientific, Virgo",
    title = "{GW170817: Observation of Gravitational Waves from a Binary Neutron Star Inspiral}",
    eprint = "1710.05832",
    archivePrefix = "arXiv",
    primaryClass = "gr-qc",
    reportNumber = "LIGO-P170817",
    doi = "10.1103/PhysRevLett.119.161101",
    journal = "Phys. Rev. Lett.",
    volume = "119",
    number = "16",
    pages = "161101",
    year = "2017"
}

@article{LIGOScientific:2018mvr,
    author = "Abbott, B. P. and others",
    collaboration = "LIGO Scientific, Virgo",
    title = "{GWTC-1: A Gravitational-Wave Transient Catalog of Compact Binary Mergers Observed by LIGO and Virgo during the First and Second Observing Runs}",
    eprint = "1811.12907",
    archivePrefix = "arXiv",
    primaryClass = "astro-ph.HE",
    reportNumber = "LIGO-P1800307",
    doi = "10.1103/PhysRevX.9.031040",
    journal = "Phys. Rev. X",
    volume = "9",
    number = "3",
    pages = "031040",
    year = "2019"
}

@article{LIGOScientific:2020ibl,
    author = "Abbott, R. and others",
    collaboration = "LIGO Scientific, Virgo",
    title = "{GWTC-2: Compact Binary Coalescences Observed by LIGO and Virgo During the First Half of the Third Observing Run}",
    eprint = "2010.14527",
    archivePrefix = "arXiv",
    primaryClass = "gr-qc",
    reportNumber = "P2000061",
    doi = "10.1103/PhysRevX.11.021053",
    journal = "Phys. Rev. X",
    volume = "11",
    pages = "021053",
    year = "2021"
}

@article{LIGOScientific:2021usb,
    author = "Abbott, R. and others",
    collaboration = "LIGO Scientific, VIRGO",
    title = "{GWTC-2.1: Deep extended catalog of compact binary coalescences observed by LIGO and Virgo during the first half of the third observing run}",
    eprint = "2108.01045",
    archivePrefix = "arXiv",
    primaryClass = "gr-qc",
    reportNumber = "LIGO-P2100063",
    doi = "10.1103/PhysRevD.109.022001",
    journal = "Phys. Rev. D",
    volume = "109",
    number = "2",
    pages = "022001",
    year = "2024"
}

@article{LIGOScientific:2025rid,
    author = "Abac, A. G. and others",
    collaboration = "LIGO Scientific, Virgo, KAGRA",
    title = "{GW250114: Testing Hawking{\textquoteright}s Area Law and the Kerr Nature of Black Holes}",
    eprint = "2509.08054",
    archivePrefix = "arXiv",
    primaryClass = "gr-qc",journal = "Phys. Rev. Lett.",
    volume = "135",
    number = "11",
    pages = "111403",
    year = "2025",
    reportNumber = "LIGO-P2500421",
    doi = "10.1103/kw5g-d732",    
}

@article{KAGRA:2021vkt,
    author = "Abbott, R. and others",
    collaboration = "KAGRA, VIRGO, LIGO Scientific",
    title = "{GWTC-3: Compact Binary Coalescences Observed by LIGO and Virgo during the Second Part of the Third Observing Run}",
    eprint = "2111.03606",
    archivePrefix = "arXiv",
    primaryClass = "gr-qc",
    reportNumber = "LIGO-P2000318",
    doi = "10.1103/PhysRevX.13.041039",
    journal = "Phys. Rev. X",
    volume = "13",
    number = "4",
    pages = "041039",
    year = "2023"
}

@article{EventHorizonTelescope:2019dse,
    author = "Akiyama, Kazunori and others",
    collaboration = "Event Horizon Telescope",
    title = "{First M87 Event Horizon Telescope Results. I. The Shadow of the Supermassive Black Hole}",
    eprint = "1906.11238",
    archivePrefix = "arXiv",
    primaryClass = "astro-ph.GA",
    doi = "10.3847/2041-8213/ab0ec7",
    journal = "Astrophys. J. Lett.",
    volume = "875",
    pages = "L1",
    year = "2019"
}

@article{EventHorizonTelescope:2019ggy,
    author = "Akiyama, Kazunori and others",
    collaboration = "Event Horizon Telescope",
    title = "{First M87 Event Horizon Telescope Results. VI. The Shadow and Mass of the Central Black Hole}",
    eprint = "1906.11243",
    archivePrefix = "arXiv",
    primaryClass = "astro-ph.GA",
    doi = "10.3847/2041-8213/ab1141",
    journal = "Astrophys. J. Lett.",
    volume = "875",
    number = "1",
    pages = "L6",
    year = "2019"
}

@article{EventHorizonTelescope:2019ths,
    author = "Akiyama, Kazunori and others",
    collaboration = "Event Horizon Telescope",
    title = "{First M87 Event Horizon Telescope Results. IV. Imaging the Central Supermassive Black Hole}",
    eprint = "1906.11241",
    archivePrefix = "arXiv",
    primaryClass = "astro-ph.GA",
    doi = "10.3847/2041-8213/ab0e85",
    journal = "Astrophys. J. Lett.",
    volume = "875",
    number = "1",
    pages = "L4",
    year = "2019"
}

@article{EventHorizonTelescope:2022wkp,
    author = "Akiyama, Kazunori and others",
    collaboration = "Event Horizon Telescope",
    title = "{First Sagittarius A* Event Horizon Telescope Results. I. The Shadow of the Supermassive Black Hole in the Center of the Milky Way}",
    eprint = "2311.08680",
    archivePrefix = "arXiv",
    primaryClass = "astro-ph.HE",
    doi = "10.3847/2041-8213/ac6674",
    journal = "Astrophys. J. Lett.",
    volume = "930",
    number = "2",
    pages = "L12",
    year = "2022"
}

@book{Chandrasekhar:1985kt,
  author = "S. Chandrasekhar",
  title = "{The mathematical theory of black holes}",
  publisher = {Clarendon Press 1998}
}

@article{Carter,
  title = {Global Structure of the Kerr Family of Gravitational Fields},
  author = {Carter, Brandon},
  journal = {Phys. Rev.},
  volume = {174},
  issue = {5},
  pages = {1559--1571},
  numpages = {0},
  year = {1968},
  month = {Oct},
  publisher = {American Physical Society},
  doi = {10.1103/PhysRev.174.1559},
  url = {https://link.aps.org/doi/10.1103/PhysRev.174.1559}
}

@article{Frolov:2017kze,
    author = "Frolov, Valeri P. and Krtous, Pavel and Kubiznak, David",
    title = "{Black holes, hidden symmetries, and complete integrability}",
    eprint = "1705.05482",
    archivePrefix = "arXiv",
    primaryClass = "gr-qc",
    doi = "10.1007/s41114-017-0009-9",
    journal = "Living Rev. Rel.",
    volume = "20",
    number = "1",
    pages = "6",
    year = "2017"
}

@article{Suzuki:1999si,
    author = "Suzuki, Shingo and Maeda, Kei-ichi",
    title = "{Signature of chaos in gravitational waves from a spinning particle}",
    eprint = "gr-qc/9910064",
    archivePrefix = "arXiv",
    reportNumber = "WU-AP-83-99",
    doi = "10.1103/PhysRevD.61.024005",
    journal = "Phys. Rev. D",
    volume = "61",
    pages = "024005",
    year = "2000"
}

@article{Kiuchi:2004bv,
    author = "Kiuchi, Kenta and Maeda, Kei-ichi",
    title = "{Gravitational waves from chaotic dynamical system}",
    eprint = "gr-qc/0404124",
    archivePrefix = "arXiv",
    reportNumber = "WU-AP-185-04",
    doi = "10.1103/PhysRevD.70.064036",
    journal = "Phys. Rev. D",
    volume = "70",
    pages = "064036",
    year = "2004"
}

@article{Bombelli:1991eg,
    author = "Bombelli, Luca and Calzetta, Esteban",
    title = "{Chaos around a black hole}",
    reportNumber = "GTCRG-91-12",
    journal = "Class. Quant. Grav.",
    volume = "9",
    pages = "2573--2599",
    year = "1992"
}

@article{Sota:1995ms,
    author = "Sota, Yasuhide and Suzuki, Shingo and Maeda, Kei-ichi",
    title = "{Chaos in static axisymmetric space-times. 1: Vacuum case}",
    eprint = "gr-qc/9505036",
    archivePrefix = "arXiv",
    reportNumber = "WU-AP-45-95",
    journal = "Class. Quant. Grav.",
    volume = "13",
    pages = "1241--1260",
    year = "1996"
}

@article{Vieira:1996zf,
    author = "Vieira, Werner M. and Letelier, Patricio S.",
    title = "{Chaos around a Henon-Heiles inspired exact perturbation of a black hole}",
    eprint = "gr-qc/9604037",
    archivePrefix = "arXiv",
    journal = "Phys. Rev. Lett.",
    volume = "76",
    pages = "1409--1412",
    year = "1996"
}

@article{Suzuki:1996gm,
    author = "Suzuki, Shingo and Maeda, Kei-ichi",
    title = "{Chaos in Schwarzschild space-time: The motion of a spinning particle}",
    eprint = "gr-qc/9604020",
    archivePrefix = "arXiv",
    reportNumber = "WU-AP-59-96",
    journal = "Phys. Rev. D",
    volume = "55",
    pages = "4848--4859",
    year = "1997"
}

@article{Cornish:1996ri,
    author = "Cornish, Neil J. and Frankel, Norman E.",
    title = "{The Black hole and the pea}",
    reportNumber = "UM-P-96-45-REV, UM-P-96-45",
    journal = "Phys. Rev. D",
    volume = "56",
    pages = "1903--1907",
    year = "1997"
}

@article{deMoura:1999wf,
    author = "de Moura, Alessandro P. S. and Letelier, Patricio S.",
    title = "{Chaos and fractals in geodesic motions around a nonrotating black hole with an external halo}",
    eprint = "chao-dyn/9910035",
    archivePrefix = "arXiv",
    journal = "Phys. Rev. E",
    volume = "61",
    pages = "6506--6516",
    year = "2000"
}

@article{Hartl:2002ig,
    author = "Hartl, Michael D.",
    title = "{Dynamics of spinning test particles in Kerr space-time}",
    eprint = "gr-qc/0210042",
    archivePrefix = "arXiv",
    journal = "Phys. Rev. D",
    volume = "67",
    pages = "024005",
    year = "2003"
}

@article{Han:2008zzf,
    author = "Han, Wenbiao",
    title = "{Chaos and dynamics of spinning particles in Kerr spacetime}",
    eprint = "1006.2229",
    archivePrefix = "arXiv",
    primaryClass = "gr-qc",
    journal = "Gen. Rel. Grav.",
    volume = "40",
    pages = "1831--1847",
    year = "2008"
}

@article{Takahashi:2008zh,
    author = "Takahashi, Masaaki and Koyama, Hiroko",
    title = "{Chaotic motion of Charged Particles in an Electromagnetic Field Surrounding a Rotating Black Hole}",
    eprint = "0807.0277",
    archivePrefix = "arXiv",
    primaryClass = "astro-ph",
    journal = "Astrophys. J.",
    volume = "693",
    pages = "472--485",
    year = "2009"
}

@article{Hashimoto:2016dfz,
    author = "Hashimoto, Koji and Tanahashi, Norihiro",
    title = "{Universality in Chaos of Particle Motion near Black Hole Horizon}",
    eprint = "1610.06070",
    archivePrefix = "arXiv",
    primaryClass = "hep-th",
    reportNumber = "OU-HET-911",
    journal = "Phys. Rev. D",
    volume = "95",
    number = "2",
    pages = "024007",
    year = "2017"
}

@article{Li:2018wtz,
    author = "Li, Dan and Wu, Xin",
    title = "{Chaotic motion of neutral and charged particles in a magnetized Ernst-Schwarzschild spacetime}",
    eprint = "1803.02119",
    archivePrefix = "arXiv",
    primaryClass = "gr-qc",
    journal = "Eur. Phys. J. Plus",
    volume = "134",
    number = "3",
    pages = "96",
    year = "2019"
}

@article{Lei:2020clg,
    author = "Lei, Yu-Qi and Ge, Xian-Hui and Ran, Cheng",
    title = "{Chaos of particle motion near a black hole with quasitopological electromagnetism}",
    eprint = "2008.01384",
    archivePrefix = "arXiv",
    primaryClass = "hep-th",
    doi = "10.1103/PhysRevD.104.046020",
    journal = "Phys. Rev. D",
    volume = "104",
    number = "4",
    pages = "046020",
    year = "2021"
}

@article{DeFalco:2020yys,
    author = "De Falco, Vittorio and Borrelli, William",
    title = "{Detection of chaos in the general relativistic Poynting-Robertson effect: Kerr equatorial plane}",
    eprint = "2001.04979",
    archivePrefix = "arXiv",
    primaryClass = "gr-qc",
    doi = "10.1103/PhysRevD.103.064014",
    journal = "Phys. Rev. D",
    volume = "103",
    number = "6",
    pages = "064014",
    year = "2021"
}

@article{DeFalco:2021uak,
    author = "De Falco, Vittorio and Borrelli, William",
    title = "{Timescales of the chaos onset in the general relativistic Poynting-Robertson effect}",
    eprint = "2105.00965",
    archivePrefix = "arXiv",
    primaryClass = "gr-qc",
    journal = "Phys. Rev. D",
    doi = "10.1103/PhysRevD.103.124012",
    month = "5",
    year = "2021"
}

@article{Dalui:2018qqv,
    author = "Dalui, Surojit and Majhi, Bibhas Ranjan and Mishra, Pankaj",
    title = "{Presence of horizon makes particle motion chaotic}",
    eprint = "1803.06527",
    archivePrefix = "arXiv",
    primaryClass = "gr-qc",
    doi = "10.1016/j.physletb.2018.11.050",
    journal = "Phys. Lett. B",
    volume = "788",
    pages = "486--493",
    year = "2019"
}

@article{Dalui:2019umw,
    author = "Dalui, Surojit and Majhi, Bibhas Ranjan and Mishra, Pankaj",
    title = "{Induction of chaotic fluctuations in particle dynamics in a uniformly accelerated frame}",
    eprint = "1904.11760",
    archivePrefix = "arXiv",
    primaryClass = "gr-qc",
    doi = "10.1142/S0217751X20500815",
    journal = "Int. J. Mod. Phys. A",
    volume = "35",
    number = "18",
    pages = "2050081",
    year = "2020"
}

@article{Das:2024iuf,
    author = "Das, Surajit and Dalui, Surojit and Samanta, Rickmoy",
    title = "{Near-horizon chaos beyond Einstein gravity}",
    eprint = "2405.09945",
    archivePrefix = "arXiv",
    primaryClass = "gr-qc",
    doi = "10.1103/PhysRevD.110.124037",
    journal = "Phys. Rev. D",
    volume = "110",
    number = "12",
    pages = "124037",
    year = "2024"
}

@article{Lu:2026kcm,
    author = "Lu, Junjie and Wu, Xin",
    title = "{Third type of spacetime with the coexistence of integrability and non-integrability}",
    eprint = "2603.12674",
    archivePrefix = "arXiv",
    primaryClass = "gr-qc",
    doi = "10.1140/epjc/s10052-026-15482-w",
    journal = "Eur. Phys. J. C",
    volume = "86",
    number = "3",
    pages = "256",
    year = "2026"
}

@article{Levin:2008yp,
    author = "Levin, Janna and Perez-Giz, Gabe",
    title = "{Homoclinic Orbits around Spinning Black Holes. I. Exact Solution for the Kerr Separatrix}",
    eprint = "0811.3814",
    archivePrefix = "arXiv",
    primaryClass = "gr-qc",
    doi = "10.1103/PhysRevD.79.124013",
    journal = "Phys. Rev. D",
    volume = "79",
    pages = "124013",
    year = "2009"
}

@article{Perez-Giz:2008ajn,
    author = "Perez-Giz, Gabe and Levin, Janna",
    title = "{Homoclinic Orbits around Spinning Black Holes II: The Phase Space Portrait}",
    eprint = "0811.3815",
    archivePrefix = "arXiv",
    primaryClass = "gr-qc",
    doi = "10.1103/PhysRevD.79.124014",
    journal = "Phys. Rev. D",
    volume = "79",
    pages = "124014",
    year = "2009"
}

@article{Jeong:2023hom,
    author = "Jeong, Soyeon and Lee, Bum-Hoon and Lee, Hocheol and Lee, Wonwoo",
    title = "{Homoclinic orbit and the violation of the chaos bound around a black hole with anisotropic matter fields}",
    eprint = "2301.12198",
    archivePrefix = "arXiv",
    primaryClass = "gr-qc",
    reportNumber = "CQUeST-2023-0718",
    doi = "10.1103/PhysRevD.107.104037",
    journal = "Phys. Rev. D",
    volume = "107",
    number = "10",
    pages = "104037",
    year = "2023"
}

@article{Li:2023bgn,
    author = "Li, Yi-Ting and Wang, Chen-Yu and Lee, Da-Shin and Lin, Chi-Yong",
    title = "{Homoclinic orbits in Kerr-Newman black holes}",
    eprint = "2302.09471",
    archivePrefix = "arXiv",
    primaryClass = "gr-qc",
    doi = "10.1103/PhysRevD.108.044010",
    journal = "Phys. Rev. D",
    volume = "108",
    number = "4",
    pages = "044010",
    year = "2023"
}

@article{Ciou:2025ygb,
    author = "Ciou, Siang-Yao and Hsieh, Tien and Lee, Da-Shin",
    title = {{Dynamics of spinning particles in Reissner-Nordstr{\"o}m black hole exterior}},
    eprint = "2503.12911",
    archivePrefix = "arXiv",
    primaryClass = "gr-qc",
    doi = "10.1088/1475-7516/2025/05/086",
    journal = "JCAP",
    volume = "05",
    pages = "086",
    year = "2025"
}

@article{Polcar:2019kwu,
    author = "Polcar, L. and Semer{\'a}k, O.",
    title = "{Free motion around black holes with discs or rings: Between integrability and chaos. VI. The Melnikov method}",
    eprint = "1911.09790",
    archivePrefix = "arXiv",
    primaryClass = "gr-qc",
    doi = "10.1103/PhysRevD.100.103013",
    journal = "Phys. Rev. D",
    volume = "100",
    number = "10",
    pages = "103013",
    year = "2019"
}

@book{KAM,
    author = "J. Guckenheimer and P. Holmes",
    title = "{Nonlinear Oscillations, Dynamical Systems, and Bifurcations of Vector Fields (Applied Mathematical Sciences, 42)}", 
    publisher = "Springer, New York 1983"
}

@book{Wigg,
    author = "S. Wiggins",
    title = "{Introduction to Applied Nonlinear Dynamical Systems and Chaos}",
    publisher = "Springer-Verlag, New York 2003"
}

@article{Hughes:2000ssa,
    author = "Hughes, Scott A.",
    editor = "Schutz, Bernard F.",
    title = "{Gravitational waves from extreme mass ratio inspirals: Challenges in mapping the space-time of massive, compact objects}",
    eprint = "gr-qc/0008058",
    archivePrefix = "arXiv",
    doi = "10.1088/0264-9381/18/19/314",
    journal = "Class. Quant. Grav.",
    volume = "18",
    pages = "4067--4074",
    year = "2001"
}

@article{Babak:2017tow,
    author = "Babak, Stanislav and Gair, Jonathan and Sesana, Alberto and Barausse, Enrico and Sopuerta, Carlos F. and Berry, Christopher P. L. and Berti, Emanuele and Amaro-Seoane, Pau and Petiteau, Antoine and Klein, Antoine",
    title = "{Science with the space-based interferometer LISA. V: Extreme mass-ratio inspirals}",
    eprint = "1703.09722",
    archivePrefix = "arXiv",
    primaryClass = "gr-qc",
    doi = "10.1103/PhysRevD.95.103012",
    journal = "Phys. Rev. D",
    volume = "95",
    number = "10",
    pages = "103012",
    year = "2017"
}

@article{Amaro-Seoane:2012lgq,
    author = "Amaro-Seoane, Pau",
    title = "{Relativistic dynamics and extreme mass ratio inspirals}",
    eprint = "1205.5240",
    archivePrefix = "arXiv",
    primaryClass = "astro-ph.CO",
    doi = "10.1007/s41114-018-0013-8",
    journal = "Living Rev. Rel.",
    volume = "21",
    number = "1",
    pages = "4",
    year = "2018"
}

@article{LISA:2017pwj,
    author = "Amaro-Seoane, Pau and others",
    collaboration = "LISA",
    title = "{Laser Interferometer Space Antenna}",
    eprint = "1702.00786",
    archivePrefix = "arXiv",
    primaryClass = "astro-ph.IM",
    month = "2",
    year = "2017"
}

@article{TianQin:2020hid,
    author = "Mei, Jianwei and others",
    collaboration = "TianQin",
    title = "{The TianQin project: current progress on science and technology}",
    eprint = "2008.10332",
    archivePrefix = "arXiv",
    primaryClass = "gr-qc",
    doi = "10.1093/ptep/ptaa114",
    journal = "PTEP",
    volume = "2021",
    number = "5",
    pages = "05A107",
    year = "2021"
}

@article{Hu:2017mde,
    author = "Hu, Wen-Rui and Wu, Yue-Liang",
    title = "{The Taiji Program in Space for gravitational wave physics and the nature of gravity}",
    doi = "10.1093/nsr/nwx116",
    journal = "Natl. Sci. Rev.",
    volume = "4",
    number = "5",
    pages = "685--686",
    year = "2017"
}

@article{Wang:2024tnk,
    author = "Wang, Bo and Li, Bichu and Xiao, Qianqian and Mo, Geyu and Cai, Yi-Fu",
    title = "{Space-based optical lattice clocks as gravitational wave detectors in search for new physics}",
    eprint = "2410.04340",
    archivePrefix = "arXiv",
    primaryClass = "gr-qc",
    doi = "10.1007/s11433-024-2573-3",
    journal = "Sci. China Phys. Mech. Astron.",
    volume = "68",
    number = "4",
    pages = "249512",
    year = "2025"
}

@article{CSST:2025ssq,
    author = "Gong, Yan and others",
    collaboration = "CSST",
    title = "{Introduction to the Chinese Space Station Survey Telescope (CSST)}",
    eprint = "2507.04618",
    archivePrefix = "arXiv",
    primaryClass = "astro-ph.IM",
    doi = "10.1007/s11433-025-2809-0",
    journal = "Sci. China Phys. Mech. Astron.",
    volume = "69",
    number = "3",
    pages = "239501",
    year = "2026"
}

@article{Bertone:2004pz,
    author = "Bertone, Gianfranco and Hooper, Dan and Silk, Joseph",
    title = "{Particle dark matter: Evidence, candidates and constraints}",
    eprint = "hep-ph/0404175",
    archivePrefix = "arXiv",
    reportNumber = "FERMILAB-PUB-04-047-A",
    doi = "10.1016/j.physrep.2004.08.031",
    journal = "Phys. Rept.",
    volume = "405",
    pages = "279--390",
    year = "2005"
}

@article{Bovy:2013raa,
    author = "Bovy, Jo and Rix, Hans-Walter",
    title = "{A Direct Dynamical Measurement of the Milky Way's Disk Surface Density Profile, Disk Scale Length, and Dark Matter Profile at 4 kpc $\stackrel{<}{\sim}$ R $\stackrel{<}{\sim}$ 9 kpc}",
    eprint = "1309.0809",
    archivePrefix = "arXiv",
    primaryClass = "astro-ph.GA",
    doi = "10.1088/0004-637X/779/2/115",
    journal = "Astrophys. J.",
    volume = "779",
    pages = "115",
    year = "2013"
}

@article{deSwart:2017heh,
    author = "de Swart, Jaco and Bertone, Gianfranco and van Dongen, Jeroen",
    title = "{How Dark Matter Came to Matter}",
    eprint = "1703.00013",
    archivePrefix = "arXiv",
    primaryClass = "astro-ph.CO",
    doi = "10.1038/s41550017-0059",
    journal = "Nature Astron.",
    volume = "1",
    pages = "0059",
    year = "2017"
}

@article{Wechsler:2018pic,
    author = "Wechsler, Risa H. and Tinker, Jeremy L.",
    title = "{The Connection between Galaxies and their Dark Matter Halos}",
    eprint = "1804.03097",
    archivePrefix = "arXiv",
    primaryClass = "astro-ph.GA",
    doi = "10.1146/annurev-astro-081817-051756",
    journal = "Ann. Rev. Astron. Astrophys.",
    volume = "56",
    pages = "435--487",
    year = "2018"
}

@article{Bertone:2018krk,
    author = "Bertone, Gianfranco and Tait, M. P., Tim",
    title = "{A new era in the search for dark matter}",
    eprint = "1810.01668",
    archivePrefix = "arXiv",
    primaryClass = "astro-ph.CO",
    doi = "10.1038/s41586-018-0542-z",
    journal = "Nature",
    volume = "562",
    number = "7725",
    pages = "51--56",
    year = "2018"
}

@article{Dubinski:1991bm,
    author = "Dubinski, John and Carlberg, R. G.",
    title = "{The Structure of cold dark matter halos}",
    doi = "10.1086/170451",
    journal = "Astrophys. J.",
    volume = "378",
    pages = "496",
    year = "1991"
}

@article{Hernquist:1990be,
    author = "Hernquist, Lars",
    title = "{An Analytical Model for Spherical Galaxies and Bulges}",
    reportNumber = "IASSNS-AST-89-63",
    doi = "10.1086/168845",
    journal = "Astrophys. J.",
    volume = "356",
    pages = "359",
    year = "1990"
}

@article{Navarro:1995iw,
    author = "Navarro, Julio F. and Frenk, Carlos S. and White, Simon D. M.",
    title = "{The Structure of cold dark matter halos}",
    eprint = "astro-ph/9508025",
    archivePrefix = "arXiv",
    doi = "10.1086/177173",
    journal = "Astrophys. J.",
    volume = "462",
    pages = "563--575",
    year = "1996"
}

@article{Navarro:1996gj,
    author = "Navarro, Julio F. and Frenk, Carlos S. and White, Simon D. M.",
    title = "{A Universal density profile from hierarchical clustering}",
    eprint = "astro-ph/9611107",
    archivePrefix = "arXiv",
    doi = "10.1086/304888",
    journal = "Astrophys. J.",
    volume = "490",
    pages = "493--508",
    year = "1997"
}

@article{Burkert:1995yz,
    author = "Burkert, A.",
    title = "{The Structure of dark matter halos in dwarf galaxies}",
    eprint = "astro-ph/9504041",
    archivePrefix = "arXiv",
    doi = "10.1086/309560",
    journal = "Astrophys. J. Lett.",
    volume = "447",
    pages = "L25",
    year = "1995"
}

@article{Salucci:2000ps,
    author = "Salucci, Paolo and Burkert, Andreas",
    title = "{Dark matter scaling relations}",
    eprint = "astro-ph/0004397",
    archivePrefix = "arXiv",
    doi = "10.1086/312747",
    journal = "Astrophys. J. Lett.",
    volume = "537",
    pages = "L9--L12",
    year = "2000"
}

@article{Jaffe:1983iv,
    author = "Jaffe, W.",
    title = "{A Simple model for the distribution of light in spherical galaxies}",
    journal = "Mon. Not. Roy. Astron. Soc.",
    volume = "202",
    pages = "995--999",
    year = "1983"
}

@article{Tremaine:1993qb,
    author = "Tremaine, Scott and Richstone, Douglas O. and Byun, Yong-Ik and Dressler, Alan and Faber, S. M. and Grillmair, Carl and Kormendy, John and Lauer, Tod R.",
    title = "{A Family of models for spherical stellar systems}",
    eprint = "astro-ph/9309044",
    archivePrefix = "arXiv",
    reportNumber = "NSF-ITP-93-122",
    doi = "10.1086/116883",
    journal = "Astron. J.",
    volume = "107",
    pages = "634",
    year = "1994"
}

@article{Dehnen,
    author = "W. Dehnen",
    title = "{A family of potential-density pairs for spherical galaxies and bulges}",
    journal = "Mon. Not. Roy. Astron. Soc.",
    volumn = "265",
    pages = {250-256},
    year = "1993",
    month = {11},
    doi = {10.1093/mnras/265.1.250}
    
}

@article{Zhao:1995cp,
    author = "Zhao, HongSheng",
    title = "{Analytical models for galactic nuclei}",
    eprint = "astro-ph/9509122",
    archivePrefix = "arXiv",
    reportNumber = "MPA-885",
    doi = "10.1093/mnras/278.2.488",
    journal = "Mon. Not. Roy. Astron. Soc.",
    volume = "278",
    pages = "488--496",
    year = "1996"
}

@article{Dutton:2014xda,
    author = "Dutton, Aaron A. and Macci{\`o}, Andrea V.",
    title = "{Cold dark matter haloes in the Planck era: evolution of structural parameters for Einasto and NFW profiles}",
    eprint = "1402.7073",
    archivePrefix = "arXiv",
    primaryClass = "astro-ph.CO",
    doi = "10.1093/mnras/stu742",
    journal = "Mon. Not. Roy. Astron. Soc.",
    volume = "441",
    number = "4",
    pages = "3359--3374",
    year = "2014"
}

@article{Graham:2005xx,
    author = "Graham, Alister W. and Merritt, David and Moore, Ben and Diemand, Juerg and Terzic, Balsa",
    title = "{Empirical models for Dark Matter Halos. I. Nonparametric Construction of Density Profiles and Comparison with Parametric Models}",
    eprint = "astro-ph/0509417",
    archivePrefix = "arXiv",
    doi = "10.1086/508988",
    journal = "Astron. J.",
    volume = "132",
    pages = "2685--2700",
    year = "2006"
}

@article{Urena-Lopez:2002ptf,
    author = "Urena-Lopez, L. A. and Matos, T. and Becerril, R.",
    title = "{Inside oscillatons}",
    doi = "10.1088/0264-9381/19/23/320",
    journal = "Class. Quant. Grav.",
    volume = "19",
    pages = "6259--6277",
    year = "2002"
}

@article{Harko:2011xw,
    author = "Harko, T.",
    title = "{Bose-Einstein condensation of dark matter solves the core/cusp problem}",
    eprint = "1105.2996",
    archivePrefix = "arXiv",
    primaryClass = "astro-ph.CO",
    doi = "10.1088/1475-7516/2011/05/022",
    journal = "JCAP",
    volume = "05",
    pages = "022",
    year = "2011"
}

@article{Begeman:1991iy,
    author = "Begeman, K. G. and Broeils, A. H. and Sanders, R. H.",
    title = "{Extended rotation curves of spiral galaxies: Dark haloes and modified dynamics}",
    doi = "10.1093/mnras/249.3.523",
    journal = "Mon. Not. Roy. Astron. Soc.",
    volume = "249",
    pages = "523",
    year = "1991"
}

@article{Moore:1999gc,
    author = "Moore, Ben and Quinn, Thomas R. and Governato, Fabio and Stadel, Joachim and Lake, George",
    title = "{Cold collapse and the core catastrophe}",
    eprint = "astro-ph/9903164",
    archivePrefix = "arXiv",
    doi = "10.1046/j.1365-8711.1999.03039.x",
    journal = "Mon. Not. Roy. Astron. Soc.",
    volume = "310",
    pages = "1147--1152",
    year = "1999"
}

@article{Retana-Montenegro:2012dbd,
    author = "Retana-Montenegro, E. and Van Hese, E. and Gentile, G. and Baes, M. and Frutos-Alfaro, F.",
    title = "{Analytical properties of Einasto dark matter haloes}",
    eprint = "1202.5242",
    archivePrefix = "arXiv",
    primaryClass = "astro-ph.CO",
    doi = "10.1051/0004-6361/201118543",
    journal = "Astron. Astrophys.",
    volume = "540",
    pages = "A70",
    year = "2012"
}

@article{Navarro:1994hi,
    author = "Navarro, Julio F. and Frenk, Carlos S. and White, Simon D. M.",
    title = "{Simulations of x-ray clusters}",
    eprint = "astro-ph/9408069",
    archivePrefix = "arXiv",
    doi = "10.1093/mnras/275.3.720",
    journal = "Mon. Not. Roy. Astron. Soc.",
    volume = "275",
    pages = "720--740",
    year = "1995"
}

@article{Schive:2014dra,
    author = "Schive, Hsi-Yu and Chiueh, Tzihong and Broadhurst, Tom",
    title = "{Cosmic Structure as the Quantum Interference of a Coherent Dark Wave}",
    eprint = "1406.6586",
    archivePrefix = "arXiv",
    primaryClass = "astro-ph.GA",
    doi = "10.1038/nphys2996",
    journal = "Nature Phys.",
    volume = "10",
    pages = "496--499",
    year = "2014"
}

@article{Shen:2009my,
    author = "Shen, Juntai and Gebhardt, Karl",
    title = "{The Supermassive Black Hole and Dark Matter Halo of NGC 4649 (M60)}",
    eprint = "0910.4168",
    archivePrefix = "arXiv",
    primaryClass = "astro-ph.CO",
    doi = "10.1088/0004-637X/711/1/484",
    journal = "Astrophys. J.",
    volume = "711",
    pages = "484--494",
    year = "2010"
}

@article{shukirgaliyev2021bound,
    author = "Shukirgaliyev, Bekdaulet and Otebay, A and Sobolenko, M and Ishchenko, M and Borodina, O and Panamarev, T and Myrzakul, S and Kalambay, M and Naurzbayeva, A and Abdikamalov, E and Polyachenko, E and Banerjee, S and Berczik, P and Spurzem, R and Just, A",
    title = "{Bound mass of Dehnen models with a centrally peaked star formation efficiency}",
    journal={A\&A},
    volume={654},
    pages={A53},
    year={2021},
    eprint = "2105.09510 ",
    archivePrefix = "arXiv",
    primaryClass = "astro-ph.GA",
    doi = "10.1051/0004-6361/202141299",
    publisher={EDP Sciences}
}

@article{Pantig:2022whj,
    author = {Pantig, Reggie C. and {\"O}vg{\"u}n, Ali},
    title = "{Dehnen halo effect on a black hole in an ultra-faint dwarf galaxy}",
    eprint = "2202.07404",
    archivePrefix = "arXiv",
    primaryClass = "astro-ph.GA",
    doi = "10.1088/1475-7516/2022/08/056",
    journal = "JCAP",
    volume = "08",
    number = "08",
    pages = "056",
    year = "2022"
}

@article{Gohain:2024eer,
    author = "Gohain, Mrinnoy M. and Phukon, Prabwal and Bhuyan, Kalyan",
    title = "{Thermodynamics and null geodesics of a Schwarzschild black hole surrounded by a Dehnen type dark matter halo}",
    eprint = "2407.02872",
    archivePrefix = "arXiv",
    primaryClass = "gr-qc",
    doi = "10.1016/j.dark.2024.101683",
    journal = "Phys. Dark Univ.",
    volume = "46",
    pages = "101683",
    year = "2024"
}

@article{Al-Badawi:2024asn,
    author = "Al-Badawi, Ahmad and Shaymatov, Sanjar and Sekhmani, Yassine",
    title = "{Schwarzschild black hole in galaxies surrounded by a dark matter halo}",
    eprint = "2411.01145",
    archivePrefix = "arXiv",
    primaryClass = "gr-qc",
    doi = "10.1088/1475-7516/2025/02/014",
    journal = "JCAP",
    volume = "02",
    pages = "014",
    year = "2025"
}

@article{UktamjonUktamov:2025emm,
    author = "Uktamov, Uktamjon and Shaymatov, Sanjar and Ahmedov, Bobomurat and Yuan, Chengxun",
    title = "{New analytical model of static black hole with a dark matter halo and parametric constraints through quasiperiodic oscillations}",
    doi = "10.1140/epjc/s10052-025-15171-0",
    journal = "Eur. Phys. J. C",
    volume = "85",
    number = "12",
    pages = "1432",
    year = "2025"
}

@article{Konoplya:2025ect,
    author = "Konoplya, R. A. and Zhidenko, A.",
    title = "{Dark matter halo as a source of regular black-hole geometries}",
    eprint = "2511.03066",
    archivePrefix = "arXiv",
    primaryClass = "gr-qc",
    doi = "10.1103/7ptp-9j1t",
    journal = "Phys. Rev. D",
    volume = "113",
    number = "4",
    pages = "043011",
    year = "2026"
}

@article{Alloqulov:2025ucf,
    author = "Alloqulov, Mirzabek and Xamidov, Tursunali and Shaymatov, Sanjar and Ahmedov, Bobomurat",
    title = "{Gravitational waveforms from periodic orbits around a Schwarzschild black hole embedded in a Dehnen-type dark matter halo}",
    eprint = "2504.05236",
    archivePrefix = "arXiv",
    primaryClass = "gr-qc",
    doi = "10.1140/epjc/s10052-025-14529-8",
    journal = "Eur. Phys. J. C",
    volume = "85",
    number = "7",
    pages = "798",
    year = "2025"
}

@article{Liang:2025vux,
    author = "Liang, Qi-Qi and Liu, Dong and Long, Zheng-Wen",
    title = "{Quasinormal modes of Schwarzschild black holes in the Dehnen-(1, 4, 5/2) type dark matter halos}",
    eprint = "2505.15540",
    archivePrefix = "arXiv",
    primaryClass = "gr-qc",
    doi = "10.1140/epjc/s10052-025-14850-2",
    journal = "Eur. Phys. J. C",
    volume = "85",
    number = "10",
    pages = "1107",
    year = "2025"
}

@article{Luo:2025xjb,
    author = "Luo, Zuting and Tang, Meirong and Xu, Zhaoyi",
    title = "{Shadows and observational images of a Schwarzschild-like black hole surrounded by a Dehnen-type dark matter halo}",
    eprint = "2505.20115",
    archivePrefix = "arXiv",
    primaryClass = "gr-qc",
    doi = "10.1088/1475-7516/2025/10/065",
    journal = "JCAP",
    volume = "10",
    pages = "065",
    year = "2025"
}

@article{Rani:2025esb,
    author = "Rani, Shamaila and Jawad, Abdul and Heydari-Fard, Malihe and Zafar, Usman",
    title = "{Thermodynamic and shadow analysis of Dehnen type dark matter Halo corrected Schwarzschild black hole surrounded by thin disk}",
    doi = "10.1140/epjc/s10052-025-14388-3",
    journal = "Eur. Phys. J. C",
    volume = "85",
    number = "6",
    pages = "677",
    year = "2025"
}

@article{Xamidov:2025prl,
    author = "Xamidov, Tursunali and Shaymatov, Sanjar and Wu, Qiang and Zhu, Tao",
    title = "{Probing the Schwarzschild black hole immersed in a dark matter halo through astrophysical tests}",
    eprint = "2507.13147",
    archivePrefix = "arXiv",
    primaryClass = "gr-qc",
    journal = "Eur. Phys. J. C",
    volume = "85",
    pages = "1193",
    year = "2025"
}

@article{Ashoorioon:2025ezk,
    author = "Ashoorioon, Amjad and Casadio, Roberto and Jafarzade, Khadije and Jahani Poshteh, Mohammad B. and Luongo, Orlando",
    title = "{Gravitational radiation reaction around a static black hole surrounded by a Dehnen type dark matter halo}",
    eprint = "2509.08569",
    archivePrefix = "arXiv",
    primaryClass = "gr-qc",
    doi = "10.1088/1475-7516/2026/02/007",
    journal = "JCAP",
    volume = "02",
    pages = "007",
    year = "2026"
}

@article{Li:2025ver,
    author = "Li, Zhi and Yu, Jiancheng",
    title = "{Observational properties of a Schwarzschild black hole surrounded by a Dehnen-type dark matter halo}",
    eprint = "2511.13156",
    archivePrefix = "arXiv",
    primaryClass = "astro-ph.HE",
    doi = "10.1140/epjc/s10052-025-14911-6",
    journal = "Eur. Phys. J. C",
    volume = "85",
    number = "10",
    pages = "1170",
    year = "2025"
}

@article{Alloqulov:2025edn,
    author = "Alloqulov, Mirzabek and Abdujabbarov, Ahmadjon and Ahmedov, Bobomurat and Yuan, Chengxun",
    title = "{Can a Dehnen-type dark matter halo affect the neutrino flavor oscillations?}",
    eprint = "2510.20563",
    archivePrefix = "arXiv",
    primaryClass = "gr-qc",
    doi = "10.1140/epjc/s10052-026-15442-4",
    journal = "Eur. Phys. J. C",
    volume = "86",
    pages = "208",
    year = "2026"
}

@article{partII,
    author = "Das, Surajit and Dalui, Surojit and Lee, Bum-Hoon and Cai, Yi-Fu",
    title = "{Extreme-Mass-Ratio Inspirals Embedded in Dark Matter Halo II: Chaos-Induced Gravitational Waves}",
    eprint = "2512.04848",
    archivePrefix = "arXiv",
    primaryClass = "gr-qc",
    year = 2025
}

@book{Van,
    author = "H. Mo and F. van de Bosch and S. White",
    title = "{Galaxy formation and evolution}",
    publisher = "Cambridge University Press, Cambridge, United Kingdom 2010"
}

@article{Bolo,
    author = "Bolokhov, S. V.",
    title = "{Revisiting black holes in dark-matter halos: on consistent solutions to the Einstein equations}",
    eprint = "2512.06930",
    archivePrefix = "arXiv",
    primaryClass = "gr-qc",
    year = 2025
}

@article{Xie:2025udx,
    author = "Xie, Zhong-Ming and Yuan, Hai-Chao and Tang, Yong",
    title = "{On equation of state of dark matter around massive black holes}",
    eprint = "2501.12574",
    archivePrefix = "arXiv",
    primaryClass = "gr-qc",
    doi = "10.1007/s11433-025-2770-8",
    journal = "Sci. China Phys. Mech. Astron.",
    volume = "69",
    number = "1",
    pages = "210412",
    year = "2026"
}

@article{Xu:2018wow,
    author = "Xu, Zhaoyi and Hou, Xian and Gong, Xiaobo and Wang, Jiancheng",
    title = "{Black Hole Space-time In Dark Matter Halo}",
    eprint = "1803.00767",
    archivePrefix = "arXiv",
    primaryClass = "gr-qc",
    doi = "10.1088/1475-7516/2018/09/038",
    journal = "JCAP",
    volume = "09",
    pages = "038",
    year = "2018"
}

@article{Matos:2003nb,
    author = "Matos, Tonatiuh and Nunez, Dario",
    title = "{The general relativistic geometry of the Navarro - Frenk - White model}",
    eprint = "astro-ph/0303594",
    archivePrefix = "arXiv",
    reportNumber = "CIEA-GR-03-35",
    journal = "Rev. Mex. Fis.",
    volume = "51",
    pages = "71--75",
    year = "2005"
}

@article{Azreg-Ainou:2014pra,
    author = "Azreg-A{\"\i}nou, Mustapha",
    title = "{Generating rotating regular black hole solutions without complexification}",
    eprint = "1405.2569",
    archivePrefix = "arXiv",
    primaryClass = "gr-qc",
    doi = "10.1103/PhysRevD.90.064041",
    journal = "Phys. Rev. D",
    volume = "90",
    number = "6",
    pages = "064041",
    year = "2014"
}

@book{Shakeshaft,
    author = "J.R. Shakeshaft",
    title = "{The formation and dynamics of galaxies (International Astronomical Union Symposia, 58)}",
    publisher = "Springer, Germany 1974"
}

@article{Matos:2000ki,
    author = "Matos, Tonatiuh and Guzman, Francisco Siddhartha and Nunez, Dario",
    title = "{Spherical scalar field halo in galaxies}",
    eprint = "astro-ph/0003398",
    archivePrefix = "arXiv",
    reportNumber = "CINVESTAV-FIS-00-28",
    doi = "10.1103/PhysRevD.62.061301",
    journal = "Phys. Rev. D",
    volume = "62",
    pages = "061301",
    year = "2000"
}

@article{Matos:2004je,
    author = "Matos, Tonatiuh and Nunez, Dario and Sussman, Roberto A.",
    title = "{The Spacetime associated with galactic dark matter halos}",
    eprint = "astro-ph/0402157",
    archivePrefix = "arXiv",
    reportNumber = "CIEA-3-02.04, ciea/3-02.04",
    doi = "10.1007/s10714-005-0061-8",
    journal = "Gen. Rel. Grav.",
    volume = "37",
    pages = "769--779",
    year = "2005"
}

@book{Poisson:2009pwt,
    author = "E. Poisson",
    title = "{A Relativist's Toolkit: The Mathematics of Black-Hole Mechanics}",
    publisher = "Cambridge University Press, Cambridge, United Kingdom 2009"
}

@article{Shen:2024qbb,
    author = "Shen, Zibo and Wang, Anzhong and Yin, Shaoyu",
    title = "{Inner radius and energy conditions of dark matter halos surrounding Schwarzschild black holes}",
    eprint = "2408.05417",
    archivePrefix = "arXiv",
    primaryClass = "gr-qc",
    doi = "10.1016/j.physletb.2025.139300",
    journal = "Phys. Lett. B",
    volume = "862",
    pages = "139300",
    year = "2025"
}

@article{Wang:2026whj,
    author = "Wang, Liubin and Wu, Xin",
    title = "{Discussion on the equivalence of two relativistic point-particle Lagrangians}",
    eprint = "2604.10876",
    archivePrefix = "arXiv",
    primaryClass = "gr-qc",
    doi = "10.1140/epjc/s10052-026-15594-3",
    journal = "Eur. Phys. J. C",
    volume = "86",
    number = "4",
    pages = "369",
    year = "2026"
}

@book{paddy,
    author = "T. Padmanabhan",
    title = "{Gravitation: Foundations and frontiers}",
    publisher = "Cambridge University Press, Cambridge, United Kingdom 2014"
}

@article{PG,
    author = "P. Painleve",
    title = "{La mecanique classique et la theorie de relativite}",
    journal = "C.R. Hebd. Seances Acad. Sci.",
    volume = "173",
    pages = "677",
    url = "https://ui.adsabs.harvard.edu/abs/1921CR....173..677P/exportcitation",
    year = "1921"
}

@article{Martel:2000rn,
    author = "Martel, Karl and Poisson, Eric",
    title = "{Regular coordinate systems for Schwarzschild and other spherical space-times}",
    eprint = "gr-qc/0001069",
    archivePrefix = "arXiv",
    doi = "10.1119/1.1336836",
    journal = "Am. J. Phys.",
    volume = "69",
    pages = "476--480",
    year = "2001"
}

@book{pg2,
    author = "T. W Baumgarte and S. L Shapiro",
    title = "{Numerical Relativity: Solving Einstein's Equations on the Computer}",
    publisher = "Cambridge University Press, Cambridge, England 2010"
}

@article{Bera:2021lgw,
    author = "Bera, Avijit and Dalui, Surojit and Ghosh, Subir and Vagenas, Elias C.",
    title = "{Quantum corrections enhance chaos: Study of particle motion near a generalized Schwarzschild black hole}",
    eprint = "2109.00330",
    archivePrefix = "arXiv",
    primaryClass = "gr-qc",
    doi = "10.1016/j.physletb.2022.137033",
    journal = "Phys. Lett. B",
    volume = "829",
    pages = "137033",
    year = "2022"
}

@book{Strogartz,
    author = "S. H Strogatz",
    title = "{Nonlinear Dynamics and Chaos: With Applications to Physics, Biology, Chemistry, and Engineering}",
    publisher = "Chapman and Hall/CRC, Abingdon, Oxon 2024"
}

@book{Stock,
    author = "H. J Stöckmann",
    title = "{Quantum Chaos, An Introduction}",
    publisher = "Cambridge University Press, Cambridge, United Kingdom 1999"
}

@book{Sandri,
    author = "M. Sandri",
    title = "{Numerical calculation of Lyapunov Exponents}",
    publisher = "University of Verona, Italy 1995"
}

@article{Maldacena:2015waa,
    author = "Maldacena, Juan and Shenker, Stephen H. and Stanford, Douglas",
    title = "{A bound on chaos}",
    eprint = "1503.01409",
    archivePrefix = "arXiv",
    primaryClass = "hep-th",
    doi = "10.1007/JHEP08(2016)106",
    journal = "JHEP",
    volume = "08",
    pages = "106",
    year = "2016"
}

@article{Gwak:2022xje,
    author = "Gwak, Bogeun and Kan, Naoto and Lee, Bum-Hoon and Lee, Hocheol",
    title = "{Violation of bound on chaos for charged probe in Kerr-Newman-AdS black hole}",
    eprint = "2203.07298",
    archivePrefix = "arXiv",
    primaryClass = "gr-qc",
    doi = "10.1007/JHEP09(2022)026",
    journal = "JHEP",
    volume = "09",
    pages = "026",
    year = "2022"
}

@article{Zhao:2018wkl,
    author = "Zhao, Qing-Qing and Li, Yue-Zhou and Lu, H.",
    title = "{Static Equilibria of Charged Particles Around Charged Black Holes: Chaos Bound and Its Violations}",
    eprint = "1809.04616",
    archivePrefix = "arXiv",
    primaryClass = "gr-qc",
    doi = "10.1103/PhysRevD.98.124001",
    journal = "Phys. Rev. D",
    volume = "98",
    number = "12",
    pages = "124001",
    year = "2018"
}

@article{Lei:2021koj,
    author = "Lei, Yu-Qi and Ge, Xian-Hui",
    title = "{Circular motion of charged particles near a charged black hole}",
    eprint = "2111.06089",
    archivePrefix = "arXiv",
    primaryClass = "hep-th",
    doi = "10.1103/PhysRevD.105.084011",
    journal = "Phys. Rev. D",
    volume = "105",
    number = "8",
    pages = "084011",
    year = "2022"
}

@article{Kan:2021blg,
    author = "Kan, Naoto and Gwak, Bogeun",
    title = "{Bound on the Lyapunov exponent in Kerr-Newman black holes via a charged particle}",
    eprint = "2109.07341",
    archivePrefix = "arXiv",
    primaryClass = "gr-qc",
    doi = "10.1103/PhysRevD.105.026006",
    journal = "Phys. Rev. D",
    volume = "105",
    number = "2",
    pages = "026006",
    year = "2022"
}

@article{Addazi:2021pty,
    author = "Addazi, Andrea and Capozziello, Salvatore and Odintsov, Sergei",
    title = "{Chaotic solutions and black hole shadow in $f(R)$ gravity}",
    eprint = "2103.16856",
    archivePrefix = "arXiv",
    primaryClass = "gr-qc",
    doi = "10.1016/j.physletb.2021.136257",
    journal = "Phys. Lett. B",
    volume = "816",
    pages = "136257",
    year = "2021"
}

@article{Addazi:2023pfx,
    author = "Addazi, Andrea and Capozziello, Salvatore",
    title = "{Black hole shadow and chaos bound violation in f(T) teleparallel gravity}",
    eprint = "2303.01956",
    archivePrefix = "arXiv",
    primaryClass = "gr-qc",
    doi = "10.1016/j.physletb.2023.137828",
    journal = "Phys. Lett. B",
    volume = "839",
    pages = "137828",
    year = "2023"
}

@article{Shen:2023erj,
    author = "Shen, Zibo and Wang, Anzhong and Gong, Yungui and Yin, Shaoyu",
    title = "{Analytical models of supermassive black holes in galaxies surrounded by dark matter halos}",
    eprint = "2311.12259",
    archivePrefix = "arXiv",
    primaryClass = "gr-qc",
    doi = "10.1016/j.physletb.2024.138797",
    journal = "Phys. Lett. B",
    volume = "855",
    pages = "138797",
    year = "2024"
}

@article{Cardoso:2021wlq,
    author = "Cardoso, Vitor and Destounis, Kyriakos and Duque, Francisco and Macedo, Rodrigo Panosso and Maselli, Andrea",
    title = "{Black holes in galaxies: Environmental impact on gravitational-wave generation and propagation}",
    eprint = "2109.00005",
    archivePrefix = "arXiv",
    primaryClass = "gr-qc",
    doi = "10.1103/PhysRevD.105.L061501",
    journal = "Phys. Rev. D",
    volume = "105",
    number = "6",
    pages = "L061501",
    year = "2022"
}

@article{Kamenshchik:2001cp,
    author = "Kamenshchik, Alexander Yu. and Moschella, Ugo and Pasquier, Vincent",
    title = "{An Alternative to quintessence}",
    eprint = "gr-qc/0103004",
    archivePrefix = "arXiv",
    doi = "10.1016/S0370-2693(01)00571-8",
    journal = "Phys. Lett. B",
    volume = "511",
    pages = "265--268",
    year = "2001"
}

@article{Bento:2002ps,
    author = "Bento, M. C. and Bertolami, O. and Sen, A. A.",
    title = "{Generalized Chaplygin gas, accelerated expansion and dark energy matter unification}",
    eprint = "gr-qc/0202064",
    archivePrefix = "arXiv",
    doi = "10.1103/PhysRevD.66.043507",
    journal = "Phys. Rev. D",
    volume = "66",
    pages = "043507",
    year = "2002"
}

@article{Scherrer:2004au,
    author = "Scherrer, Robert J.",
    title = "{Purely kinetic k-essence as unified dark matter}",
    eprint = "astro-ph/0402316",
    archivePrefix = "arXiv",
    doi = "10.1103/PhysRevLett.93.011301",
    journal = "Phys. Rev. Lett.",
    volume = "93",
    pages = "011301",
    year = "2004"
}

@article{Dymnikova:2015yma,
    author = "Dymnikova, Irina and Khlopov, Maxim",
    title = "{Regular black hole remnants and graviatoms with de Sitter interior as heavy dark matter candidates probing inhomogeneity of early universe}",
    eprint = "1510.01351",
    archivePrefix = "arXiv",
    primaryClass = "gr-qc",
    doi = "10.1142/S0218271815450029",
    journal = "Int. J. Mod. Phys. D",
    volume = "24",
    number = "13",
    pages = "1545002",
    year = "2015"
}

@article{Balakin:2003tk,
    author = "Balakin, Alexander B. and Pavon, Diego and Schwarz, Dominik J. and Zimdahl, Winfried",
    title = "{Curvature force and dark energy}",
    eprint = "astro-ph/0302150",
    archivePrefix = "arXiv",
    doi = "10.1088/1367-2630/5/1/385",
    journal = "New J. Phys.",
    volume = "5",
    pages = "85",
    year = "2003"
}

@article{Sahni:2002dx,
    author = "Sahni, Varun and Shtanov, Yuri",
    title = "{Brane world models of dark energy}",
    eprint = "astro-ph/0202346",
    archivePrefix = "arXiv",
    doi = "10.1088/1475-7516/2003/11/014",
    journal = "JCAP",
    volume = "11",
    pages = "014",
    year = "2003"
}

@article{Gondolo:1999ef,
    author = "Gondolo, Paolo and Silk, Joseph",
    title = "{Dark matter annihilation at the galactic center}",
    eprint = "astro-ph/9906391",
    archivePrefix = "arXiv",
    reportNumber = "MPI-PHT-99-10, OUAST-99-9",
    doi = "10.1103/PhysRevLett.83.1719",
    journal = "Phys. Rev. Lett.",
    volume = "83",
    pages = "1719--1722",
    year = "1999"
}

@article{Sadeghian:2013laa,
    author = "Sadeghian, Laleh and Ferrer, Francesc and Will, Clifford M.",
    title = "{Dark matter distributions around massive black holes: A general relativistic analysis}",
    eprint = "1305.2619",
    archivePrefix = "arXiv",
    primaryClass = "astro-ph.GA",
    doi = "10.1103/PhysRevD.88.063522",
    journal = "Phys. Rev. D",
    volume = "88",
    number = "6",
    pages = "063522",
    year = "2013"
}

@article{Speeney:2022ryg,
    author = "Speeney, Nicholas and Antonelli, Andrea and Baibhav, Vishal and Berti, Emanuele",
    title = "{Impact of relativistic corrections on the detectability of dark-matter spikes with gravitational waves}",
    eprint = "2204.12508",
    archivePrefix = "arXiv",
    primaryClass = "gr-qc",
    doi = "10.1103/PhysRevD.106.044027",
    journal = "Phys. Rev. D",
    volume = "106",
    number = "4",
    pages = "044027",
    year = "2022"
}
\end{document}